\pdfoutput=1 
\RequirePackage{amsmath}
\documentclass[runningheads]{llncs}
\usepackage[svgnames]{xcolor}
\usepackage{longtable}

\usepackage{graphicx}
\usepackage{vdmlisting}
\usepackage{hyperref}
\usepackage{times}
\usepackage{url}
\usepackage{tcolorbox}

\lstdefinelanguage{html}{ 
  backgroundcolor={\color[gray]{1}},
  basicstyle=\small\ttfamily,
  morekeywords={code}, 
  sensitive=true, 
  morestring=[s]{"}{"}, 
  style=HtmlStyle 
} 
\lstdefinestyle{HtmlStyle}{ 
}

\usepackage{colortbl}

\newcommand{\eg}{e.\,g.,\ }
\newcommand{\ie}{i.\,e.,\ }

\usepackage{soul}

\usepackage[acronym]{glossaries} 

\usepackage{chngcntr}

\usepackage[capitalise,nameinlink]{cleveref}
\crefname{section}{Sect.}{Sect.}
\Crefname{section}{Section}{Sections}
\crefname{lstlisting}{Listing}{Listing}
\Crefname{lstlisting}{Listing}{Listing}

\usepackage{listings} 
\usepackage{times}    

\lstdefinelanguage{json}{
    basicstyle=\ttfamily\small, 
    numbers=left,
    stepnumber=1,
    numbersep=8pt,
    breaklines=true,
    frame=single,
    xleftmargin=.11\textwidth, 
    xrightmargin=.11\textwidth
}

\definecolor{stringcolor}{rgb}{0.06, 0.10, 0.98} 
\lstdefinelanguage{Gherkin}{
    morekeywords={Feature, Scenario, Outline, Given,
        When, Then,
        Examples  },
    sensitive=true, 
    morecomment=[l]{//}, 
    morecomment=[s]{/*}{*/}, 
    morestring=[b]", 
    keywordstyle=\bfseries\color{blue},
    stringstyle=\ttfamily\color{red!50!brown}, 
} %

\lstdefinelanguage{vdmpp}
  {morekeywords={RESULT,\#act,\#active,\#else,\#endif,\#fin,\#ifdef,\#ifndef,\#req,\#waiting,abs,all,always,and,async,atomic,be,be st,bool,by,card,cases,char,class,comp,compose,conc,dcl,def,dinter,div,do,dom,dunion,elems,else,elseif,end,
  eq,error,errs,exists,exists1,exit,ext,false,floor,for,for all,forall,from,functions,hd,if,in,in set,inds,init,inmap,instance,instance variables,int,inter,inv,inverse,iota,is,is not yet specified,is subclass of,is subclass responsibility,is\_,isofbaseclass,isofclass,lambda,len,let,map,measure,merge,mk\_,mod,mu,munion,mutex,narrow\_,
  nat,nat1,new,nil,not,not in set,obj\_,of,operations,or,ord,others,per,periodic,post,power,pre,private,protected,psubset,public,pure,rat,rd,real,rem,
  responsibility,return,reverse,rng,samebaseclass,sameclass,self,seq,seq1,set,set1,skip,specified,sporadic,st,start,
  startlist,static,stop,stoplist,subclass,subset,sync,then,thread,threadid,tixe,tl,to,token,traces,trap,true,types,undefined,
  union,values,variables,while,with,wr,yet
 },
   sensitive,
   morecomment=[l]--,
   morestring=[b]",
   morestring=[b]',
  }[keywords,comments,strings]
\usepackage{nicematrix}
\usepackage{adjustbox}
\usepackage{tcolorbox}
\tcbuselibrary{skins,breakable}
\usepackage{tikz}
\usetikzlibrary{calc}
\newenvironment{codebox}{%
    \tcolorbox[%
    empty,
    parbox=false,
    noparskip,
    enhanced,
    boxrule=0.7pt,
    no shadow,
    sharp corners,
    colback=white,
    left=.03in, 
    top=-.04in, bottom=-.04in,
    before skip=.2in,
    after skip=.2in]
}{\endtcolorbox}

\definecolor{plantgreen}{rgb}{0.97,1,0.85}

\definecolor{codegreen}{rgb}{0,0.6,0}
\definecolor{codegray}{rgb}{0.5,0.5,0.5}
\definecolor{codeorange}{rgb}{0.75,0.5,0}
\definecolor{codepurple}{rgb}{0.68,0,0.62}
\lstdefinestyle{mystyle}{
    commentstyle=\color{codegreen},
    keywordstyle=\color{codepurple},
    numberstyle=\tiny\color{codegray},
    stringstyle=\color{codeorange},
    basicstyle=\ttfamily\footnotesize,
    breakatwhitespace=false,         
    breaklines=true,                 
    captionpos=b,                    
    keepspaces=true,                 
    numbersep=5pt,                  
    showspaces=false,                
    showstringspaces=false,
    showtabs=false,                  
    tabsize=2
}
\lstdefinestyle{consolestyle}{
    backgroundcolor=\color{black},   
    commentstyle=\color{white},
    keywordstyle=\color{white},
    numberstyle=\tiny\color{white},
    stringstyle=\color{white},
    basicstyle=\color{white}\ttfamily\scriptsize,
    breakatwhitespace=false,         
    breaklines=true,                 
    captionpos=b,                    
    keepspaces=true,                 
    numbersep=5pt,                  
    showspaces=false,                
    showstringspaces=false,
    showtabs=false,                  
    tabsize=2
}
\definecolor{darkblue}{rgb}{0.0,0.0,0.6}
\definecolor{cyan}{rgb}{0.0,0.6,0.6}
\lstdefinelanguage{XML}
{
  morestring=[b]",
  morestring=[s]{>}{<},
  morecomment=[s]{<?}{?>},
  stringstyle=\color{black},
  identifierstyle=\color{darkblue},
  keywordstyle=\color{cyan},
  morekeywords={xmlns,version,type}
}

\usepackage[capitalise,nameinlink]{cleveref}
\crefname{section}{Sect.}{Sect.}
\Crefname{section}{Section}{Sections}
\crefname{lstlisting}{Listing}{Listing}
\Crefname{lstlisting}{Listing}{Listing}

\usepackage{subcaption}

\usepackage{graphicx}
\usepackage{colortbl}
\usepackage{algorithm}
\usepackage{algpseudocode}
\usepackage{pseudocode}

\usepackage{tikz}
\usetikzlibrary{shapes,fit,arrows,calc,arrows,arrows.meta,positioning,decorations.text}

\usepackage{pgfplots}
\pgfplotsset{compat=newest}
\usetikzlibrary{pgfplots.groupplots, pgfplots.units}
\pgfplotsset{
	every axis label/.append style={font=\normalsize},
	tick label style={font=\small},
	/pgfplots/enlargelimits=false,
    legend style={legend pos=north east, font=\small},
    legend cell align=left,
    xlabel near ticks,
    ylabel near ticks,
	axis on top,
    highlight/.code args={#1:#2}{
        \fill [every highlight] ({axis cs:#1,0}|-{rel axis cs:0,0}) rectangle ({axis cs:#2,0}|-{rel axis cs:0,1});
    },
    /tikz/every highlight/.style={
        on layer=\pgfkeysvalueof{/pgfplots/highlight layer},
        red!10
    },
    /tikz/highlight style/.style={
        /tikz/every highlight/.append style=#1
    },
    highlight layer/.initial=axis background
}%
\definecolor{maroon}{rgb}{0.5,0,0}
\definecolor{darkgreen}{rgb}{0,0.5,0}
\definecolor{ao}{rgb}{0.0, 0.5, 0.0}
\definecolor{mycolor1}{rgb}{0.0, 0.53, 0.74}%
\definecolor{mycolor2}{rgb}{0.21783,0.72504,0.61926}%
\definecolor{mycolor4}{rgb}{0.93, 0.53, 0.18}%
\definecolor{plum}{rgb}{0.56, 0.27, 0.52}
\definecolor{pinegreen}{rgb}{0.0, 0.47, 0.44}
\definecolor{pthaloblue}{rgb}{0.0, 0.06, 0.54}
\definecolor{saffron}{rgb}{0.96, 0.77, 0.19}

\usepackage{cprotect}
\usepackage{multirow}
\usepackage{graphicx}
\usepackage[T1]{fontenc}
\usepackage{hyperref}
\usepackage{xspace}
\usepackage{paralist}
\usepackage{fancybox}
\usepackage{calc}
\usepackage{verbatim}
\usepackage{marginnote}
\usepackage{todonotes}

\usepackage{booktabs}
\usepackage[scaled]{helvet}

\usepackage{times}

\usepackage{listingsVDM}

\definecolor{mygray}{rgb}{0.5,0.5,0.5}

\newcommand{\keywords}[1]{\par\addvspace\baselineskip
  \noindent\keywordname\enspace\ignorespaces#1}

\usepackage{listings}

\colorlet{punct}{red!60!black}
\definecolor{background}{HTML}{EEEEEE}
\definecolor{delim}{RGB}{20,105,176}
\colorlet{numb}{magenta!60!black}

\usepackage{relsize}
\usepackage{xspace}
\newcommand\CC{C\nolinebreak[4]\hspace{-.05em}\raisebox{.4ex}{\relsize{-3}{\textbf{++}}}\xspace}

\lstdefinelanguage{json}{
	basicstyle=\scriptsize\ttfamily,
	numbers=left,
	numberstyle=\scriptsize,
	stepnumber=1,
	numbersep=8pt,
	showstringspaces=false,
	breaklines=true,
	frame=lines,
	backgroundcolor=\color{white},
	literate=
	*{0}{{{\color{numb}0}}}{1}
	{1}{{{\color{numb}1}}}{1}
	{2}{{{\color{numb}2}}}{1}
	{3}{{{\color{numb}3}}}{1}
	{4}{{{\color{numb}4}}}{1}
	{5}{{{\color{numb}5}}}{1}
	{6}{{{\color{numb}6}}}{1}
	{7}{{{\color{numb}7}}}{1}
	{8}{{{\color{numb}8}}}{1}
	{9}{{{\color{numb}9}}}{1}
	{:}{{{\color{punct}{:}}}}{1}
	{,}{{{\color{punct}{,}}}}{1}
	{\{}{{{\color{delim}{\{}}}}{1}
	{\}}{{{\color{delim}{\}}}}}{1}
	{[}{{{\color{delim}{[}}}}{1}
	{]}{{{\color{delim}{]}}}}{1},
}

\definecolor{native}{named}{LightCoral}

\begin{document}

\makeatletter

\let\origcite\cite
\let\origbibitem\bibitem
\let\origref\ref
\let\origpageref\pageref
\let\origlabel\label
\let\origgls\gls
\let\origacronym\newacronym
\newcommand\locallabels[1]{%
    \renewcommand\cite[1]{%
        \foreach \i in {##1}
        {%
        \origcite{#1\i}%
        }}%
  \renewcommand\bibitem[1]{\origbibitem{#1##1}}%
  \renewcommand\label[1]{\origlabel{#1##1}}%
  \renewcommand\ref[1]{\origref{#1##1}}%
  \renewcommand\pageref[1]{\origpageref{#1##1}}%
  \renewcommand\gls[1]{\origgls{#1##1}}%
  \renewcommand\newacronym[3]{\origacronym{#1##1}{##2}{##3}}%
}

\makeatother
\title{Proceedings of the 20$^{th}$ International Overture Workshop}
\institute{}
\author{Hugo Daniel Macedo \and Ken Pierce (Editors)}
\authorrunning{}   
%

\maketitle

\setcounter{page}{1}

\chapter*{Preface}
\markboth{Preface}{Preface}

The 20$^{th}$ in the ``Overture'' series of workshops on the Vienna Development
Method (VDM), associated tools and applications was held as a hybrid event both
online at in person at Aarhus University on July 5, 2022. VDM is one of the
longest established formal methods, and yet has a lively community of
researchers and practitioners in academia and industry grown around the
modelling languages (VDM-SL, VDM++, VDM-RT) and tools (VDM VSCode Extension, VDMTools,
VDMJ, ViennaTalk, Overture, Crescendo, Symphony, and the INTO-CPS chain).
Together, these provide a platform for work on modelling and analysis
technology that includes static and dynamic analysis, test generation,
execution support, and model checking.

Research in VDM is driven by the need to precisely describe systems. In order
to do so, it is also necessary for the associated tooling to serve the current
needs of researchers and practitioners and therefore remain up to date. The
20$^{th}$ Workshop reflected the breadth and depth of work supporting and applying
VDM. This technical report includes two papers on translating requirements into
models. Then, two works on the tooling supporting the methodology. The last
paper is related to Cyber-Physical Systems and design space exploration. 

In addition to the talks related to the papers here published, the workshop
included an invited talk on VDMJ by Nick Battle and a round table discussion.
The topic of the invited talk was an introduction to the VDMJ extension plugin
architecture and its envisaged evolution. The talk provided insight for the
many newcomers interested in extending it, and it sparked the discussion for
the round table, where educational concerns and users views were 
discussed.

We would like to thank the authors, PC members, reviewers and participants for
their help in making this a valuable and successful workshop, and we look forward
together to meeting once more in 2023.

\medskip
\begin{flushright}\noindent
Hugo Daniel Macedo, Aarhus\\
Ken Pierce, Newcastle\\
\end{flushright}

\tableofcontents

\chapter*{Organization}

\section*{Programme Committee}
\begin{longtable}{p{0.3\textwidth}p{0.7\textwidth}}
Alessandro Pezzoni & Anaplan, UK\\[12pt]

Fuyuki Ishikawa & National Institute of Informatics, Japan\\[12pt]

Hugo Daniel Macedo & Aarhus University, Denmark\\[12pt]

Keijiro Araki & National Institute of Technology, Kumamoto College, Japan\\[12pt]

Ken Pierce & Newcastle University, UK\\[12pt]

Kenneth G. Lausdahl & AGROCorp International and Aarhus University, Denmark\\[12pt]

Marcel Verhoef & European Space Agency, The Netherlands\\[12pt]

Mirgita Frasheri & Aarhus University, Denmark\\[12pt]

Nick Battle & Newcastle University, UK\\[12pt]

Paolo Masci & National Institute of Aerospace (NIA), USA\\[12pt]

Peter Gorm Larsen & Aarhus University, Denmark\\[12pt]

Tomo Oda & Software Research Associate Incorporated, Japan\\[12pt]

Victor Bandur & McMaster University, Canada

\end{longtable}

\mainmatter              

\titlerunning{20th Overture Workshop, 2022}

\setcounter{page}{5}

\clearpage

\begingroup
\renewcommand\theHchapter{1-Oda:\thechapter}
\renewcommand\theHsection{1-Oda:\thesection}
\locallabels{1-Oda:}
\setcounter{footnote}{0}
\setcounter{chapter}{0}
\setcounter{lstlisting}{0}

\makeatletter
\def\input@path{{1-Oda/}}
\makeatother

\graphicspath{{1-Oda}}

\title{VDM-SL in action: A FRAM-based approach to contextualise formal specifications}


\author{
Tomohiro Oda\inst{1} \and 
Shigeru Kusakabe\inst{2} \and
Han-Myung Chang\inst{3} \and
Peter Gorm Larsen\inst{4}
}

\authorrunning{ }

\institute{Software Research Associates, Inc.~(\email{tomohiro@sra.co.jp})
\and University of Nagasaki~(\email{kusakabe@sun.ac.jp})
\and Nanzan University~(\email{chang@nanzan-u.ac.jp})
\and Aarhus University, DIGIT, Department of Electrical and Computer Engineering,~(\email{pgl@ece.au.dk}) 
}

\maketitle

\begin{abstract}
Software development is a collaborative effort by stakeholders from different domains of expertise and knowledge.
Developing a useful system needs to take different views on the system's functionalities into account from the system's internal properties, the individual user's tasks at hand, and an organisation's mission as a whole.
Participation of a wide range of individual and organisational stakeholders is crucial in the specification phase.
This paper proposes a use of the FRAM diagram as a pre-formal notation to describe the system's roles in individual users' tasks and organisational activities.
A tool that links public operations in VDM-SL and activities in FRAM is introduced to demonstrate and discuss how a FRAM diagram contextualises a formal notation.
\end{abstract}

\section{Background}
\label{sec:Background}

Software development is often driven by  multidisciplinary collaboration 
and the developed system is also often operated by people from different areas of expertise. 
The specification phase aims to explicitly define what the system to be developed shall be able to perform.
A rigorous specification provides a basis to create a software system that works as planned.
However, just working as planned is not enough to be  useful.
The plan needs to fit with various activities of stakeholders that may vary over time.
Participation of a wide range of stakeholders is thus crucial to bring out an appropriate plan.

The stakeholders desire to foresee how the system to be developed will work in their activities.
It is important for successful software development that the system's functionalities are understood by the stakeholders.
Although a rigorous specification in a formal notation, including VDM-SL~\cite{Larsen&13b}, concisely defines the system's functionalities without ambiguity, it is difficult for many stakeholders to understand how the specified system will work in their activities.
One difficulty is the notation.
Although the rapid spread of programming education may relax this difficulty, reading fluency of formal notation requires a certain time and effort.
Another difficulty is contextualisation.
The set of functionalities defined in the specification needs to make sense in each stakeholder's own expertise.
The formal specification rigorously defines a functionality with inputs, outputs and internal states without ambiguity.
However, the inputs, outputs and internal states are defined in terms of the system's internal structure and properties although they need to be grounded and explained in the stakeholder's activities.

The third difficulty is variability.
One obvious source of variability comes from the user.
A user, or a group of people, sometimes performs unexpected actions on the system, or they do not perform expected actions in time.
An external server also may fail to respond in time.
The variability is not only from failures but also from good reasons.
Data may become available earlier than expected because the person in charge of the previous action has outperformed the expectation.
The data may need tentative storage until the originally expected time.
Such variability emerges from actions happening out of the system, and is needed to understand how the system will work in action with foreseen variabilities.

The authors has been addressing the difficulty of notation using specification animation.
Specification animation is a technique to simulate the execution of the specified system using interpreters.
The stakeholders can preview the functionality of the system before the implementation without understanding the details in the formal specification.
The animation-based approach, however, does not solve the difficulty of contextualisation.

In this paper, a FRAM-based approach to define and understand the contextual information of the system's functionalities.
FRAM (Functional Resonance Analysis Method) will be briefly explained in Section \ref{sec:FRAM}.
Section \ref{sec:Related-Work} explains existing use of FRAM in the requirement elicitation.
Section \ref{sec:FRAM-and-VDMSL} proposes our approach to develop formal specifications in VDM-SL with pre-formal analysis using FRAM.
An example development of review bidding system will be explained in Section \ref{sec:Example} and discussed in Section \ref{sec:Discussions}.
Section \ref{sec:Concluding-Remarks} will conclude this paper.

\section{Function Resonance Analysis Method}
\label{sec:FRAM}

FRAM (Function Resonance Analysis Method)~\cite{Hollnagel17,Hollnagel18} was originally introduced by E.  Hollnagel to unfold how a complex, dynamic and socio-technical system works and its anticipated variability in operations.
The basic idea of FRAM is to identify functions or activities in a collaborative system, and analyse influences between them based on six aspects.
To avoid confusion with VDM-SL's functions, we denote {\em activity} instead of function.

Activities can be categorised into three types according to their performers: organisational activities, human activities, and technological activities.
organisational activities are those carried out by a group of people.
Human activities include,  but are not limited to, the end users' tasks at hand.
Series of activities that directly or indirectly influences the system's activities are also identified and included in human activities.
Technological activities are those performed by the developing system and also existing systems.
The difference among the three types is often apparent in their variability.
Human activities often vary more significantly from their original plans than technological activities.
Organisational activities may vary even further than human activities.
To develop a useful system, such variabilities need to be taken into account.

Once activities are identified, each of them is analysed with regard to six aspects as follows:

\begin{enumerate}
\item Input: what the activity processes the input, or what starts the activity.
\item Precondition: what must be done before the activity is performed.
\item Resource: what the activity needs or consumes.
\item Time: when the activity starts and/or finishes.
\item  Control: what monitors, supervises or regulates the process.
\item Output: what the activity yields.
\end{enumerate}

Input, precondition, resource, time and control are aspects that may receive influences of other activities while output may give influence to others. 
Functional couplings among activities are identified by matching the name of aspects between the first five aspects of an activity and outputs of other activities.

For example, assume that code review is mandatory after coding and a quality control team aggregates review reports from reviews.
The quality control team then supplies review advise for subsequent code reviews.
In this scenario, coding, code review and quality control are identified as activities.
The coding activity has an output called source code, which is also input of code review.
Similarly, a review report is output of  the code review activity, and it is also input of the quality control activity.
The quality control activity outputs review advisory, which controls the code review activity.
Fig.~\ref{fig:FRAM-code-review} shows a FRAM diagram that visualises the identified activities and functional couplings among them.

\begin{figure}
\begin{center}
\includegraphics[width=0.9\textwidth]{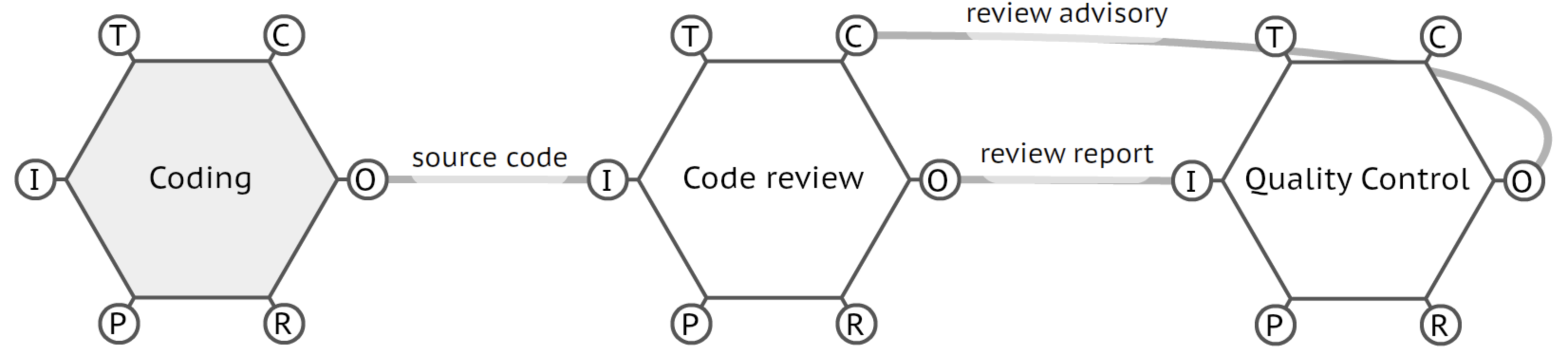}
\end{center}
\caption{An example FRAM diagram of code review}
\label{fig:FRAM-code-review}
\end{figure}

In Fig.~\ref{fig:FRAM-code-review}, participants in the development can see that code review should be scheduled after writing source code, and may required to revise the code.
For the coder, writing a code always needs code review, and scheduling a review session might be a bottleneck of coding in the long run.
The quality control team does not directly discuss coding issues with the coders because they send review advise to reviewers for subsequent use.

\begin{figure}
\begin{center}
\includegraphics[width=0.9\textwidth]{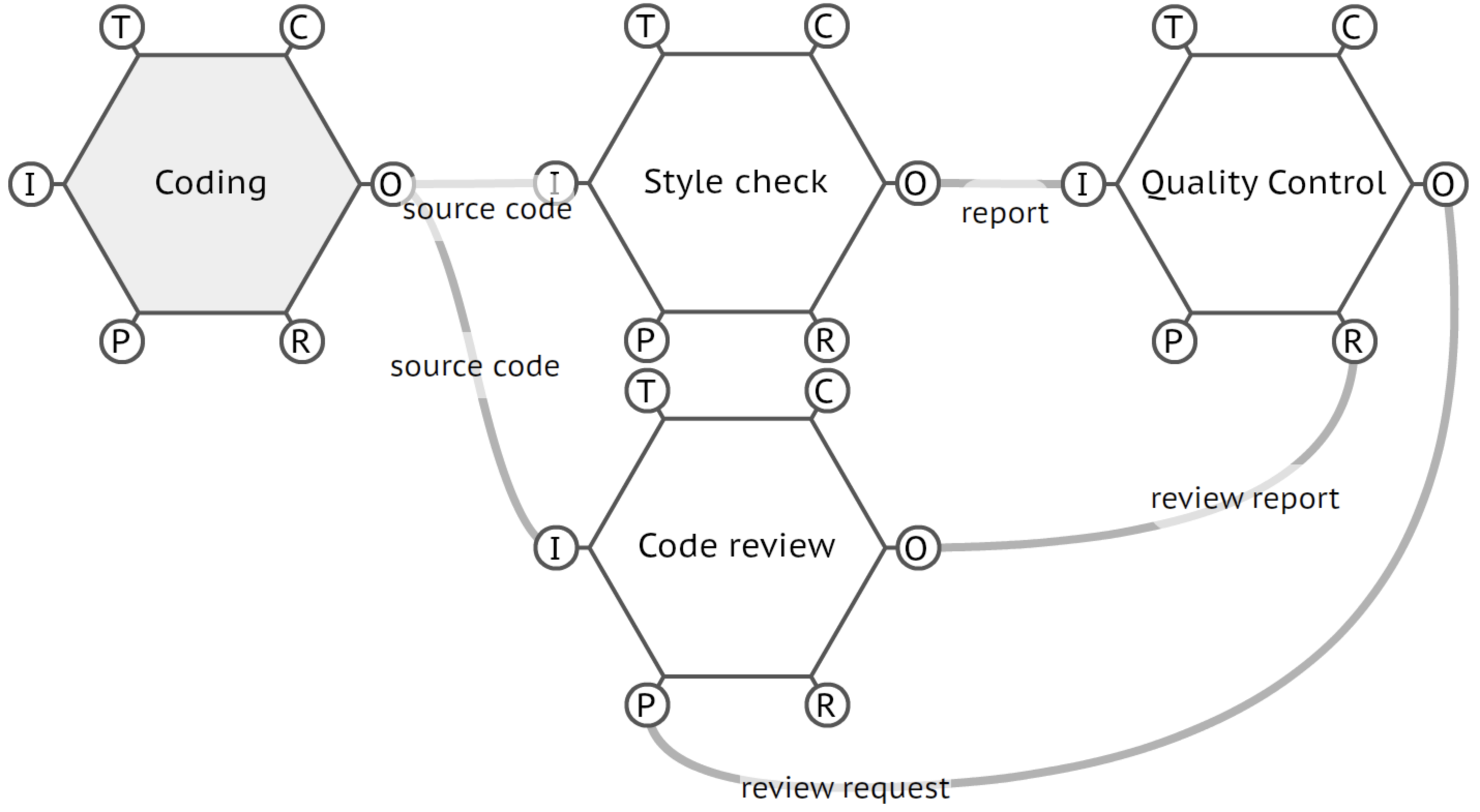}
\end{center}
\caption{An example FRAM diagram of code review with style checker}
\label{fig:FRAM-style-checker}
\end{figure}

Fig. \ref{fig:FRAM-style-checker} illustrates the modified code review by introducing an automated style checker driven by the Continuous Integration (CI) server.
With the FRAM diagram shown in Fig. \ref{fig:FRAM-style-checker},  the developers can see how the style checker will be used in the development project.
The style checker analyses the source code and creates a report that identifies code-smells.
The quality control team monitors the reports and issues review requests for suspicious code with code-smells.
Please note that the review request is a precondition of the code review activity, which means that a code review session starts only if the quality control team requests it.
For the coders, the style checker is not solely a quick code review, but rather it eliminates the delay caused by scheduling of review sessions.
By introducing the style checker, the coders will not need to wait for code review sessions.
The quality control team can use the style checker as a means to directly communicate with the coders through review requests.

The FRAM diagram does not provide detailed information and control structure within the collaborated activities.
It instead illustrates a context of the collaborations that the system to be developed will participate in.
In conjunction with formal specification, the stakeholders can preview how the system will work before the implementation.
Our approach to use FRAM diagram along with VDM-SL specification will be described in the section below.

\section{Related Work}
\label{sec:Related-Work}

De Carvalho introduced four phases to elicit functional and non-functional requirements using FRAM~\cite{deCarvalho21}.
Experiments to elicit functional and non-functional requirements using FRAM and BPMN (Business Process Modeling Notation) were conducted,  and the use of FRAM for identifying was confirmed promising.
Our approach is based on the four phases performed in the experiment, namely contextualisation, aggregation, transformation, and specification.

While stakeholders requirements elicitation phase in general starts with identifying stakeholders and collecting requirements to the system, the first two phases in the experiment were to analyse how the stakeholders collaborate without the system, followed by the last two phases to address the found difficulties.
In the contextualisation phase,  the scope of the domain area, work processes, business goals are analysed by ethnographic observation, interviews and collecting documents.
A catalog of activities in the business is produced in the contextualisation phase.
In the aggregation process, the catalog of activities is further analysed in the perspective of solutions.
Functional coupling between activities and potential difficulties in the activities are identified.
Transformation is the phase to produce a solution to address difficulties in the current Work-as-Done.
A list of expected activities to solve the difficulties are created.
In the specification phase,  main features and restrictions in a natural language are identified to realise the expected activities identified in the transformation phase.
The four phases go from the top to the bottom in a one-way manner.

Our approach is to produce a formal specification in VDM-SL from the FRAM diagrams created in the transformation phase.
We expect the formal specification will reveal unidentified activities and hidden couplings among activities, and such feedback should be reflected to the transformation phase to form a cyclic process to develop the specification.
Formal engineers can also use the FRAM diagram to produce walk-through scenarios to test the formal specification.

\section{FRAM diagram to VDM-SL specification}
\label{sec:FRAM-and-VDMSL}

FRAM was originally developed for analysing complex, dynamic and socio-technical systems.
A FRAM diagram depicts how organisations, individuals and technological entities including computer systems work together, and its applications include requirements elicitation~\cite{deCarvalho21}.
This paper proposes to use FRAM diagrams as a pre-formal notation to identify functionalities of the system, and specifies them in VDM-SL in a traceable manner.
By {\em pre-formal}, we mean documents whose parts or the whole would later be specified in a formal notation.
A pre-formal document could be an input to the specification phase, a reason of practical validity of the specification, or a supplemental explanation to the formal specification.
Our approach takes the following steps.
\begin{enumerate}
\item Modeling WAD (Work as Done) in a FRAM diagram that describes how the work is currently carried out.  Identify which activities need the support of a new system.  If the work is not present yet, start with the next step.
\item Modeling transformation in a FRAM diagram that describes how the work will be carried out with the support of a new system. Introduce new technological activities that the new system will provide.
\item Specifying the new system in VDM-SL that corresponds to the technological activities. Use annotations to place traceability information to the FRAM diagram.
\item Providing feedback.  Specify extra activities discovered or emerged in the VDM-SL specification, and place them in the FRAM diagram.
\end{enumerate}

The first two steps correspond to the stakeholders requirements elicitation.
While stakeholders requirements elicitation, in general, starts with collecting requirements for the system from stakeholders, the first step of our approach is to analyse how the stakeholders collaborate.
Then, in the second step, the difficulties found in the first step will then be addressed by introducing a system.

The last two steps form a loop, and thus the traceability between an activity in FRAM and its counterpart in VDM-SL is important.
Annotations are a mechanism to embed additional information into a VDM-SL specification\cite{Lausdahl&17}\cite{Battle&20}.
An annotation takes the form of {\tt --@} which is processed as a comment by interpreters by default but can also be processed by tools that support annotations.
Table \ref{table:annotations} shows the forms of annotations that we introduced for traceability with FRAM.
The annotations for FRAM traceability are implemented in ViennaTalk~\cite{Oda&17a}.
From a FRAM diagram, ViennaTalk can generate the annotations with a template definition of an operation for each activity.
A specification with FRAM annotations can also be merged into an existing FRAM diagram as feedback.

\begin{table}
\caption{Annotations for traceability with FRAM}
\label{table:annotations}
\begin{tabular}{l l l | l }
prefix & identifier & free text & description\\
\hline
{\tt --@FRAM} & {\tt Function} & {\it activity name} & specifies the name of a corresponding activity\\
{\tt --@FRAM} & {\tt Input} & {\it aspect name} & specifies the name of corresponding aspect\\
{\tt --@FRAM} & {\tt Precondition} & {\it aspect name} &  specifies the name of corresponding aspect\\
{\tt --@FRAM} & {\tt Resource} & {\it aspect name} &  specifies the name of corresponding aspect\\
{\tt --@FRAM} & {\tt Time} & {\it aspect name} &  specifies the name of corresponding aspect\\
{\tt --@FRAM} & {\tt Control} & {\it aspect name} &  specifies the name of corresponding aspect\\
{\tt --@FRAM} & {\tt Output} & {\it aspect name} &  specifies the name of corresponding aspect\\
\end{tabular}
\end{table}

\section{Example: Paper review}
\label{sec:Example}
In this section, we describe an experimental development of a support tool for academic paper review.
Assume that an academic community wants to build a system that supports the paper review.
For each submitted paper, a PC chair assigns a certain number of reviewers to evaluate the quality of the paper and makes the final call to either accept or reject.
Some candidate reviewers are registered.
A reviewer may receive a review request from the PC chair, read the paper, and send a review report to the PC chair.
If the review reports of a paper do not agree, the reviewers will start a discussion, and the PC chair is responsible to make the final call.
In the rest of this section, we follow the four steps described in Section \ref{sec:FRAM-and-VDMSL} to develop a formal specification of a bidding system for reviewer assignment.

\subsection{Modeling WAD (Work as Done) in FRAM}

Fig. \ref{fig:FRAM-paperreview-ASIS} shows a FRAM diagram of the review process.
The activity at the top row is the {\tt Submit a paper} activity to be carried out by an author.
The  activities at the middle row are those carried out by the PC chairs who are responsible to assign reviewers to submitted papers,  make the final calls, and notify them.
The activity at the bottom row is carried out by the reviewers. 
The {\tt Assign reviewers} activity requires much effort and is time consuming to find a good combination of topics of papers and the expertise of reviewers.
A delay of the {\tt Assign reviewers} activity may shorten the {\tt Review} activity that requires enough duration to make a fair judgement to the quality of the paper.
The {\tt Make a call} activity  is also difficult to complete in time when the discussion among reviewers does not converge, but less couplings with other activities are present than the {\tt Assign reviewer} activity.

\begin{figure}[tb]
\begin{center}
\includegraphics[width=1\textwidth]{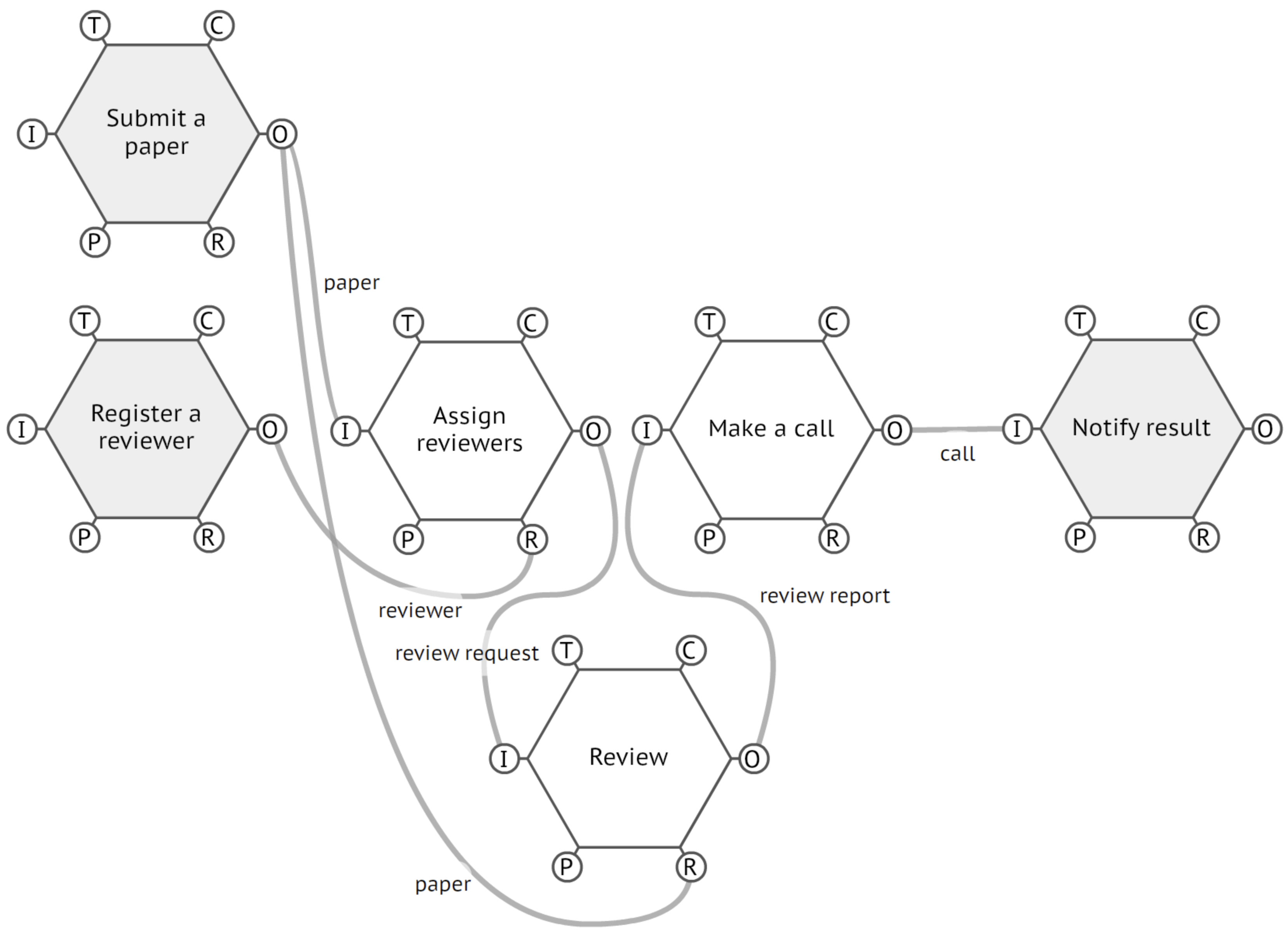}
\end{center}
\caption{An example FRAM diagram of paper review}
\label{fig:FRAM-paperreview-ASIS}
\end{figure}

\subsection{Modeling Transformation in FRAM}

The FRAM diagram shown in Fig.~\ref{fig:FRAM-paperreview-withBidding} is a solution to solve the difficulty of reviewer assignment.
A bidding system is introduced to assist the reviewer assignment.
Evaluating the matching between a paper and a reviewer is delegated to the reviewer.
Each reviewer takes a look at submitted papers and make a {\it bid} to declare which paper the reviewer would be confident to review according to the reviewer's own expertise.
With the bids, the system can propose to the PC chair several good combinations, and the PC chair chooses one.

\begin{figure}[tb]
\begin{center}
\includegraphics[width=1\textwidth]{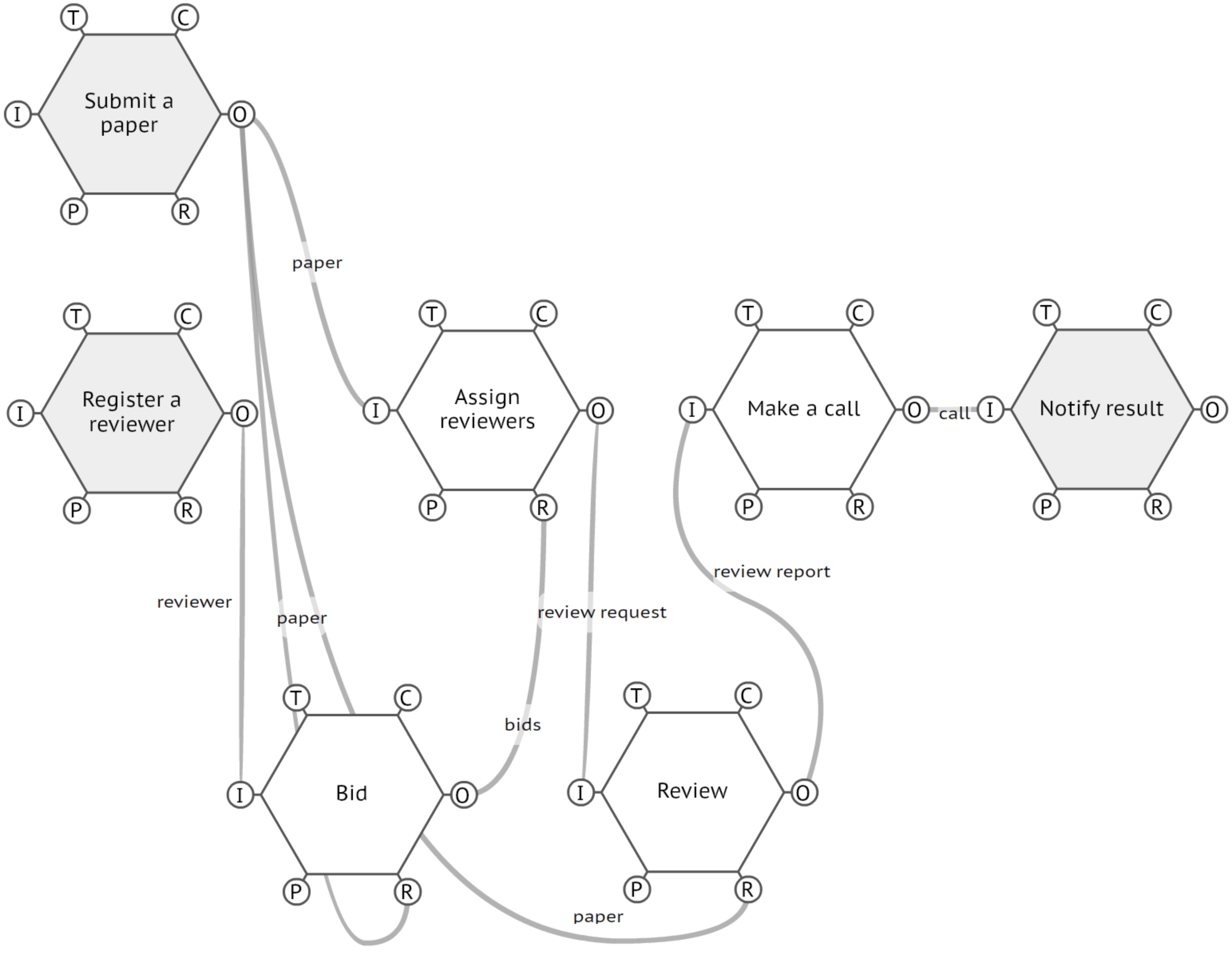}
\end{center}
\caption{An example FRAM diagram of paper review with bidding}
\label{fig:FRAM-paperreview-withBidding}
\end{figure}

\subsection{Specifying the system in VDM-SL}

The next step is to specify the bidding system.
Fig.~\ref{fig:VDM-bidding-types-values-state} shows the definitions of types, values and state required to define the operations listed in the FRAM diagram in Fig.~\ref{fig:FRAM-paperreview-withBidding}.
Each submitted paper is given a review status of either {\tt <SUBMITTED>}, {\tt <IN\_REVIEW>},  and so on.
The bidding system manages the review status of all papers in the state variable {\tt calls}.
A bid is a mapping of each paper and preference in five grades, and the preference is specified {\tt <CONFLICT>} if the reviewer has a conflict of interests.
The system holds the bids by reviewers in the state variable {\tt bids}.

Fig.\ref{fig:VDM-bidding-fromFRAM} shows definitions of the operations that corresponds to the technological activities {\tt Submit a paper}, {\tt Register a reviewer}, {\tt Bid} and {\tt Make a call}.
The lines that begin with {\tt --@FRAM} are annotations to keep traceability information with the original FRAM diagram.
The operation {\tt submit} has two annotations that tells that the operation corresponds to the  {\tt Submit a paper} activity and the activity should have an output aspect named {\tt paper}.
The contextual information in the FRAM diagram may help the specifier understand when and how the operation would be used, what inputs are given, and what effects are expected.

\begin{figure}
\begin{vdmsl}
types
     ID = nat;
     Reviewer = ID;
     Paper = ID;
     BidWeight = nat1 inv l == l <= 5;
     Bid = map Paper to (BidWeight| <CONFLICT>);
     Assignment = map Paper to set of Reviewer;
     Call =
        <SUBMITTED>| <IN_REVIEW>| 
        <ACCEPTED>| <REJECTED>| <RETRACTED>;

values
     DEFAULT_BIDWEIGHT = 3;
     REVIEWS_PER_PAPER = 3;

state Bidding of
  calls : map Paper to Call
  bids : map Reviewer to Bid
init s ==
  s
  = mk_Bidding({|->}, {|->})
end
\end{vdmsl}
\caption{Types, values and state definition for the bidding system}
\label{fig:VDM-bidding-types-values-state}
\end{figure}

\begin{figure}
\begin{vdmsl}
operations
  --@FRAM Function Submit a paper
  --@FRAM Output paper
  registerPaper : () ==> Paper
  registerPaper() ==
    (dcl newPaper:Paper 
      := if calls = {|->} then 0 else max(dom calls) + 1;
    calls(newPaper) := <SUBMITTED>;
    bids := {r |-> ({newPaper|->DEFAULT_BIDWEIGHT} ++bids(r))
            | r in set dom bids};  
    return newPaper);
  
  --@FRAM Function Register a reviewer
  --@FRAM Output reviewer
  registerReviewer : Reviewer ==> ()
  registerReviewer(reviewer) ==
    if
      reviewer not in set dom bids
    then
      bids(reviewer) 
        := {p |-> DEFAULT_BIDWEIGHT | p in set dom calls};
  
  --@FRAM Function Bid
  --@FRAM Input reviewer
  --@FRAM Resource paper
  --@FRAM Output bids
  bid : Reviewer * Bid ==> ()
  bid(reviewer, bid) ==
    if
      reviewer in set dom bids
    then
      bids(reviewer)
        := dom calls <: 
          ({p |-> DEFAULT_BIDWEIGHT | p in set dom calls}
           ++ bid);

  --@FRAM Function Make a call
  --@FRAM Input review report
  --@FRAM Output judgement
  changeCall : Paper * Call ==> ()
  changeCall(paper, call) == calls := calls ++ {paper |-> call};
\end{vdmsl}
\caption{Operations derrived from FRAM diagram}
\label{fig:VDM-bidding-fromFRAM}
\end{figure}

\begin{figure}
\begin{vdmsl}
  --@FRAM Function View bids
  --@FRAM Output bids
  --@FRAM Resource bids
  --@FRAM Input reviewer
  viewBid : Reviewer ==> [Bid]
  viewBid(reviewer) ==
    return if reviewer in set dom bids 
        then bids(reviewer) 
        else nil;
  
  --@FRAM Function View assignments
  --@FRAM Input bids
  --@FRAM Output assignments
  --@FRAM Resource paper
  viewAssignments : () ==> set of Assignment
  viewAssignments() ==
    (dcl
      reviewSlots:seq of [ID] := [],
      assignments:map Assignment to real := {|->},
      submitted:seq of Paper := [];
    for all p in set dom (calls :> {<SUBMITTED>}) do 
      submitted := submitted ^ [p];
    while len reviewSlots < len submitted  * REVIEWS_PER_PAPER
    do for all r in set dom bids do 
        reviewSlots := reviewSlots ^ [r];
    for all reviewers in set perms(reviewSlots) do
      let
        assignment : Assignment =
          {submitted(index)
          |-> {reviewers(r)
            | r
            in set {(index - 1)  * REVIEWS_PER_PAPER + 1,
            ..., index  * REVIEWS_PER_PAPER}}
          | index in set {1, ..., len submitted}},
        score = scoreByBids(assignment)
      in
        if
          score <> nil
        then
          assignments 
            := assignments munion {assignment |-> score};
    return dom (assignments :> {max(rng assignments)}));
\end{vdmsl}
\caption{Additional operations for bidding}
\label{fig:VDM-bidding-additional}
\end{figure}

\begin{figure}[tb]
\begin{center}
\includegraphics[width=1\textwidth]{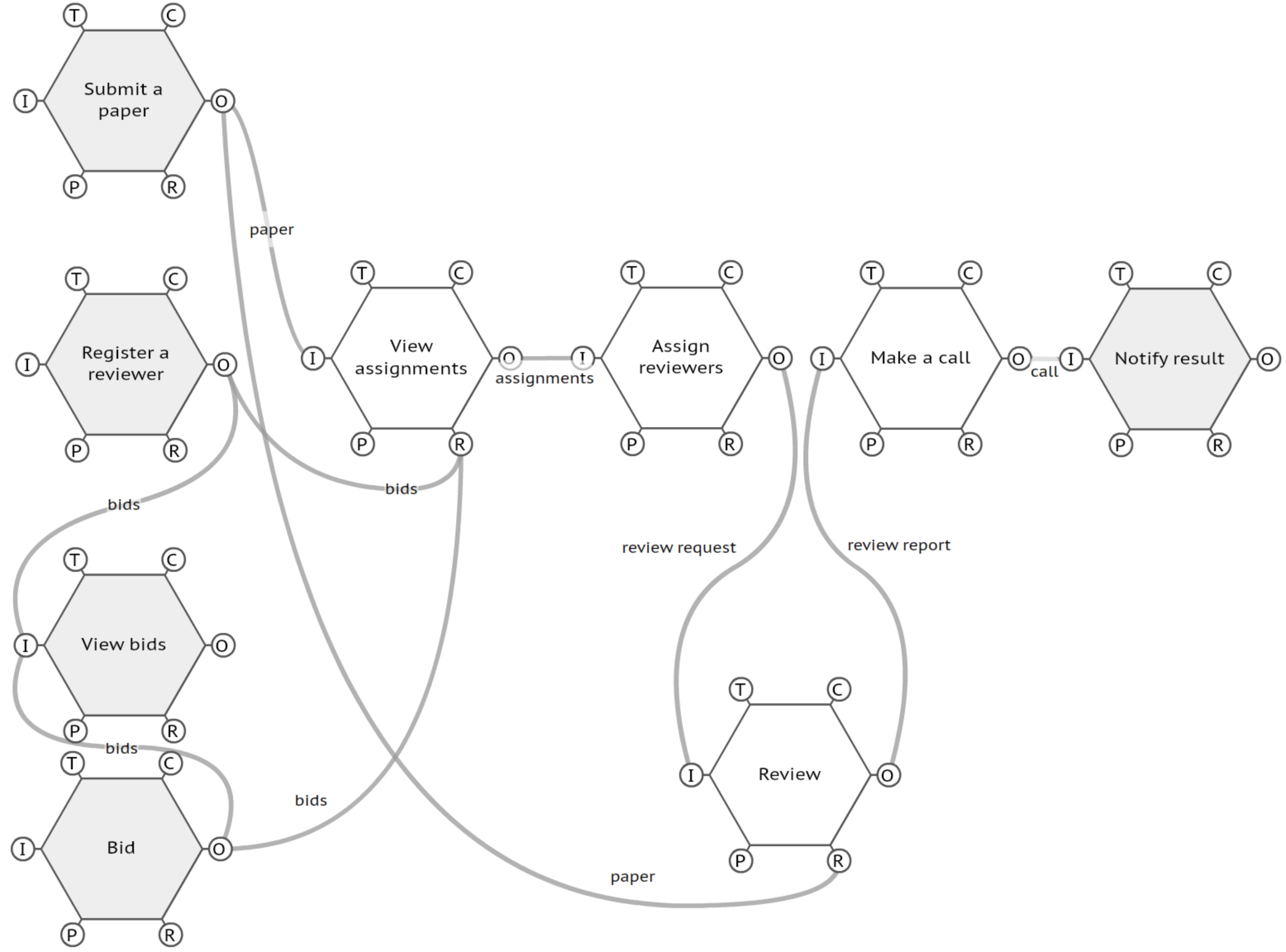}
\end{center}
\caption{An example FRAM diagram with feedback from VDM specification}
\label{fig:FRAM-paperreview-withBiddingVDM}
\end{figure}

The FRAM diagram is drawn and analysed from the perspective of collaborative activities among organisations, people and technologies.
On the other hand, the VDM specification is written from the perspective of rigorous information processing.
Some extra operations may be discovered in the formal specification phase.

Fig. \ref{fig:VDM-bidding-additional} shows the definitions of operations found possibly necessary along with the operations in Fig.~\ref{fig:VDM-bidding-fromFRAM}.
The {\tt viewBid} and {\tt viewAssignments} operations would be needed in the construction of the user interface of the {\tt Bid} and {\tt Assign reviewer} activities.
The {\tt viewAssignments} operation is a core part of this system to propose assignments with the highest matching to bids.
These additional activities from the system's perspective should be reflected into the FRAM diagram.

Fig. \ref{fig:FRAM-paperreview-withBiddingVDM} shows the FRAM diagram that incorporated the additional activities.
One can see that a reviewer can see the own bids (the {\tt View bids} activity) when the reviewer is newly registered (the {\tt Register reviewer} activity by the PC chair) and also after making bids (the {\tt Bid} activity by the reviewer).
The FRAM diagram also indicates that the PC chair will see the proposed assignments of reviewers (the {\tt View assignments} action), and then decide the assignment (the {\tt Assign reviewers} action).
The stakeholders including the research community members can see the resulting FRAM diagram in Fig.\  \ref{fig:FRAM-paperreview-withBiddingVDM} to confirm whether or not the whole work flow would fit with the actual paper review process.

\section{Discussions}
\label{sec:Discussions}

The objective of our approach to using the FRAM diagram as a pre-formal notation is the involvement of a wider range of stakeholders.
The expected use of FRAM diagrams in our approach is as an input to the specification phase, as a model to confirm the practical validity of the specification, or as a supplemental explanation to understand the formal specification.
The use of the FRAM diagram is not limited to before the formal specification but also continues after the specification phase.

To support the use of FRAM diagrams in association with VDM-SL specification, a toolchain that handles both notations is required.
FMV (FRAM Model Visualiser)\footnote{\url{https://zerprize.co.nz/home/FRAM}} is a commonly used diagram editor for FRAM users.
FMV uses XML format to file a FRAM diagram, and thus collaboration with FMV is technically possible via files.
We have extended ViennaTalk to read, modify, and write FRAM models so that technological activities in a FRAM diagram can be incorporated into a VDM model, and vice versa.

One difficulty in keeping track of associations between FRAM activities and VDM operations is vocabulary mismatching.
FRAM and VDM have different scopes of the model.
FRAM overviews the collaborative work as a whole while VDM specifies the functionality within a system.
The difference in modelling scopes may cause mismatching of vocabularies.
An activity in a FRAM diagram is worded from the perspective of the collaborative workers while its corresponding operation is named according to its functionality in the system.
For example, the {\tt Submit a paper} in Fig.~\ref{fig:FRAM-paperreview-withBidding} corresponds to the operation named {\tt registerPaper}.

Our approach uses annotations to embed traceability information into a VDM specification.
Annotations are a special form of comment intended to be processed as a comment by the standard interpreter but can place tool dependent information.
Lausdahl et. al. used annotation to declare an interface with FMUs~\cite{Lausdahl&17}.
Battle et. al. used annotation as an extension to perform directives such as probing data and dispatching plugin routines at runtime~\cite{Battle&20}.
We consider traceability information to be another usage of annotation not limited to FRAM but also applicable to other notations.

Our approach also introduces a cyclic process between formalisation by VDM and user feedback by FRAM.
We believe such a cycle is highly encouraged especially in the early stages of the specification phase~\cite{Oda&17a,Oda&15b}.
Fraser and Pezzoni used EAC (Executable Acceptance Criteria) to document and enforce system requirements~\cite{Fraser&21}.
EAC plays a key role in stakeholder involvement in Behaviour Driven Specification.
We consider EAC is also documentation looking at the system's functionality from an external point of view.
We expect more investigations to make more use of external views on VDM specification so that more stakeholders' interests can be reflected in the rigorous specification in VDM.

\section{Concluding Remarks}
\label{sec:Concluding-Remarks}
Formal specification serves as a blueprint of a system's internal design and structural constructions, and hardly describes how it looks to the users and other stakeholders.
The authors have been investigating the way to make formal specifications understood by a wider range of stakeholders through the development of ViennaTalk.
ViennaTalk makes use of specification animation to make a formal specification understandable to the non-engineering stakeholders.
The user still needs to understand the way how the system will be used.
This paper focuses on FRAM as a semi-formal notation to describe the exterior views of the system in the perspective of how and why the stakeholders will use the system.
We demonstrated that annotations in a VDM specification can glue the formal specification and semi-formal notations such as FRAM.
Although FRAM is still in the engineering domain, its diagrammatic nature has the potential to make sense for a wider range of stakeholders.
We expect further investigation including automated generation of use scenarios from FRAM diagram and translation to VDM-SL test cases using annotations.

\section*{Acknowledgements}
The authors thank Keijiro Araki for valuable and fruitful discussion on pre-formal notations.
We would also like to thank anonymous reviewers for their valuable feedback.
A part of this work was supported by JSPS KAKENHI Grant Number 20K11759. 
\bibliographystyle{splncs03}
 \newcommand{\noop}[1]{}



\clearpage
\endgroup

\begingroup
\renewcommand\theHchapter{2-Villadsen:\thechapter}
\renewcommand\theHsection{2-Villadsen:\thesection}
\locallabels{2-Villadsen:}
\newacronym{ast}{AST}{Abstract Syntax Tree}
\newacronym{ct}{CT}{Combinatorial Testing}
\newacronym{czt}{CZT}{Community Z Tools}
\newacronym{dap}{DAP}{Debug Adapter Protocol}
\newacronym{dbgp}{DBGP}{Common DeBugGer Protocol}
\newacronym{gui}{GUI}{Graphical User Interface}
\newacronym{ide}{IDE}{Integrated Development Environment}
\newacronym{lsp}{LSP}{Language Server Protocol}
\newacronym{poc}{PoC}{Proof of Concept}
\newacronym{pog}{POG}{Proof Obligation Generation}
\newacronym{po}{PO}{Proof Obligation}
\newacronym{slsp}{SLSP}{Specification Language Server Protocol}
\newacronym{vdm}{VDM}{Vienna Development Method}
\newacronym{vscode}{VS Code}{Visual Studio Code}
\newacronym{iot}{IoT}{Internet of Things}
\newacronym{rsg}{RSG}{Requirements-Specification Gap}
\newacronym{bdd}{BDD}{Behaviour-Driven Development}
\newacronym{bds}{BDS}{Behaviour-Driven Specification}
\newacronym{dsl}{DSL}{Domain Specific Language}
\newacronym{tdd}{TDD}{Test-Driven Development}
\newacronym{bdm}{BDM}{Behaviour-Driven Modelling}
\newacronym{qa}{QA}{Quality Assurance}
\newacronym{eac}{EAC}{Executable Acceptance Criteria}
\newacronym{aac}{AAC}{Abstract Acceptance Criteria}
\newacronym{ides}{IDEs}{Integrated Development Environments}

\setcounter{footnote}{0}
\setcounter{chapter}{0}
\setcounter{lstlisting}{0}

\makeatletter
\def\input@path{{2-Villadsen/}}
\makeatother

\graphicspath{{2-Villadsen/}}
\title{Bridging the Requirements-Specification Gap using Behaviour-Driven Development}
\titlerunning{Bridging the Requirements-Specification Gap using BDD}

\author{ Kristoffer Stampe Villadsen
\and  Malthe Dalgaard Jensen
\and Peter Gorm Larsen
\and Hugo Daniel Macedo
}
\authorrunning{ }

\institute{
DIGIT, Aarhus University, Department of Electrical and Computer Engineering, \\
Finlandsgade 22, 8200 Aarhus N, Denmark\\
\email{\{villadsen67,malthedj\}@hotmail.com,\{pgl,hdm\}@ece.au.dk}
}
			
\maketitle
\begin{abstract}
How is it possible to bridge the gap between requirement elicitation and formal specification? This paper proposes a solution that involves the methodology Behaviour-Driven Modelling inspired by the agile methodology Behaviour-Driven Development and Behaviour-Driven Specification. To support this methodology, we developed a tool that enables users to define behaviours as executable scenarios. A scenario is executed on a formal model through the tool to validate requirements. The tool is developed as a Visual Studio Code Extension. The tooling uses VDMJ to read and interpret formal specifications and Cucumber to discover and execute Gherkin features and scenarios. We believe that the methodology will increase readability of requirements for formal specifications, reduce the amount of misunderstandings by stakeholders, and facilitate the process of developing formal specifications in an agile process. 
\end{abstract}

\keywords{Behaviour-Driven Development, VDM, Agile Methods, Formal Methods}

\section{Introduction}
\label{sec:intro}

Formal modelling involves the transformation of a set of elicited requirements listed in a natural language into an executable model in the form of a specification written in a formal language \cite{Macedo&08}. There are well defined methods for writing formal models, e.g.: \gls{vdm} \cite{Jones90}, Z \cite{Spivey92} and B \cite{Abrial05}. The same happens with requirements elicitation e.g.: use-cases diagram, entity-relationship modelling, and user stories. 

In most of the approaches, the practice is to transform the requirements into the specification as a jump which consists of a mental exercise including the iteration of two steps: first understand the requirements, then formulate them as a formal model. A conceptual illustration of the metaphoric jump can be seen in figure \ref{fig:jump}. This jump bridges what we will metaphorically describe as the \gls{rsg}. The writing of adequate specifications becomes an art that experienced formal modellers acquire by performing several transformations (different jumps across the \gls{rsg} in several projects). 

\begin{figure}[!htb]
    \centering
    \includegraphics[width=\linewidth]{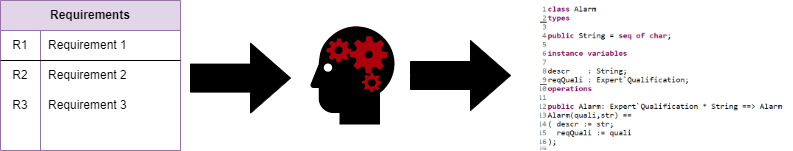}
    \caption{An abstract list of requirements are shown on the left side. These are interpreted by the modeller, which is illustrated by the pictogram in the middle. The modeller translates the interpreted requirements into a model, shown on the right side. Thereby, performing the metaphoric jump.}
    \label{fig:jump}
\end{figure}



The existing tool support for working with formal specification includes editing, interpreting, debugging and testing through various libraries and \gls{ides} \eg VDMTools\cite{Larsen01}, Overture \cite{Larsen&10a}, VDM-VSCode\cite{Rask&20}, Rodin\cite{Abrial&10}, and VDMJ \cite{Battle09}. Testing of specifications can also be performed within VDMJ by utilising combinatorial testing or VDMJUnit. However, no domain-agnostic tool exists which supports bridging the \gls{rsg}.

In this paper, we show how and propose a tool to bridge that gap and ease the translation task for modellers by assimilating and providing tool support for \gls{bdm}, a methodology inspired by the \gls{bdd} agile approach. \gls{bdd} prescribes a methodology to develop software focusing on exemplifying requirements into concrete natural language examples, increasing readability and reducing misunderstandings.

Behaviour-Driven Modelling adds a step between requirement elicitation and specification development, where requirements are translated into behaviours. The behaviours are written in a constrained subset of natural language with the goal of becoming more understandable by stakeholders. A prototype of a \gls{vscode} extension has been developed and is described throughout this paper. The prototype uses Cucumber\footnote{\url{https://cucumber.io/}} as a \gls{bdd} test runner to discover and execute the behaviours. The tool maps behaviours to concrete operations within a formal specification. This mapping is made possible using the VDMJ interpreter and VDMJ Annotations, such that operations can be annotated within the VDM++ specification language. The extension allows for a modeller to create a \gls{bdm} project and use the \gls{bdd} test runner to validate behaviours against the specification. The goal of the extension is to be incorporated as part of the Overture tooling support for the VDM languages. 

We believe this methodology and tooling will contribute to closing the RSG and increase readability of requirements for formal specifications, reduce the amount of misunderstandings by stakeholders, and facilitate the process of developing formal specifications in an agile process.

The outline of this paper is as follows: Section \ref{sec:background} gives an introduction to relevant topics necessary to understand the paper. Section \ref{sec:bdm} and \ref{sec:tecnicalSolution} describes the proposed approach and technical solution respectively. An example project uses the proposed solution in section \ref{sec:results}. The paper concludes upon the findings in section \ref{sec:conclusion} together with a paragraph describing the future work to deploy the tool in the \gls{vscode} marketplace.

\section{Background}
\label{sec:background}
In agile development the workload is broken down into smaller pieces which can be worked on simultaneously and iteratively\cite{Fraser&21}. This allows for smaller teams to work together improving the performance of development\cite{HOEGL2005209}. Agile methods ensure that the best practices, are applied and the correctness of the engineering processes is upheld. Agile methods help both stakeholders and software engineers in building, deploying, and maintaining complex software with its associated changing requirements~\cite{8409911}.

BDD builds on the concepts of \gls{tdd} with another approach to system analysis; how the different entities of a software system interact based on the domain model\cite{rose2015cucumber}. \gls{bdd} utilises a \gls{dsl} for specifying requirements, which are more readable for stakeholders. The specified requirements are formulated as concrete examples of system behaviour\footnote{http://behaviour-driven.org/}. 

An agile development tool which supports \gls{bdd} through acceptance testing is Cucumber. It provides the \gls{dsl}
Gherkin, which is used to specify, in a restricted subset of plain natural language, the desired behaviour of a system. A decisive advantage of formulating behaviours using Gherkin is that not only is a stakeholder more likely to understand it, but importantly a computer is able to.
Cucumber aids developers in writing concrete examples of behaviours. This is accomplished by defining a requirement as a feature of a system. This feature is then exemplified through a scenario as seen in listing \ref{lst:examplefeature}. By having concrete examples which defines acceptance tests allows for easier discovery of edge cases\cite{rose2015cucumber}. Working with Cucumber and following the \gls{bdd} methodology gives developers an opportunity to get early feedback from the system. 

\begin{lstlisting}[basicstyle=\small,frame=tb,language=Gherkin,label=lst:examplefeature,caption=An example Gherkin feature and scenario,captionpos=b]
Feature: feature-name
  feature-description

 Scenario: scenario-name
   Given a pre-condition
   When an action occurs
   Then post-condition is satisfied
\end{lstlisting}

Cucumber groups scenarios in a feature to provide tracing of requirements for the system. This tracing provides stakeholders with an overview of which requirements that are currently implemented.
Each scenario consists of steps which are either \textit{Given}, \textit{When} or \textit{Then}. These steps are conceptually equivalent to Hoare triples~\cite{Hoare69}.  

\begin{table}[!hbt]
\resizebox{\textwidth}{!}{
\begin{tabular}{|
>{\columncolor[HTML]{EFEFEF}}l |l|l|}
\hline
\cellcolor[HTML]{C0C0C0}{\color[HTML]{333333} } & \cellcolor[HTML]{C0C0C0}{\color[HTML]{333333} Cucumber}                                                                                                                                                                                                       & \cellcolor[HTML]{C0C0C0}{\color[HTML]{333333} Vienna Development Method}                                                                                                                                                                                                              \\ \hline
Pre-condition                                   & \begin{tabular}[c]{@{}l@{}}Cucumber utilises  the \textit{Given} annotation\\ to identify the pre-condition of the scenario,\\ within the pre-condition the state is initialised.\end{tabular}                                                                        & {\color[HTML]{343434} \begin{tabular}[c]{@{}l@{}}The pre-condition in VDM is utilised within \\ the operation scope noted by the pre keyword\\ at the end of the operation, VDM does not initialise any state.\\ VDM checks if the input adhere to a defined constraint\end{tabular}} \\ \hline
Action                                          & \begin{tabular}[c]{@{}l@{}}Cucumber utilises the \textit{When} annotation\\ to identify the action performed by program \\ based on the state and the desired feature.\end{tabular}                                                                                    & \begin{tabular}[c]{@{}l@{}}Within an operation of VDM, the model is acted on by\\ the input based on the state of the model.\end{tabular}                                                                                                                                             \\ \hline
Post-condition                                  & \begin{tabular}[c]{@{}l@{}}Cucumber utilises the \textit{Then} annotation to identify\\ the post-condition of the scenario, within the post-condition\\ cucumber asserts on the state of the program to determine \\ if it verifies the desired behaviour.\end{tabular} & \begin{tabular}[c]{@{}l@{}}The post-condition in VDM is utilised within\\ the operation scope noted by the post keyword\\ at the end of the operation, VDM verifies that the output\\ of the function adheres to a defined constraint\end{tabular}                                    \\ \hline
\end{tabular}}
\caption{Description of how the Hoare triples are applied in Cucumber and VDM}
\label{tab:HoareTriplesInCucumberVsVDM}
\end{table}

\textit{Given} corresponds to a pre-condition, which in \gls{bdd} also initialise the state. \textit{When} corresponds to an action that is performed. \textit{Then} corresponds to a post-condition, which is asserting if the new state is as expected. A table describing the differences and similarities between Hoare triples and \gls{bdd} steps can be seen in table \ref{tab:HoareTriplesInCucumberVsVDM}. 

In listing \ref{lst:exampleGivenStep} a definition of a \textit{Given} step is provided. This is defined as a step definition, which is a function that is mapped to a step within a scenario. For a step definition to be mapped to a step, the function should be annotated with a step annotation. Cucumber runs through each scenario and finds a matching annotated function for each step.  Cucumber executes this step definition and repeats this for all steps in a scenario.
A scenario passes if step definitions executes correctly and all assertions are satisfied.  

\begin{lstlisting}[basicstyle=\small,frame=tb,language=Java,label=lst:exampleGivenStep,caption=An example Java step definition,captionpos=b]
@Given("A pre-condition")
  public void A_pre_condition() {
    pending();
  }
\end{lstlisting}

Simon Fraser et al.(2021) presented an approach called \gls{bds}\cite{Fraser&21}.
\gls{bds} applies \gls{bdd} on formal specifications by validation of the formal specification against requirements formulated as \gls{eac}. The Azuki platform\footnote{\url{https://github.com/anaplan-engineering/azuki}} - a generic framework that assists in using \gls{bds}. The framework aids in analysing legacy systems which results in human readable acceptance criteria. Furthermore, the framework creates \gls{eac} which it runs against the specification to verify the system. \gls{eac} are \gls{bds} equivalent to executable scenarios from \gls{bdd}. Mapping these directly to the behaviours of the desired system will result in validation of the specification against the requirements. The framework is designed and developed to solve a specific problem case for a company as stated in Simon Fraser et al.(2021)\cite{Fraser&21}. The framework has been developed constrained by a specific domain, therefore, challenges will inherently arise when applying this framework more widely and domain agnostic.

\section{Behaviour-Driven Modelling}
\label{sec:bdm}
Traditional formal modelling consists of a modeller performing a transformation of elicited requirements into an executable model. This is a two-step iterative process: understand the requirement, then formulate it as a formal model. Performing the transformation will map the requirements into a specification which seeks to prove the properties and be used as a \gls{qa} metric of the system. The understanding of the requirements and formulation of the specification is at the mercy of the modeller. This means that the properties and the \gls{qa} is determined by the modeller. Attempts at adapting this traditional formal approach through merging of formal methods and agile methodologies have been proven to be possible \cite{Wolff12a,olszewska2016using}.
The results of the conceptual approaches include an iterative process focusing on smaller parts of the system.

The conceptual difference between the traditional approach and the agile formal modelling approach is shown in figure \ref{fig:approach}. In this paper, the Agile formal modelling approach has been extended with tool support for the \gls{bdd} approach, and the approach will be described as \gls{bdm}.
\begin{figure}[!htb]
    \centering
    \includegraphics[width=\linewidth]{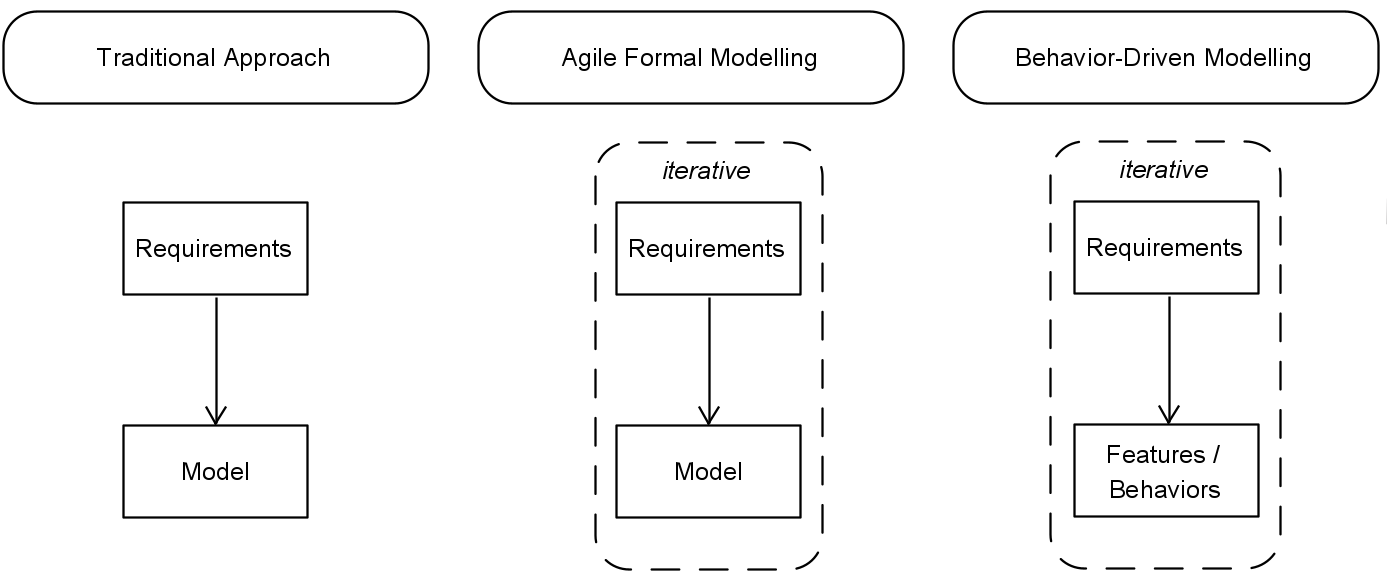}
    \caption{This figure shows the evolution of \gls{rsg}. On the left is the traditional approach of a direct transformation from requirements into a specification. In the middle is the conceptual depiction of the agile formal merging approach. On the right is the approach introduced in this paper.}
    \label{fig:approach}
\end{figure}
\noindent \gls{bdm} introduces the additional step which bridges the gap from requirements to the specification within an agile approach as seen in figure \ref{fig:approach}. 

The \gls{rsg} is bridged by focusing on requirement analysis to define behaviours in terms of features and scenarios. The behaviours act as guidance for both developers and stakeholders to capture requirements as artifacts that can be traced as they are being implemented in the formal model. The tracing allows for quality assurance teams to be part of the early development process. It allows for modellers and developers to be part of requirement elicitation in a way that can result directly in step definitions, which can run against both the model as well as the production code.

The tooling for \gls{bdm} focuses on requirement elicitation through stakeholder communication to define behaviours of the desired system. The \gls{bdm} approach is illustrated in figure \ref{fig:iterative}, firstly, the requirements elicitation is performed. After the requirements have been elicited and analysed, the behaviours are defined. These behaviours are then mapped to step definitions. The \gls{bdm} tool allows for mapping of the properties of the model to the step definitions. Lastly, the specification is validated against the requirements through these step definitions using a \gls{bdd} test runner. The \gls{bdd} test runner performs the mapping and validation of behaviours against the specification. This approach is then repeated for each requirement that is defined and can be applied directly in an agile process.

\begin{figure}[!htb]
    \centering
    \includegraphics[width=\linewidth]{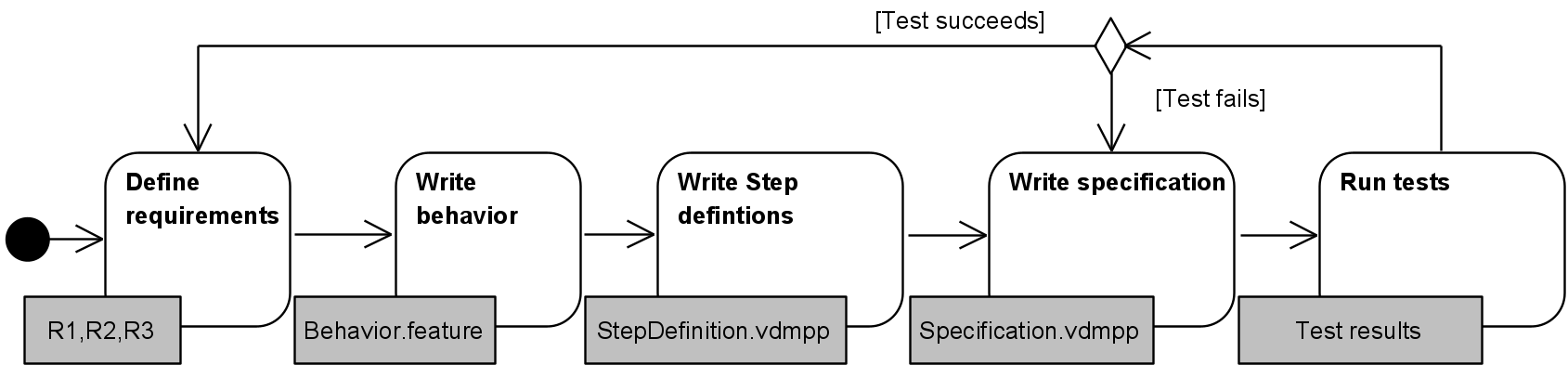}
    \caption{This diagram shows an overview of the proposed \gls{bdm} methodology. The process repeats the steps illustrated until all requirements have been satisfied. }
    \label{fig:iterative}
\end{figure}

A screenshot of the developed tooling can be seen in figure \ref{fig:screenshotExtension}. The figure shows the tooling from the user's perspective. At the current stage of the tooling, three areas are of interest, these are noted with the three numbered red boxes on the figure:
\begin{itemize}
    \item \textbf{Box \#1} - Shows the project file structure depicting the location of the specification files, features, and step definitions.  
    \item \textbf{Box \#2} - Shows the file that is currently being edited, in this case, it is a feature file describing a scenario with different input values. 
    \item \textbf{Box \#3} - Shows the output of running the executable scenario described in the feature. Here, it shows that three tests have been executed successfully. 
\end{itemize}

The execution of tests utilises VDMUnit to read and interpret the specification to validate it against the features.
 
\begin{figure}[!htb]
    \centering
    \includegraphics[width=.73\linewidth]{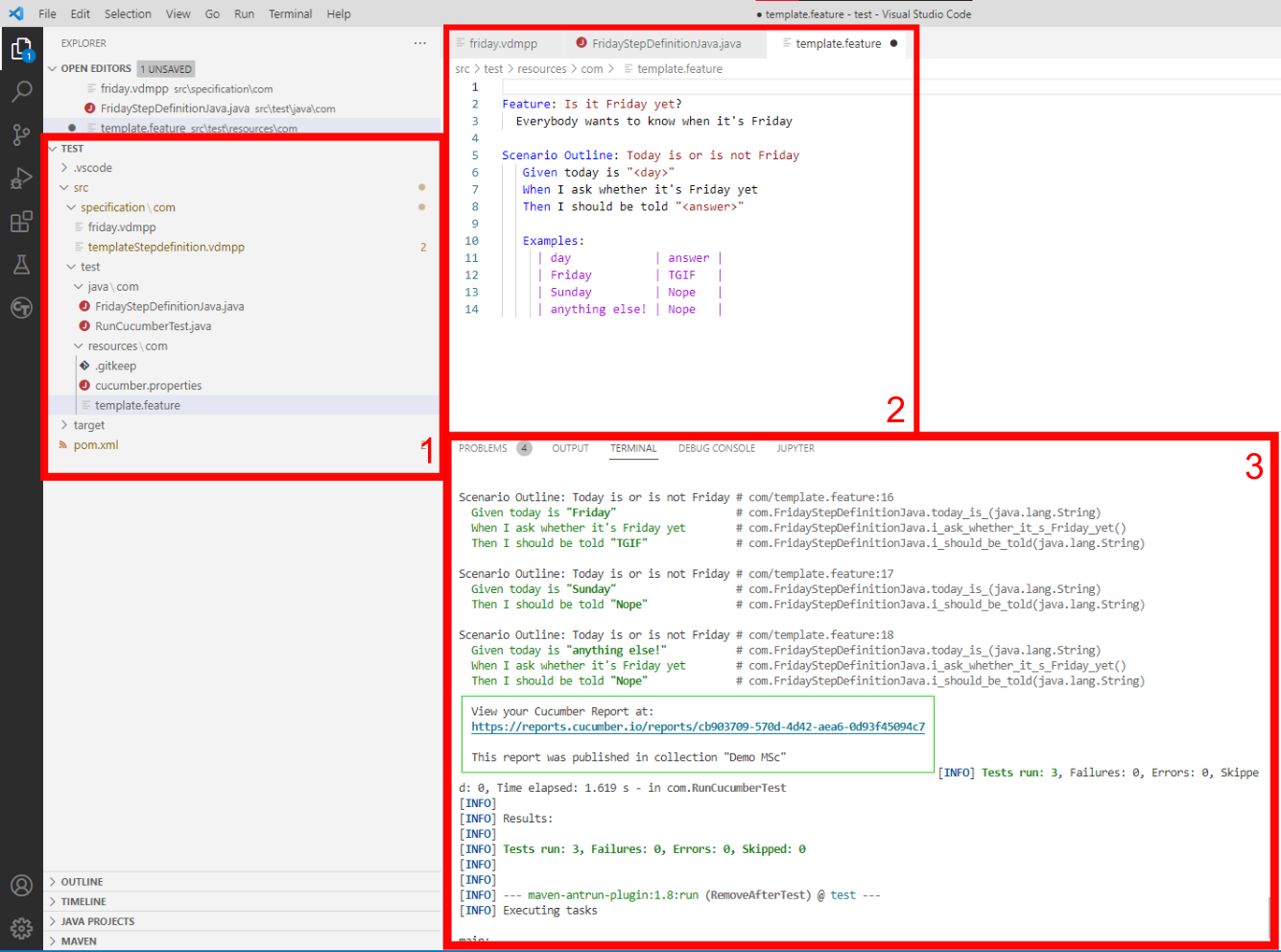}
    \caption{Screenshot of the \gls{vscode} extension.}
    \label{fig:screenshotExtension}
\end{figure}

\section{Technical Solution}
\label{sec:tecnicalSolution}

This section describes the tool developed to enable users to follow the approach described in section \ref{sec:bdm}. An overview of the proposed solution can be seen in figure \ref{fig:overview} illustrating how the Cucumber test engine finds and executes scenarios through generation of Java step definitions.

The user of the tool defines scenarios and VDM++ specifications. The VDMJ interpreter performs type checking of the defined VDM source. During this operation, the tool looks for annotations for VDM definitions and generates Java step definitions based on the annotations detected. The generated Java step definitions are provided to the cucumber test engine together with user defined scenarios. When the Cucumber test engine executes a scenario with the corresponding generated java step definitions, it will return either passed, failed, or pending. A pending test result means that the scenario is not yet implemented. 

\begin{figure}[!htb]
    \centering
    \includegraphics[width=.8\linewidth]{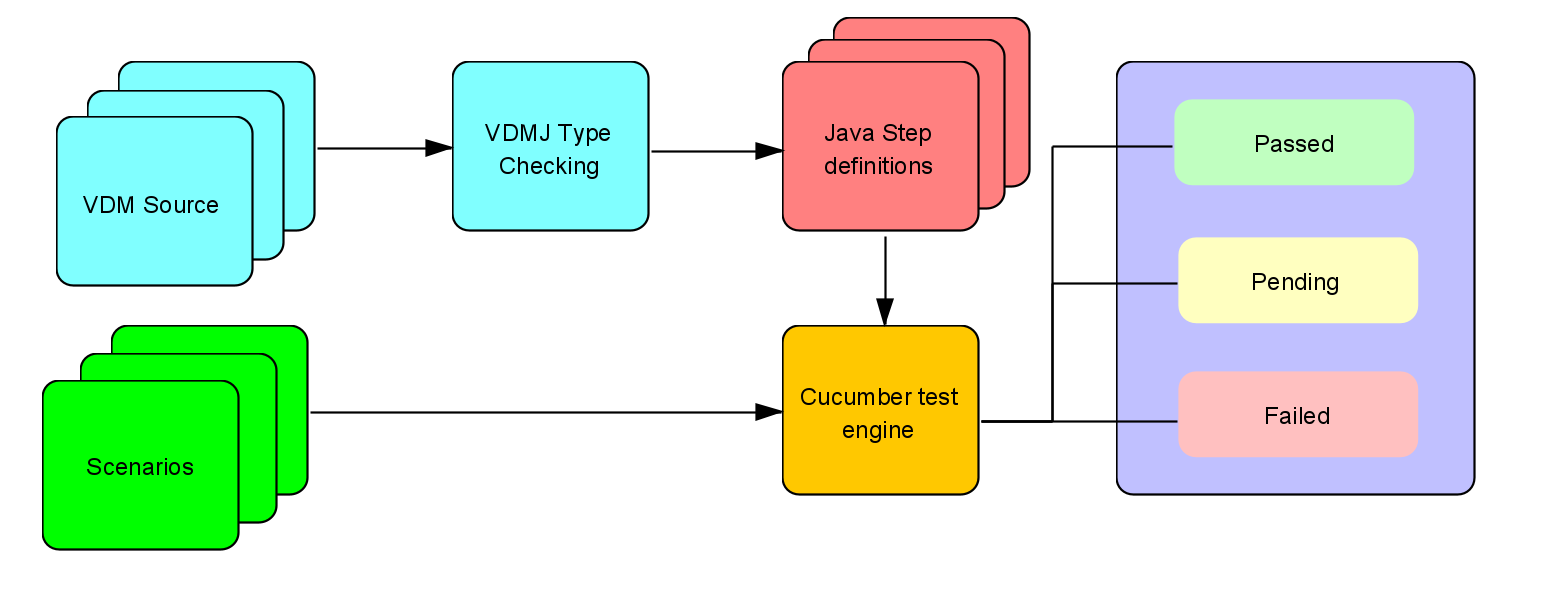}
    \caption{An overview of how the Cucumber test engine finds and executes scenarios and step definitions created through the \gls{bdm} tool. }
    \label{fig:overview}
\end{figure}

To effectively apply the methodology, the tool provides support for the following Cucumber annotations in VDM: \textit{Given}, \textit{When} and \textit{Then}. Additionally, the tool also supports an annotation called \textit{StepDefinition}, which is used to annotate the VDM++ class containing the operations annotated with the Cucumber specific annotations. These four annotations provide the means to effectively define step definitions in the VDM++ language according to the format that the Cucumber test engine can understand and execute.

A \gls{bdm} project is required to conform to a defined directory structure. The project directory structure can be seen in figure \ref{fig:project_structure} and shows how the different types of files should be located. The root project folder contains two directories, src and target. The target sub-directory contains generated files from the BDM project. The src folder contains the VDM source files within the specification sub-directory. The VDM source files is both the specification and step definitions. Java source files is located within the Java sub-directory inside the Test directory. The only Java file for the project is \textit{RunCucumberTest.java}, which is required to run the Cucumber test engine. Scenario files is located within the Resources sub-directory inside the Test directory.

\begin{figure}[!htb]
    \centering
    \includegraphics[width=.9\linewidth]{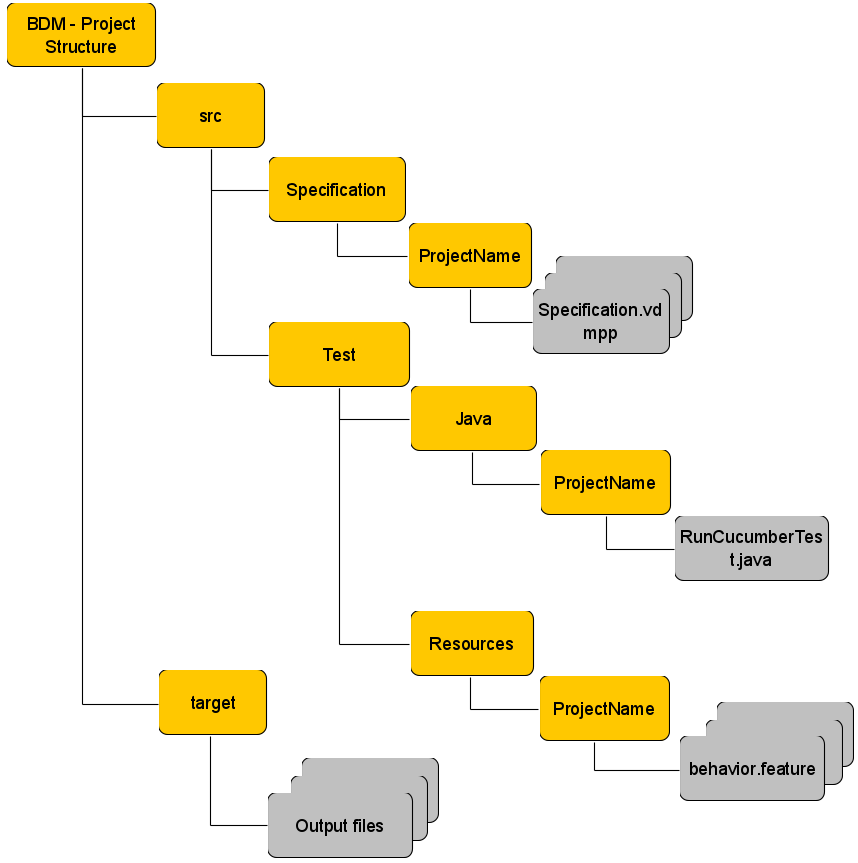}
    \caption{An example of a \gls{bdm} project structure.}
    \label{fig:project_structure}
\end{figure}

A conceptual package diagram depicting the implementation and dependencies of the tool developed can be seen in figure \ref{fig:package_big}. The package diagram displays the BDMAnnotations package containing the four different annotations: \texttt{Given}, \texttt{When}, \texttt{Then}, and \texttt{Step Definition} and their dependency to \textit{BDM.Lib}. This library contains utility functionality through the VDMUtility class and functionality to build Java step definitions through the StepDefinitionsBuilder. \textit{BDM.lib} has a dependency to VDMJ and VDMJUnit in order to read and interpret VDM++ specifications. Furthermore, the package has a dependency to Javassist to provide functionality that generates java classes at runtime.  

\begin{figure}[!htb]
    \centering
    \includegraphics[width=.9\linewidth]{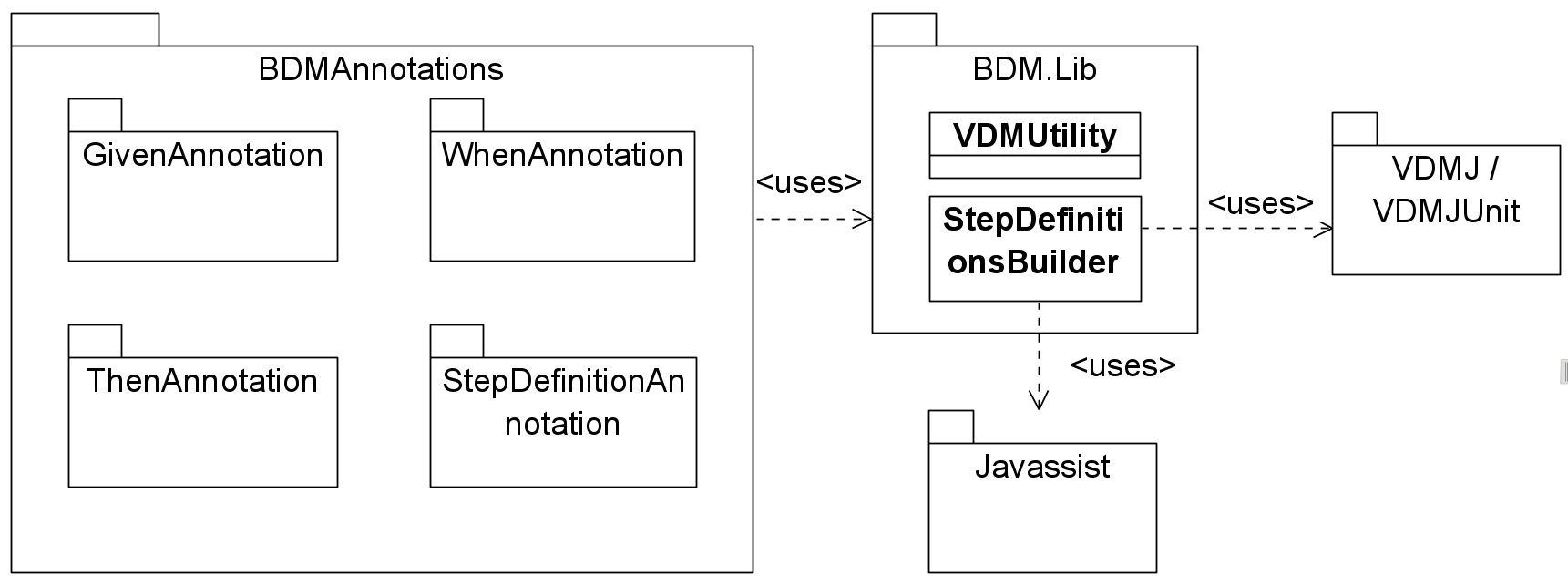}
    \caption{Package diagram depicting implementation and dependencies of the BDM tool.}
    \label{fig:package_big}
\end{figure}
The tool extends upon the VDMJ annotations library to enable support for defining step definitions in the VDM++ language that can be understood and executed with the Cucumber test engine as the \gls{bdd} test runner. The Cucumber test engine handles discovery, selection, and execution of Cucumber scenarios. The extension of the VDMJ Annotations library was implemented with respect to VDMJ Annotations documentation\cite{Battle09}.  When an annotation is found the tool generates a corresponding Cucumber Java step definition that uses the VDMJUnit library to read specifications and execute the corresponding operations in the VDM++ step definitions  \cite{Battle09}. 

The VDMJ Annotations library provides the ability to affect different aspects of the VDMJ operation: the parsing, the type checking, the interpreting and the proof obligation generation. For the annotations described above, the VDMJ type checking operation is affected to generate step definitions in the Java language. When implementing an extension to the VDMJ Annotations, it is required to know which VDM constructs that should be annotated and whether the functionality should be executed before or after the VDMJ operation. The developed tool performs the generation of classes before the VDMJ type checker performs its operation. To illustrate the steps performed by the tool, a sequence diagram can be seen in figure \ref{fig:sequence_given} which displays the steps performed when a \textit{Given} annotation is found. 

To build step definitions in the Java language, a third-party library is used called Javassist. This enables bytecode manipulation, meaning that Javassist can perform alterations and creations of programs or constructs within a program such as a class\cite{chiba2000load}. The created Java step definition is generated through the Javassist library and implements the Cucumber step definition functions that is executed by the Cucumber test engine. To effectively read specifications and execute functions or operations defined in a VDM++ model, the library VDMJUnit is used. The created Cucumber Java step definitions inherits from the \textit{VDMJUnitTestPP} Java class and implements the \textit{Given}, \textit{When}, and \textit{Then} step definition functions. The functions are annotated as specified in the corresponding VDM++ annotations. 

\begin{figure}[!htb]
    \centering
    \includegraphics[width=\linewidth]{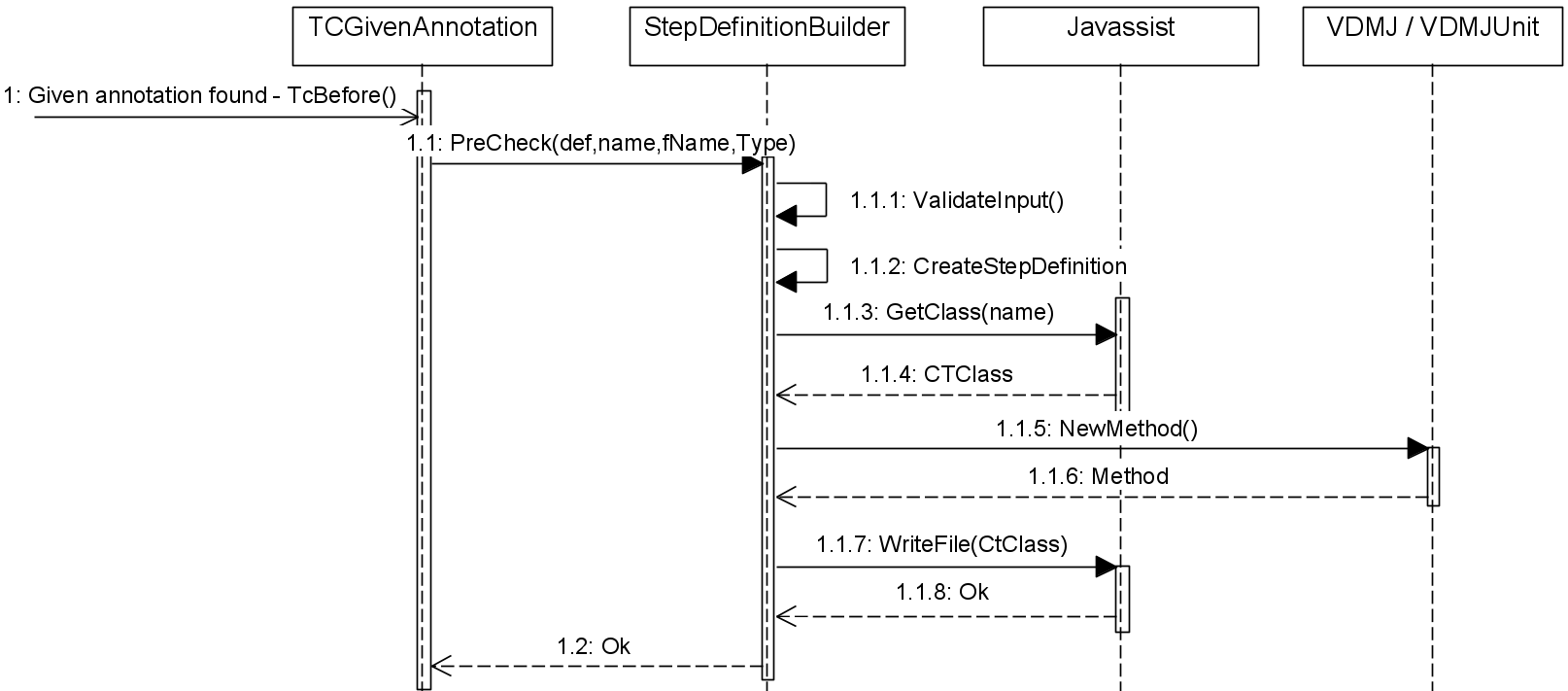}
    \caption{Sequence diagram depicting the steps performed by the tool for the \textit{Given} annotation.}
    \label{fig:sequence_given}
\end{figure}

When a \textit{Given} annotation is found by VDMJ, it calls the \textit{PreCheck} function from the \textit{StepDefinitionBuilder} inside the \textit{BDM.Lib} library. The function performs validation of the input values to ensure that the operation name and parameters are defined as expected by the \textit{Given} annotation. The input values are:

\begin{itemize}
    \item \textbf{def} - The definition from the VDMJ type checker. 
    \item \textbf{name} - The name of the Cucumber behaviour, used to map requirements and step definitions together. 
    \item \textbf{fName} - The name of the function annotated.
    \item \textbf{Type} - The type of Cucumber annotation.
\end{itemize}

When the input values have been validated, the builder then generates the Java step definition function within the step definition class. The builder makes sure that the generated function is annotated correctly according to the standards given by the Cucumber test engine.  

The prototype for the tool is developed as a \gls{vscode} extension which handles creation of \gls{bdm} projects and execution of cucumber tests. The architecture for the extension can be seen in figure \ref{fig:BDMVSCodeExtension}. A detailed description of the architecture can be seen in section \ref{sec:conclusion}.

\section{Demonstration}
\label{sec:results}
To demonstrate the developed tool and approach described throughout this paper, an example project is used. The example is a very simple project inspired by the Cucumber "10 Minute Tutorial"\footnote{\url{https://cucumber.io/docs/guides/10-minute-tutorial/}}. The requirements for the project are defined as the following:
\begin{itemize}
    \item \textbf{R1} - The system must output "TGIF" if today is Friday.
    \item \textbf{R2} - The system must output "Nope" if today is not Friday.
\end{itemize}

The first step in the proposed approach is to define behaviours based upon these requirements. These behaviours are defined according to the traditional \gls{bdd} approach, by defining Cucumber scenarios within a feature file, and can be seen in listing \ref{lst:behaviors}. \clearpage

\begin{lstlisting}[basicstyle=\small,frame=tb,language=Gherkin,label=lst:behaviors,caption=The Cucumber behaviour definition for the case project.,captionpos=b]
Feature: Is it Friday yet?
  Everybody wants to know when it's Friday
    Scenario Outline: Today is or is not Friday
    Given today is "<day>"
    When I ask whether it's Friday yet
    Then I should be told "<answer>"
    Examples:
     | day            | answer |
     | Friday         | TGIF   |
     | Sunday         | Nope   |
     | anything else! | Nope   |
\end{lstlisting}

The next step involves defining step definition operations in VDM++, that should be annotated with one of the step definition operations from the defined behaviour, such as \textit{Given}, \textit{When}, or \textit{Then}. The annotation takes two parameters; the first is the name of the temporary variable that is used to maintain the state of the object, and the second is a string containing the text defined for the step definition operation within the feature file. On listing \ref{lst:stepdefinition}, the step definition operations can be seen for the defined behaviour. An annotation is written according to the VDMJ documentation, within a comment and with an \textit{"@"} as a prefix. The assertion of the behaviour happens in the operation annotated with \textit{Then}, where the tool will assert that the post condition is satisfied. If the post condition is satisfied the test will succeed, if not, it will fail.
 
\begin{lstlisting}[basicstyle=\small,frame=tb,language=vdmpp,label=lst:stepdefinition,caption=Step definition operations for the defined behaviours written in VDM++,captionpos=b]
-- @Given("f1","today is {string}")
public Today_is: seq of char ==> ()
Today_is(str) ==
(
  friday := new Friday(str);
);
-- @When("f1","I ask whether it's Friday yet")
public I_ask_whether_it_is_friday: () ==> ()
I_ask_whether_it_is_friday() ==
(
    actualResult := friday.IsItFriday();
);
-- @Then("f1","I should be told {string}")
public ResultingThen: seq of char ==> bool
ResultingThen(expected) == 
 return expected = actualResult
post RESULT = true;
\end{lstlisting}

When the Cucumber test engine execute the tests generated based on the behaviour descriptions, it will find and execute these VDM++ operations through a generated Java step definition class that utilises VDMJUnit to access the VDM++ step definition operations. The generated Java functions can be seen on listing \ref{lst:stepdefinitionjava}. Each of the functions contains a call to a helper function called \textit{checkLocalVariable}, this function checks whether the local variable that are given as parameter actually exists. If the variable does not exist, the function creates it. 

\begin{lstlisting}[basicstyle=\small,frame=tb,language=java,label=lst:stepdefinitionjava,caption=Step definition functions  for the defined behaviours written in Java,captionpos=b]
@Given("today is {string}")
  public void Today_is(String paramString) {
    checkLocalVariable();
    run("f1.Today_is(\"" + paramString + "\"" + ") ");
  }
@When("I ask whether it's Friday yet")
  public void I_ask_whether_it_is_friday() {
    checkLocalVariable();
    run("f1.I_ask_whether_it_is_friday() ");
  }
@Then("I should be told {string}")
  public void ResultingThen(String paramString) {
    checkLocalVariable();
    run("f1.ResultingThen(\"" + paramString + "\"" + ") ");
  }
\end{lstlisting}

The result of utilising this approach is the ability to define requirements in a more stakeholder-readable language, which are converted to executable scenarios using step definition annotations. These executable scenarios are executed by the Cucumber for Java test engine. This tool and approach make it possible to perform BDM of VDM++ specifications.

\section{Concluding Remarks and Future Work}
\label{sec:conclusion}
This paper introduces the \gls{bdm} approach to bridge the \gls{rsg} by extending the traditional two steps with an extra step. The additional step includes definition of behaviours based on requirements. To ease the translation task, a tool is introduced that supports \gls{bdm}.

The tool described in section \ref{sec:tecnicalSolution} is still under development but, at its current state, it is possible to use it to perform \gls{bdm} as described in section \ref{sec:results}. The current state of the tool is the ability to define step definition operations within a VDM++ class and use the Cucumber test runner to execute these operations. The tool requires a defined directory structure to locate specification files and feature files. A minimal \gls{vscode} \cite{Rask&20a,Rask&21} extension have been developed to generate a new \gls{bdm} project that conforms to the requirements of the tool. The extension also supports execution of tests through the Cucumber test engine.

\gls{bdm} supports formal validation through testing of operations in the defined specification. \gls{bdm} extends this by supporting validation of the requirements through the tool support described in section \ref{sec:tecnicalSolution}. Given that the testing is handled through the Cucumber test engine allows for validation of production code corresponding to the formal specification - with respect to constraints. The testing of the formal specification cannot run directly on production code, however, the behaviours can be mapped to step definitions whereas both the production code and the formal specification satisfies the requirements. It is noteworthy to mention that this approach does not eliminate the need for verification through unit testing of individual elements of a formal specification.

\begin{table}[]
\resizebox{\textwidth}{!}{
\begin{tabular}{|
>{\columncolor[HTML]{EFEFEF}}l |l|l|l|}
\hline
\cellcolor[HTML]{C0C0C0}{\color[HTML]{333333} Feature support}                        & \cellcolor[HTML]{C0C0C0}{\color[HTML]{333333} Traditional approach} & \cellcolor[HTML]{C0C0C0}{\color[HTML]{333333} \begin{tabular}[c]{@{}l@{}}Agile Merging Methodology.\\  e.g. Scrums goes formal, FormAgi..\end{tabular}} & \cellcolor[HTML]{C0C0C0}Behaviour-Driven Modelling \\ \hline
Formal specification                                                                  & Supported                                                           & {\color[HTML]{343434} Supported}                                                                                                                        & Supported                                          \\ \hline
Iterative process                                                                     &                                                                     & Supported                                                                                                                                               & Supported                                          \\ \hline
Early feedback                                                                        &                                                                     & Supported                                                                                                                                               & Supported                                          \\ \hline
\begin{tabular}[c]{@{}l@{}}Tool support for\\ specification validation\end{tabular}   &                                                                     &                                                                                                                                                         & Supported                                          \\ \hline
\begin{tabular}[c]{@{}l@{}}Tool support for \\ implementation validation\end{tabular} &                                                                     &                                                                                                                                                         &                                                    \\ \hline
\end{tabular}
}
\caption{Feature supported by different approaches.}
\label{tab:featuresupport}
\end{table}

The approach presented in this paper differs from previous attempts at bridging the gap, in a few different ways. \gls{bdm} obtains the same advantages as the approaches which merge the agile methods and formal methods. \gls{bdm} utilises the iterative nature from \gls{bdd}, which other approaches have gained from e. g. SCRUM \cite{Wolff12a}. The early feedback and validation of \gls{bdd} is still maintained with \gls{bdm}. An overview of the features supported by \gls{bdm} is shown in table \ref{tab:featuresupport}. However, \gls{bds} applies a similar approach as \gls{bdm}. Both methodologies applies \gls{bdd} for development of formal specifications. The main difference stems from their associated tool, which have been developed for different intended users and problem domains. 

A valuable extension of the developed tooling, would include the functionality of validating formal specification and the corresponding implementation against the same requirements. The extension would aid in the process of discovering differences between them. The consequences of a change in requirement would be accentuated in the validation process for formal specification and implementation.
This would provide the stakeholders with insight about the sometimes costly consequences of changing requirements.   

The proposed solution described by this paper is not the only solution for the problem definition. An alternative solution would be to implement native support for the VDM dialects directly within Gherkin and Cucumber. This would eliminate the dependencies for generation of Java classes that can be understood by the Cucumber Java tool. This solution was considered when developing the proposed solution but were deemed too comprehensive. 
This paper have introduced the methodology of \gls{bdm} and shown that it can be applied to a simple case project. More research is needed in order to determine the feasibility of the methodology and associated tool for bridging the \gls{rsg}. This research should include more case studies on more complex projects in order to gain insight in the scope of applicability and the limitations of adapting \gls{bdm}. However, the presented methodology attempts at defining a formal process for translation of requirements into formal specifications.

\begin{figure}[!htb]
    \centering
    \includegraphics[width=\linewidth]{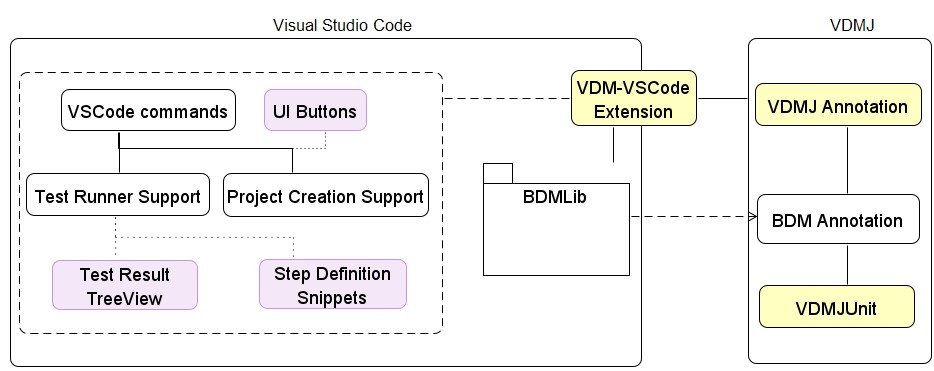}
    \caption{Overview of the \gls{vscode} extension architecture.}
    \label{fig:BDMVSCodeExtension}
\end{figure}

The tool itself is under development with several features in the pipeline. As mentioned above, a \gls{vscode} extension is being developed to provide project creation and testing capabilities within \gls{vscode}. The extension utilises a modified Maven Archetype\footnote{https://maven.apache.org/guides/introduction/introduction-to-archetypes.html}, which has been modified to create a new project that has the necessary structure and files to use \gls{bdm}. This project structure can be seen in figure \ref{fig:project_structure}. The testing capabilities are handled in this extension through a modified maven test command. An overview of the architecture of this extension is shown in figure \ref{fig:BDMVSCodeExtension}. The current version of the extension supports the white elements on the figure, the package \textit{BDM.lib} and the element named \textit{BDM Annotations} are described in section \ref{sec:tecnicalSolution}. The yellow boxes are third party systems. The extension is not yet ready for release while writing this submission but we expect it to be complete by the time of the workshop. The extension should be extended with user interface buttons and a testing view, which are depicted in pink in figure \ref{fig:BDMVSCodeExtension}, since the extension only provides its functionality through commands executed in the terminal and through the \gls{vscode} command line interface. The extension should also provide the user with snippets to step definition since this is standard practice within the Cucumber tooling.  Additionally, the tool should be extended to support the two other dialects of the VDM; VDM-SL and VDM-RT, the tool only supports the VDM++ dialect. 

\subsubsection*{Acknowledgments}
We would like to thank Simon Fraser and Alessandro Pezzoni for inspiring us to approach agile formal methods with \gls{bdd} and for providing information about the Azuki framework. We would also like to thank the Overture community for providing information about extension development in \gls{vscode} and the VDMJ library.

\clearpage


 \newcommand{\noop}[1]{}

\clearpage
\endgroup

\begingroup
\renewcommand\theHchapter{3-Rask:\thechapter}
\renewcommand\theHsection{3-Rask:\thesection}
\locallabels{3-Rask:}
\setcounter{footnote}{0}
\setcounter{chapter}{0}
\setcounter{lstlisting}{0}
\newacronym{ast}{AST}{Abstract Syntax Tree}
\newacronym{ct}{CT}{Combinatorial Testing}
\newacronym{czt}{CZT}{Community Z Tools}
\newacronym{dap}{DAP}{Debug Adapter Protocol}
\newacronym{dbgp}{DBGP}{Common DeBugGer Protocol}
\newacronym{gui}{GUI}{Graphical User Interface}
\newacronym{ide}{IDE}{Integrated Development Environment}
\newacronym{lsp}{LSP}{Language Server Protocol}
\newacronym{poc}{PoC}{Proof of Concept}
\newacronym{pog}{POG}{Proof Obligation Generation}
\newacronym{po}{PO}{Proof Obligation}
\newacronym{slsp}{SLSP}{Specification Language Server Protocol}
\newacronym{vdm}{VDM}{Vienna Development Method}
\newacronym{vscode}{VS Code}{Visual Studio Code}

\makeatletter
\def\input@path{{3-Rask/}}
\makeatother

\graphicspath{{3-Rask/}}

\title{Advanced VDM Support in Visual Studio Code}

\author{ Jonas Kjær Rask\inst{1}
\and Frederik Palludan Madsen\inst{1} 
\and Nick Battle\inst{2} 
\and Leo Freitas\inst{3}
\and Hugo Daniel Macedo\inst{1}
\and Peter Gorm Larsen\inst{1}}
\authorrunning{ }

\institute{
DIGIT, Aarhus University, Department of Engineering, \\
Finlandsgade 22, 8200 Aarhus N, Denmark\\
\email{\{jkr, fpm, hdm, pgl\}@ece.au.dk}
\and
Independent, \email{nick.battle@acm.org}
\and
Newcastle University, \email{leo.freitas@newcastle.ac.uk}
}
			
\maketitle
\begin{abstract}
The VDM VSCode extension has made its way to the daily practice of several engineers and students.  Recent development furnished the extension with many of the features previously only available in the Eclipse based tools. With the new additions, we are able to label the VDM VSCode extension to be the most up to date, preferred and recommended tool support for the VDM practitioner. In addition to keeping up with the previous features, recent developments brought new ones available in VSCode only (e.g.: Code Lenses), advances in proof support, and advances on the interpreter that equip the user with a modern tool suite. This paper provides an update on the current features and lays down the discussion of the future developments.
\end{abstract}

\keywords{Overture, IDEs, VDM, Language Server Protocol, Debug Adapter Protocol}

\section{Introduction}
\label{sec:intro}

The \gls{vdm} is one of the approaches to follow, when
applying formal methods during the development of computer systems. The method has been widely used
both in industrial contexts and academic ones covering several domains of the field: Security 
\cite{Kulik&20,Kulik&21a}, Fault-Tolerance \cite{Nilsson&18}, Medical Devices \cite{Macedo&08}, among others, and its application /acceptance depends on usable and extendable tool support.

The standard open-source tool
support for \gls{vdm}, the Overture tool, has migrated from an Eclipse rich-client
platform application to a \gls{vscode} extension, which supported basic 
editor features as reported in \cite{Rask&20}.
As described in  \cite{Rask&21}, \gls{vscode} is an editor, not an Integrated
Development Environment (IDE), yet with the introduction of the \gls{lsp} and the \gls{dap}, the editor becomes indistinguishable from an IDE. Our implementation of those protocols is thoroughly described in
\cite{Rask&20a} with a comparison of legacy features in the Eclipse based tool and the ones in the new \gls{vscode} extension. 

The prototype matured into a stable tool which was enriched with both the missing legacy features, proof-support and modern features. We report on the new features, and claim that, with the additions, the
VDM VSCode extension has become a fully fledged \gls{vdm} tool support. In addition to the new features, the claim is based on the fact that the extension transitioned from prototype to classroom and daily
practice, which is reflected on its usage metrics (e.g. the extension has over a thousand installs). 

In this paper we report on the recent addition of previously existing features in the Eclipse based tooling and missing in the VDM VSCode extension (legacy features), namely:
\begin{itemize}
        \item Import Examples - Allowing quick demos and facilitating training (e.g.: classroom exercises).
        \item Java Code Generation - The conversion of a VDM object oriented dialects specification into a Java project. 
        \item Add Library - The possibility to select and add libraries to a VDM project.
        \item Remote Control - Extending the console based running and debugging with GUIs.        
        \item Test Coverage - Highlights the parts of the specification that has been exercised.
        \item Proof Support - The option to generate the boilerplate Isabelle theory and spec definitions to be used in proofs.
\end{itemize}
Moreover, we have also augmented the features available to VDM practitioners, namely:
\begin{itemize}
        \item Code snippets - Provide a template/skeleton of a specification primitive by typing a prefix and a triggering the features. 
        \item Code Lenses - A popular feature in Visual Studio Code that we use to allow the launching and debugging of specification components by clicking on links that are interspersed throughout the specification.
        \item Dependency Graph Generation - Providing the visualisation of the class dependency of a object oriented VDM dialect in graphical form.
\end{itemize}
        
Given the completion of added legacy features and its transitioning into practice, we are now able to declare the VDM VSCode extension as a de facto tool support for \gls{vdm}. The outline of this paper is as follows: Section \ref{sec:background} provides an overview of the current \gls{vscode} extension components and features. Section \ref{sec:advanced} describes the advanced features added to the initial extension as reported in \cite{Rask&20,Rask&20a,Rask&21}. Section \ref{sec:proof} contains the details of the recently added proof support.  Section \ref{sec:language} provides insight on the changes made to the previous language server to support the new/modern editing features.  Finally, in Section \ref{sec:conclusion}, we report on the insights and work done so far, while laying out the features under implementation.

\section{Background}
\label{sec:background}
In this section the VDM VSCode extension is described and in this context \gls{vscode} is also briefly introduced. Lastly, the implementation of the language feature support in the extension is made tangible by a concrete example. 

\subsection{The VDM VSCode Extension}
The VDM VSCode extension is developed for the rich extensibility model found in the free source code editor \gls{vscode} which is designed to run on multiple platforms including web and to be lightweight with additional functionality being added by the user through extensions. Amongst other functionality an extension can be used to provide support for a given programming language through the use of a language server that adheres to the \gls{lsp} protocol and potentially also the \gls{dap} protocol if debugging capabilities are needed. 
Both protocols are developed by Microsoft in relation to \gls{vscode} which has native support for much of the front-end parts of the programming language features that are supported by the protocols such as debugging, type checking, go-to definition and more. Thus, \gls{vscode} is a perfect candidate for a modern development tool for \gls{vdm}.
The VDM VSCode extension can, as other extensions, be installed directly from within \gls{vscode} itself which enables the user to quickly and easily add support for \gls{vdm} in the editor.

The extension is continually being developed and since it was last presented \cite{Rask&20} a number of new features and functionality have been made available to the users of the extension.
Its core design implements a client-server architecture that uses the \gls{slsp}\cite{Rask&21}, \gls{lsp}, and \gls{dap} protocols for communication. With the \gls{slsp} protocol being an extension to the \gls{lsp} protocol to enable support for specification language features.

The client implementation in the extension is tightly coupled to VSCode, whereas the server implementation is fully decoupled from the client. The \gls{vdm} language support is provided by the server as an implementation of VDMJ\cite{Battle09} with extended capabilities to support the aforementioned protocols. As a result, the server can be reused for any clients that support the protocols, and vice versa. Figure \ref{fig:ExtensionArchitecture} provides an overview of the high level components in the current extension architecture.

\begin{figure}
    \centering
    \includegraphics[width=0.9\textwidth]{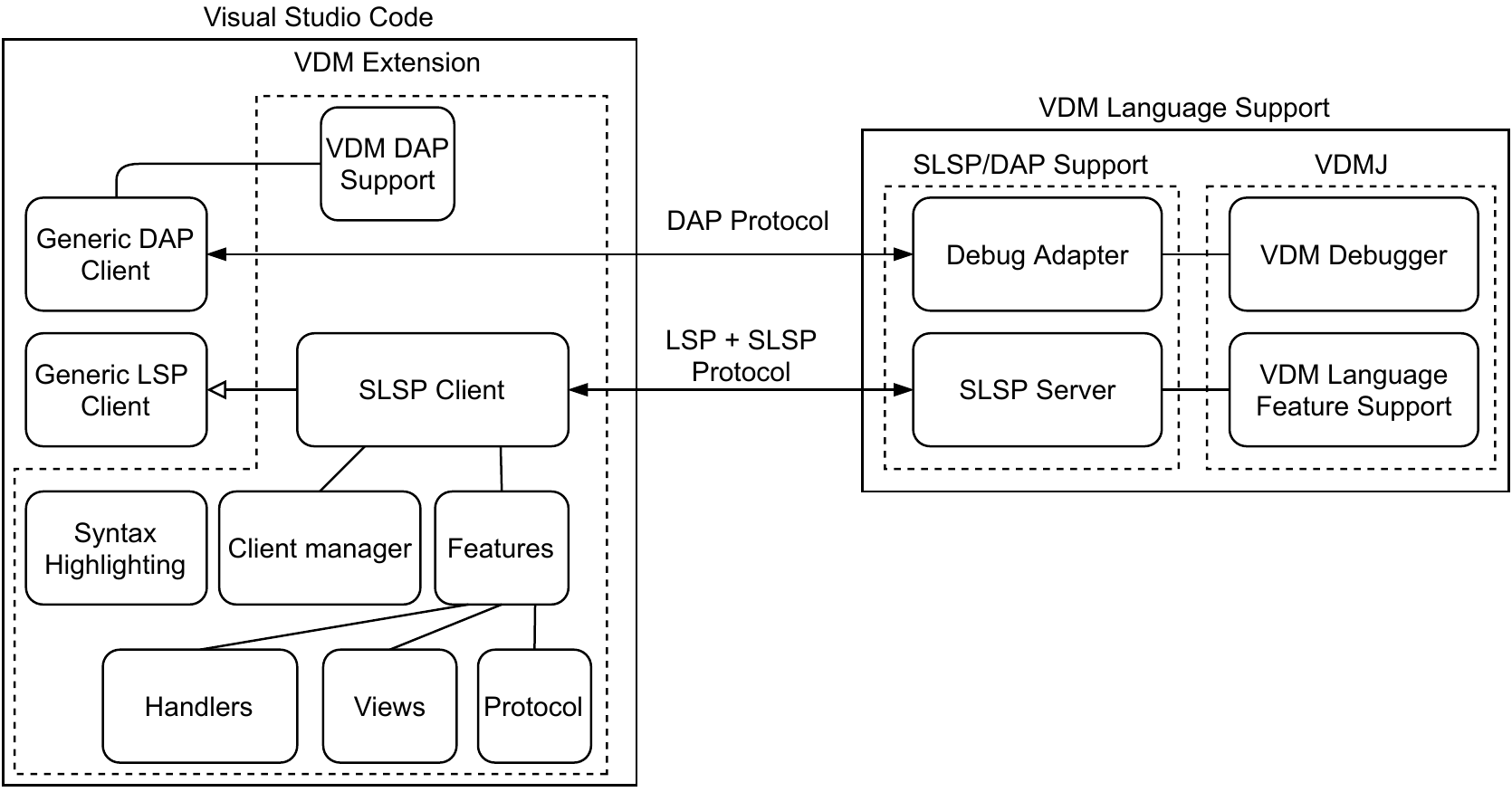}
    \caption{Architecture of the VDM VSCode extension components.}
    \label{fig:ExtensionArchitecture}
\end{figure}

Following is a short description highlighting the purpose and composition of each of the high level components found in Figure \ref{fig:ExtensionArchitecture}.

\begin{itemize}
    \item \textbf{Visual Studio Code:}
    \begin{itemize}
        \item         Generic DAP Client: Native \gls{dap} component in VSCode. It handles the integration of the features enabled by the DAP protocol and provides the messaging infrastructure to handle the protocol on the client side.
        \item         VDM DAP Support: Extends the Generic DAP Client to connect to the Debug Adapter on the server and handles any specific logic for initiating a VDM debugging session.
        \item         Generic LSP Client: Native \gls{lsp} component in VSCode. Handles the integration of the features enabled by the \gls{lsp} protocol and provides the messaging infrastructure to handle the protocol on the client side.
        \item        SLSP Client: Extends the Generic LSP Client to support features that utilises the \gls{slsp} protocol.
        \item        Client Manager: Manages client instances and features for each workspace that the user has opened in a given session.
        \item        Features: A collection of language feature functionality components that utilises the \gls{lsp} protocol. This includes features such as \gls{pog}, \gls{ct} and language translations.
        \item        Views: A collection of view logic for each feature component where user interaction is necessary. These does not directly depend on the \gls{slsp} protocol.
        \item        Protocol: Logic that describes the SLSP protocol messages.
        \item        Handlers: Contains all the components to support functionality that does not directly rely on the \gls{slsp} protocol. This includes functionality such as Java code generation, displaying code coverage, adding \gls{vdm} libraries and more.
        \item        Syntax Highlighting: Uses TextMate grammars to highlight VDM keywords.
    \end{itemize}
    \item \textbf{VDM Language Support:}
    \begin{itemize}
        \item          Debug Adapter: Wraps the VDM Debugger to provide the debug functionality using the \gls{dap} protocol.
        \item          VDM Debugger: Provides the VDM debugger functionality.
        \item          SLSP Server: Wraps the VDMJ language support to provide it using the \gls{slsp} and \gls{lsp} protocols.
         \item         VDM Language Feature Support: Provides the VDM language support functionality.
    \end{itemize}
\end{itemize}

\subsection{SLSP Client Features}

The client side architecture of the extension is designed such that feature functionality which rely on the \gls{slsp} protocol is decoupled from the protocol itself. Thus, the interface elements (buttons, web views, tree views) and their supporting logic in the client are not directly coupled to a client-server architecture.

A concrete example of the decoupling is provided in Figure \ref{fig:POGFCD} which shows a class diagram for the \gls{pog} feature, with the other \gls{slsp} features of the extension following the same design.

Following is a description of key elements of the class diagram relevant for understanding the decoupling and also the integration of \gls{slsp} features in the client in more technical detail. 

As depicted in the diagram the class \texttt{extension} has an instance of the class \texttt{ClientManager} which in turn has multiple instances, one for each workspace, of the class \texttt{SpecificationLangaugeClient}. This class extends the \gls{vscode} class \texttt{LangaugeClient} to enable the use of the functionality that is provided by the \gls{vscode} API. The multiplicity is needed since a client and server instance is created for each workspace folder that is opened in the editor.
In the \texttt{LangaugeClient} class it is possible to register features that implement one of the interfaces \texttt{StaticFeature} or \texttt{DynamicFeature}. Both of these interfaces describe the functions that must be implemented to handle the initialisation of the feature between client and server. However, the features that are supported by the \gls{slsp} protocol currently only supports static registration, hence they implement the \texttt{StaticFeature} interface. For the \gls{pog} feature this is implemented in the class \texttt{ProofobligationGenerationFeature} which handles communication to and from the server and is responsible for providing a data provider for the \texttt{ProofObligationPanel} class which in turn provides the \gls{gui} elements to be displayed in the \gls{pog} view in the editor. Only one instance of the \texttt{ProofObligationPanel} class is created by the extension, as all the workspace folders share the same \gls{po} view. To be able to support all the workspace folders the panel can have multiple \texttt{ProofObligationProvider}s (one for each workspace folder) which provide the data that is shown in the \gls{po} view and subscribe to events that are associated with the feature. Finally, the \texttt{ProofObligationPanel} is also responsible for registering the handlers for the commands relating to \gls{pog} that the VDM VScode extension provides.

\begin{figure}
    \centering
    \includegraphics[width=0.9\textwidth]{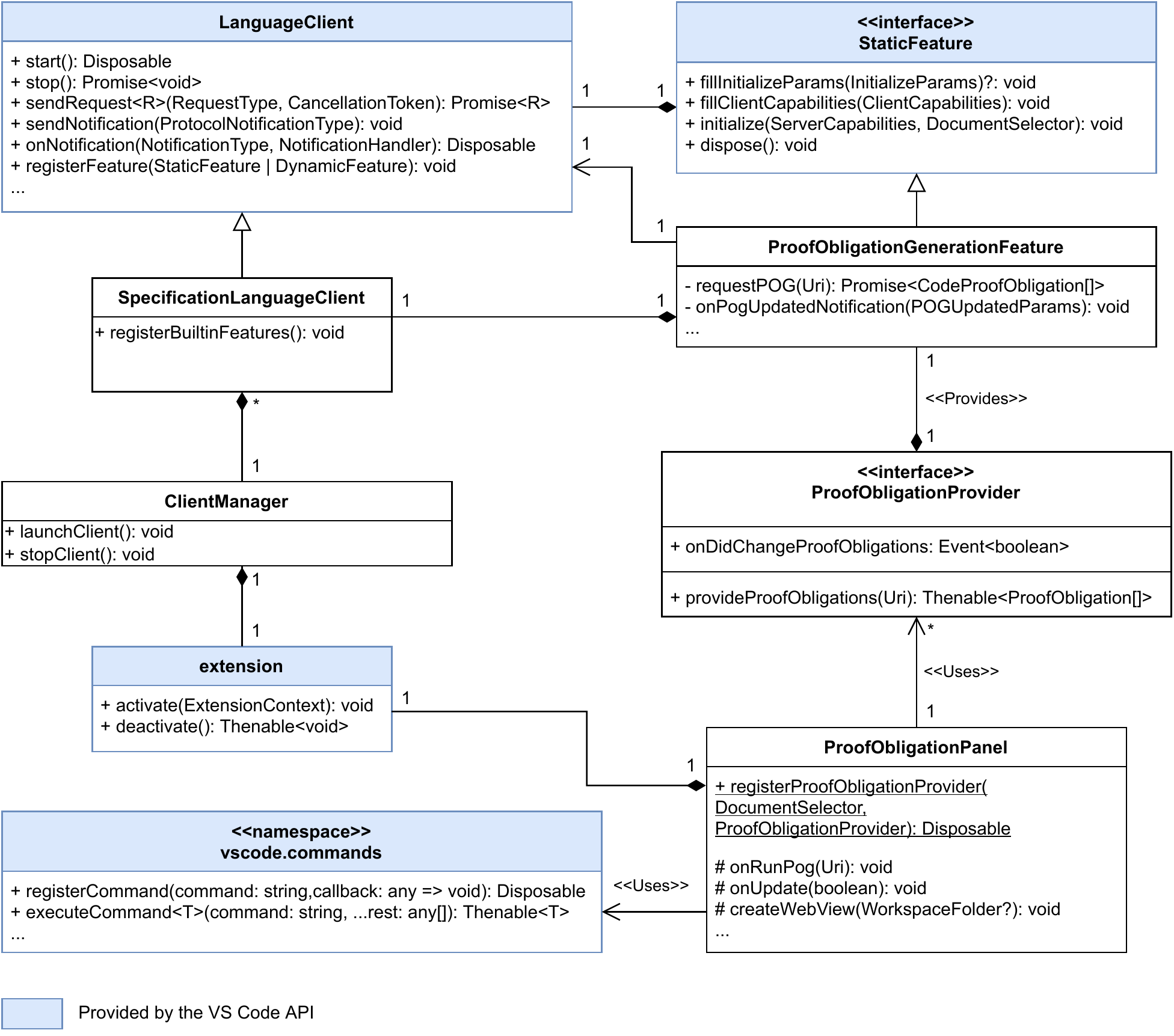}
    \caption{Class diagram for the \gls{pog} feature. For several of the classes, only the functions that are mainly used or help in understanding the diagram are included in the class description.}
    \label{fig:POGFCD}
\end{figure}

\section{Adding Legacy Features}
\label{sec:advanced}


\paragraph{Import Examples} This addition makes it possible to automatically import  project from a large collection of existing examples. As it was the case in the previous Eclipse based IDE, users can experiment with a running example project.

\paragraph{Java Code Generation} It is possible to generate Java code for a large subset of VDM-SL and VDM++ models. In addition
to Java, C and C++ code generators are currently being developed. Both these code generators are
in the early stages of development. For comparison, code generation of VDM-SL and VDM++
specifications to both Java and C++ is a feature that is available in VDMTools [Java2VDMMan,
CGMan, CGManPP]. The majority of this chapter focuses solely on the Java code generator avail-
able in VDM VSCode Extension.

\paragraph{Add Library Support} 
It is possible to add existing standard libraries. This can be done by right-clicking on the Explorer
view where the library is to be added and then selecting Add VDM Library. That will make a new
window as shown in Figure 4.2. Here the different standard libraries provide different standard
functionalities.

\paragraph{Remote Control}        
In some situations, it may be valuable to establish a front end (for example a GUI or a test har-
ness) for calling a VDM model. This feature corresponds roughly to the CORBA based API from
VDMTools [APIMan].
Remote control should be understood as a delegation of control of the interpreter, which means
that the remote controller is in charge of the execution or debug session and is responsible for
taking action and executing parts of the VDM model when needed. When finished, it should return
and the session will stop. When a Remote controller is used, the Overture debugger continues
working normally, so for example breakpoints can be used in debug mode. Moreover, all dialects
(VDM-SL, VDM++ and VDM-RT) support Remote Control. A new configuration with the use of
a remote controller can be started by (see Figure 13.1 for more details):
1. Clicking on the button ”Add Configuration...”
2. Selecting ”VDM Debug: Remote Control (VDM-SL/++/RT)
Then a new snippet (see Figure 13.2) will be created with the remoteControl option. And you
simply have to write the full package/class name of the Remote Control.

\paragraph{Code snippets}
VSCode templates can be particularly useful when you are new to writing VDM models. If you
press CTRL+space after typing the first few characters of a template name, Overture will offer
a proposal. For example, if you type ”fun” followed by CTRL+space, the IDE will propose the
use of an implicit or explicit function template as shown in Figure 5.1. The IDE includes several
templates: cases, quantifications, functions (explicit/implicit), operations (explicit/implicit) and
many more. The use of templates makes it much easier for users to create models, even if they are
not deeply familiar with the VDM syntax

\paragraph{Code Lenses}
    
To make execution and debugging easy using the extension we have added code lenses for all public operations and functions. These are shown above their respective definitions. When pressing Launch or Debug a launch configuration is generated and launched immediately. If the operation/function has parameters you will be prompted to input these, the same applies if the class that the operation belongs to has a constructor that takes parameters.

Example of a lens-generated launch configuration

\begin{lstlisting}
{
  "name": "Lens config: Debug Test1`Run",
  "type": "vdm",
  "request": "launch",
  "noDebug": false,
  "defaultName": "Test1",
  "command": "p new Test1().Run()"
}
\end{lstlisting}

Notice that the name starts with Lens config: if you remove this the launch configuration will not be overwritten the next time you activate a lens.

\section{Proof Support within the VSCode Extension}
\label{sec:proof}



Original proof support for VDM existed within the Mural~\cite{JJLM91} theorem proving system. Since then, theoretical development demonstrated that it was possible to soundly prove VDM theorems in other logical systems~\cite{DBLP:conf/iceccs/WoodcockF08,leomarktoberdorf}. Moreover, a translation strategy from VDM to Isabelle/HOL\footnote{\url{https://isabelle.in.tum.de}} was developed as part of a deliverable in the AI4FM project led by Cliff Jones~\cite{DBLP:conf/fm/FreitasW14}. This includes extended proof support for VDM-style expressions.    

The current VDM proof support stems from the combination of these theoretical results and Isabelle translation strategy\footnote{\url{https://github.com/leouk/VDM_Toolkit}}. It comprises four parts:~1)~VDM VSCode plugin integration;~2)~VDMJ plugins extension;~3)~VDMJ proof annotations;~and 4)~Isabelle VDM toolkit.  

\subsection{VDM VSCode Proof Support Plugin}

VSCode proof support plugin hooks with the translation feature described in Section~\ref{subsec:Features}. It provides access to VDMJ's (native) plugins, through the language server console as an additional user-command, as well as through the IDE menu. 
The VSCode proof plugin also provides translation and proof support configuration information available for users to choose. Practically, it is a wrapping interface to the native VDMJ plugin described next. This separation of concerns is crucial for maintenance and stability of the tool chain.  

\subsection{VDMJ Plugins Extension}


Through the LSP, users can access the VDMJ native plugins interface. This gives access to the VDM AST, typechecker, POG and other tools. The VDM to Isabelle/HOL translation plugin is divided in four parts: 

\paragraph{1.~\textsf{exu}.}~This plugin analyses a type checked VDM AST looking for:~a)~unsupported features (\eg\ Isabelle requires declaration before use);~b)~specification consistency (\eg\ function call graphs ought to also participate in the corresponding function specification);~and c)~``proof-friendly'' VDM style (\eg\ keep type invariants compartmentalised and as close to their defining type as possible), which enables higher automation in predicting how proof unfolding will take place. These checks ensure that there will be no Isabelle type errors as a result of translation. They also help improve proof script automation. For example, if a function \textsf{f} calls \textsf{g}, then its good practice that \textsf{f}'s precondition also call \textsf{g}'s precondition. Of course, this might be spurious, yet can increase proof automation, if present. Similarly, instead of defining type for a numeric sequence ensuring all elements are negative, it is preferable to define a type for negative numbers then make a sequence of that.    

\paragraph{2.~\textsf{vdm2isa}.}~This plugin compiles a type checked VDM AST into a Isabelle/HOL translation tree, which is then executed to generate the corresponding Isabelle theory file. Its execution does not depend on \textsf{exu}, yet lingering \textsf{exu} errors/warnings will entail \textsf{vdm2isa} errors and likely Isabelle/HOL type errors. This stage of the translation strategy transform each VDM top-level definition (\eg\ types, functions, etc.) to their corresponding Isabelle construct. That is, for every VDM module \textsf{M}, a corresponding Isabelle theory file \textsf{M.thy} is generated. Moreover, it also ensures that the (implicit) VDM rules are checked within the translated Isabelle specification. For example, type invariants are implicitly checked as part of the precondition of translated functions.   

\paragraph{3.~\textsf{isapog}.}~This plugin compiles the POG POs into corresponding Isabelle theorems to be proved. Beyond translating the PO predicate itself, the compilation also groups POs within Isabelle local context principle (\ie \textsf{locale}) to ensure that all proofs within that context \textbf{must} be discharged, and to compartmentalise/stratify their proof scripts. That is, for every translated Isabelle theory file \textsf{M.thy} from VDM module \textsf{M}, a further Isabelle theory file \textsf{M\_PO.thy} is generated. The plugin also enables the user to extend the POG with lemmas that might be useful for discharging POs. This is achieved through VDM annotations, described in the next subsection.       

\paragraph{4.~\textsf{isaproof}.}~This plugin analyses the VDM translation tree to generate tentative proof scripts for each of the \textsf{isapog}-generated POs. That is, for every VDM POs Isabelle theory file \textsf{M\_PO.thy}, a proof strategy theory file \textsf{M\_PS.thy} is generated containing tentative proof scripts for each of the POG POs. This obviously imply the plugin depends on successful execution of both the \textsf{vdm2isa} and \textsf{isapog} plugins to access translated VDM definitions and POs. 

\subsection{VDM Annotations}

VDMJ allows the user-defined annotations. These can be processed by either the parser, type-checker, interpreter or POG. For proof support, we include two new types of annotations:

\begin{itemize}
    \item \textsf{@Theorem} and \textsf{@Lemma}.
    
    These annotations enable the user to extend the VDMJ POG with new user-defined proof obligations about the model being developed. There is no semantic difference between lemmas and theorems (\ie they are both boolean expressions that must be true). Theorem expressions must have a unique (global) name and be type correct. If the expression is executable, the interpreter will determine whether the theorem is true (or not); otherwise, the POG will generate the named PO associated with the theorem. 
    
    Beyond documenting specific properties wanted of the model, user-defined lemmas can also be useful to document stepping stones in the later proofs associated with either POG POs or user-defined theorems. These new POs are processed by the \textsf{isapog} and \textsf{isaproof} plugins, as described above. 
    
    \item \textsf{@Witness}.

    This annotation enables explicit user-defined values for types, or specific function/operation calls. The type checker ensures that witness expressions are correct, and the interpreter evaluate them to ascertain whether the witness is valid (\eg\ a positive number as a witness for a new \textsf{nat} sub-type). That is, the witness chosen is executable and satisfy any associated specification~\cite{Jacobs2018}.
    
    These witness expressions can then be used in proofs involving existential introduction. That is, a witness to a record type will be type checked and interpreted; if that succeeds, this is akin to an existentially quantified variable witness, which will be useful for later generation of translated POs proof scripts. 
     
\end{itemize}

\section{Going Beyond The Previously Existing Tools}
\label{sec:language}

\subsection{Analysis Plugins}

The primary purpose of the VDM VSCode language server is to respond to Client requests via the Language Server Protocol or the Debug Adapter Protocol. The Client facing components of the server accept RPC message requests and dispatch these to the a workspace manager for each protocol. The workspace managers maintain the current state of the specification, for example by applying edits when received from the Client. But to perform more advanced language processing, like checking for syntax or type errors, the workspace managers delegate processing to \emph{Analysis Plugins}:

\begin{itemize}
\item
Analysis plugins isolate the language specific processing from the workspace managers. In particular, they hide the difference between modular (VDM-SL) and class based processing (VDM++ and VDM-RT).
\item
Plugins encapsulate the AST specialised for one particular analysis type. VDMJ maps the original AST from the parser into specialised ASTs comprised of classes that perform one analysis, such as type checking, PO generation or execution.
\item
Plugins register themselves and may involve other plugin services while processing, though this is usually coordinated via the workspace managers.
\item
Plugins can be added to the system via configuration, and allow it to be extended with new analyses (such as the Isabelle translator \ref{sec:proof}).
\end{itemize}

The server has three analysis plugins that \emph{must} be available to open any VDM project. These are built into the server:

\begin{itemize}
\item{AST:}
This is the plugin for the basic parse of the specification text. It supplies the AST tree to the other plugins for further processing and it offers "on the fly" syntax error reporting to the workspace manager as text is entered. The plugin also has basic symbol location services that can be invoked if there are type errors and the TC plugin cannot do this.
\item{TC:}
The type checker plugin manages the TC specialised tree and its lifecycle. It returns type related warning and error messages, as well as building the "outline" of the symbols defined by the specification. It also allows definitions to be located (e.g., for "Go-to definition" requests).
\item{IN:}
The interpreter plugin manages the IN specialised tree and its lifecycle. It enables the DAP workspace manager to obtain an interpreter for executing expressions in the specification.
\end{itemize}

In addition to these essential plugins, two further plugins are shipped with the VSCode extension, since they support Client-side UI enhancements:
\begin{itemize}
\item{PO:}
The proof obligation analysis plugin manages the PO specialized tree. It offers a service to build a list of proof obligations, either for the entire specification or for just one file. It is complemented by a Client panel to display the list of obligations and navigate from an obligation to its location in the specification.
\item{CT:}
The combinatorial test analysis plugin manages the CT specialised tree and, in combination with the IN plugin, enables the server to generate and execute combinatorial tests. The server feature is complemented by a UI panel to allow the generated tests and results to be explored.
\end{itemize}

Further analysis plugins can be added by implementing the \emph{AnalysisPlugin} interface. These are instantiated by the LSPX workspace manager (which handles extensions to LSP), by reading a system property that defines the classes to load. Such dynamic plugins register themselves like any other plugin. They typically create and manage their own tree of specialised classes, using the VDMJ \emph{ClassMapper}.

The LSPX workspace manager handles non-standard LSP method requests by asking the registered plugins whether they can process the request. The expectation is that a new Client feature will send new SLSP requests that are handled by a new user plugin.

\subsection{Code Lenses}

The LSP protocol offers language servers the ability to label parts of the source text with a clickable tag that then invokes an arbitrary command on the Client. These are called "code lenses". This request is offered to each of the analysis plugins when a Client code lens request is received.

Only one code lens is currently generated, by either the AST or TC plugins (depending on whether there are type errors). The lens offers "Launch" and "Debug" tags for executable functions or operations and links these with a Client side command that allows the user to quickly build and launch or debug the definition.

But this is a powerful feature and we expect new plugins to add their own lenses. For example, the Isabelle translation plugin may be able to link VDM definitions to the equivalent in the Isabelle environment.

\subsection{User Libraries}

A library is comprised of one or more VDM sources, in a variety of dialects, with any associated native Java code. These are packaged into a regular Java "jar" file, with one additional file in the "META" folder, called "library.json". The meta-file defines the source files that should be included for a given VDM dialect as well as dependencies on other libraries. The VDMJ "stdlib" meta-file is a good example (below).

The extension allows new libraries to be included in a workspace and selectively added to projects.

\begin{verbatim}
{
    "vdmsl": [
        {
            "name": "IO",
            "description": "The IO library",
            "files": ["IO.vdmsl"],
            "depends": []
        },
        {
            "name": "MATH",
            "description": "The MATH library",
            "files": ["MATH.vdmsl"],
            "depends": []
        },
        ...
    ],
    "vdmpp": [
        {
            "name": "IO",
            "description": "The IO library",
            "files": ["IO.vdmpp"],
            "depends": []
        },
        {
            "name": "VDMUnit",
            "description": "The VDMUnit library",
            "files": ["VDMUnit.vdmpp"],
            "depends": ["IO"]
        },
        ...
    ],
    "vdmrt" : [
        ...
    ]
}
\end{verbatim}

\paragraph{Dependency Graph Generation}

The VDM VScode extension allows the generation of a dependency graph that represents the dependencies of the classes/objects in a specification. Thus, it is
considered as a directed graph where each node points to the node on which it depends. In some
cases you can add some conditions set on the different connections between the nodes. Moreover,
each shape represents a node (usually ellipses or circles) and each connector, composed of one
or two arrow heads, indicates the direction of the dependencies. You have also the possibility to
add labels on connectors to specify the relation between two nodes. To finish, the main usage
of a dependency graph consists of describing processes to make it easier for the developer to
understand, reuse and maintain his code. An example of such a dependency graph is shown on Figure \ref{fig:DependencyGraph}

\begin{figure}[htbp]
    \centering
        \includegraphics[width=0.53\textwidth]{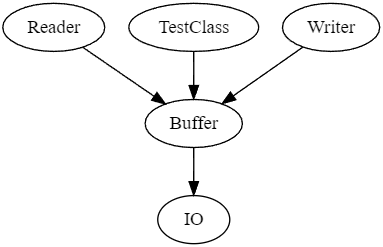}
    \caption{Example of a visualisation of a dependency graph file generated by the extension. The visualisation is performed by an external tool.}
    \label{fig:DependencyGraph}
\end{figure}

\section{Concluding Remarks and Future Work}
\label{sec:conclusion}

In this paper, we report on the recent additions to the VDM VSCode extension taking the current support of VDM to a stage where most of the legacy features are supported and new ones are added. The latter are specially important, as they allow the VDM practitioner to use features, which are available in the integrated development environments they are used to, but were not available while writing VDM specifications.

\subsection{Features Completeness}\label{subsec:Features}

VDM VSCode is now labelled as the most up to date, preferred and recommended tool support for the \gls{vdm} practitioner. To provide a brief overview of the features of VDM VSCode a comparison to Overture, which can then be considered as the prior state of the art tool, is made in Table \ref{tab:featureCompare}.
As illustrated, some features found in Overture is still missing in VDM VSCode, however these are actively being developed. In addition, VDM VScode has support for features like `Translate to Word` and `VDM to Isabelle` which are not supported by Overture.
The features `VDM to LLVM` and `VDM to Python` are not supported by either Overture or VDM VSCode. But, they are being considered for future development and implementation in VDM VSCode.

\begin{table}[htbp]
\centering
\caption{Feature comparison between VDM VSCode and Overture.}
\label{tab:featureCompare}
\begin{tabular}{|l|c|c|c|}
\hline
\textbf{Features}         & \textbf{Overture} & \textbf{VDM VSCode} \\ \hline
Syntax highlighting                 & X                & X                 \\ \hline
Syntax check                        & X                & X                 \\ \hline
Type check                          & X                & X                 \\ \hline
Evaluation                          & X                & X                 \\ \hline
Debugging                           & X                & X                 \\ \hline
Proof obligation generation         & X                & X                 \\ \hline
Combinatorial testing               & X                & X                 \\ \hline
Translate to LaTex                  & X                & X                 \\ \hline
Translate to Word                   &                  & X                 \\ \hline
Extract VDM from Word               & X                & X                 \\ \hline
Code completion                     & (X)              & (X)               \\ \hline
Template expansion                  & X                & X                 \\ \hline
Standard library import             & X                & X                 \\ \hline
Go-to definition                    & (X)              & X                 \\ \hline
Coverage display                    & X                & X                 \\ \hline
VDM to Java                         & X                & (X)               \\ \hline
VDM to XMI (UML)                    & X                &                   \\ \hline
VDM to Isabelle                     &                  & (X)               \\ \hline
VDM Example Import                  & X                & X                 \\ \hline
Real-time log viewer for VDM-RT     & X                & X                 \\ \hline
Launch in VDMTools                  & X                & X                 \\ \hline
FMU wrapper                         & X                & X                 \\ \hline
VDM to C                            & (X)              &                   \\ \hline
VDM to LLVM                         &                  &                   \\ \hline
VDM to Python                       &                  &                   \\ \hline
\end{tabular}
\end{table}

\subsection{Future Work}

\paragraph{VDM-RT Log Viewer.} As described by Fitzgerald et al. \cite{Fitzgerald&07h} real-time traces, which can be generated when executing a VDM-RT model, can provide insight into the ordering and timing of exchange of messages, activation of threads and invocation of operations and combined with validation conjectures\cite{Fitzgerald&07h} enables explicit consideration of system-level properties during the modelling process. As such, a tool for displaying the traces and validation conjectures can provide valuable information. There is therefore ongoing work on implementing a log viewer tool in the VDM VSCode extension similar to the log viewer tool found in Overture today.

\paragraph{Behaviour Driven Development (BDD).} VDM modelling involves the transformation of a set of elicited requirements listed in a natural language into an executable model. the practice is to transform the requirements into the specification as a jump which consists of a mental exercise including the iteration of two steps: first understand the requirements, then formulate them as a formal model. It is possible to aid professionals in this jump by adopting the approach of BDD, where requirements are translated into behaviours. The behaviours
are written in a constrained subset of natural language with the goal of becoming more
understandable by stakeholders. Tool support has been developed in both \cite{Fraser&21} showing its feasibility and recent efforts are in place to develop a VSCode extension supporting BDD in VDM \cite{Villadsen&22}.

\paragraph{UML/VDM translation.} The UML connection, allowing the bidirectional translation between object oriented VDM models and UML diagrams, a legacy feature supported in the Eclipse based tools is still not supported in the VSCode extension. The legacy UML transformations were developed with the informal modelling tool, Modelio, in mind, which require the installation of external tools. New and modern UML tools have been developed and are integrated in the VSCode marketplace since the Modelio based translations, and we are working to re-purpose the connection inside the ecosystem. A preliminary account of the new connection can be found in \cite{Lund&22}.

\paragraph{FMI Support.} Though, currently, the VSCode extension allows the generation of Function Mock-Up Units (FMUs), its implementation is still dependent on the source code from the existing Eclipse based solution, given that the generated FMUs wrap the Eclipse based interpreter. This solution allows 
the usage of tested and reliable code, but introduces the potential for discrepancies between the results when the model is launched/debugged using the VDMJ based LSP-server and the results produced by executing the FMU, given that two different interpreters are used. To simplify codebase maintenance and make sure results are ran with the same interpreter, we plan to overall the VDMJ scheduler and create a new FMI implementation directly in it, thus completely deprecating the Eclipse-based solution.

\subsubsection*{Acknowledgements}
We acknowledge the Poul Due Jensen Foundation for funding the project Digital Twins for Cyber-Physical Systems (DiT4CPS).

 \newcommand{\noop}[1]{}

\clearpage
\endgroup

\begingroup
\renewcommand\theHchapter{4-Lund:\thechapter}
\renewcommand\theHsection{4-Lund:\thesection}
\locallabels{4-Lund:}
\setcounter{footnote}{0}
\setcounter{chapter}{0}
\setcounter{lstlisting}{0}

\makeatletter
\def\input@path{{4-Lund/}}
\makeatother

\graphicspath{{4-Lund/}}

\title{Towards UML and VDM Support in the VS Code Environment}

\author{Jonas Lund \and  Lucas Bjarke Jensen \and Hugo Daniel Macedo \and Peter Gorm Larsen }
\authorrunning{ }

\institute{
DIGIT, Aarhus University, Department of Engineering, \\
Finlandsgade 22, 8200 Aarhus N, Denmark\\
\email{\{201906201,201907355\}@post.au.dk, \{hdm,pgl\}@ece.au.dk}
}
			
\maketitle
\setcounter{footnote}{0} 
\begin{abstract}
The coupling between the object-oriented dialects of VDM (VDM++ and VDM-RT) and UML, found on The Overture Tool has been left behind due to the shift in focus onto VDM VSCode. The paper first presents how the coupling is reestablished in VS Code. However, it is subsequently proposed that UML visualisations of VDM models can be further improved by using the text-based diagram tool called PlantUML, as it is supported as a VS Code extension and offers several advantages compared to other UML tools. The best integration of PlantUML requires a direct translation between VDM and PlantUML, which is made possible by class mapping introduced in VDMJ 4.

\end{abstract}


\section{Introduction}
\label{sec:intro}

The modelling of critical and embedded systems is a complex task with many parts. One of the most popular and mature languages designed for modelling such systems is the Vienna Development Method (VDM) which is extended by the two object-oriented (OO) dialects, VDM++ and VDM Real-Time (VDM-RT) \cite{Fitzgerald&05}. Some advantages of modelling using the OO dialects are that encapsulation of methods and data along with abstraction of the internal state of these classes allows for higher degrees of model complexity. 

The task of modelling these systems is complex and requires a deep understanding of the modelling tools in use. The widely used Overture Tool built on top of the Eclipse IDE allows for easier modelling integrating several debugging tools and plugins. One of these plugins is the transformation from and to Unified Modelling Language (UML) files \cite{Lausdahl&08}. This connection was developed because UML class diagrams are a fitting and intuitive way of visualising OO models.

The Eclipse version of Overture is slowly growing obsolete as a new contender rises, the VDM VS Code extension, letting Visual Studio Code (VS Code) function as an IDE for modelling in languages from the VDM language family. Many of the features from Eclipse have been ported to the extension already, but the UML connection has yet to be incorporated. 

The UML transformations for The Overture Tool culminated in the coupling with the visual modelling tool, Modelio. However, the coupling has its flaws and shortcomings with regards to visualising the generated UML models and exporting new ones. New tools that integrate more closely with modeling tools have been developed and it may be worthwhile looking into whether the connection between VDM and UML can be repurposed for such tools. 

This paper will be tackling two tasks. The first is covering the work that has gone into porting the already existing UML connection from the Eclipse version of Overture to the new VDM VSCode extension. The second is discussing other tools for visualising UML and designing a new transformation plugin for this tool.



\section{Background}
\label{sec:background}


\subsection{Vienna Development Method and Unified modelling Language}

The Vienna Development Method (VDM) is a leading formal method, focused on the modelling of computer-based systems while ensuring safety and security in the software before deployment \cite{initiative}. There exist two dialects that extend the original VDM language, the VDM Specification Language (VDM-SL). These are the object-oriented extensions called VDM++ and VDM Real-Time (VDM-RT), which support concurrency and real time modelling as well. Both VDM++ and VDM-RT use classes to structure their models. A visual representation of a model helps in designing and communicating the architecture of the system. This is at its most effective via class diagrams using the Unified modelling Language (UML) \cite{UML251}. UML is specified by the Object Management Group (OMG) and uses a semi-formal visual language to give an abstract representation of object-oriented systems. 

A set of transformations between VDM and version 2 of UML exist on the Overture Project as a plugin for the Overture Tool. \cite{initiative} These transformations were made with Modelio in mind as the UML modelling tool, both for import of generated UML files from VDM and for export of class diagrams to be translated to VDM. 

\subsection{From The Overture Tool to Visual Studio Code}

The Overture Tool (Overture) is the most mature and popular platform for VDM modelling. However, the focus of the Overture project has shifted to the Visual Studio Code (VS Code) platform in recent years with the development of the VDM VS Code extension developed by Jonas Rask and Frederik Palludan Madsen \cite{Rask&20}. 

This extension establishes a connection to a language server, letting VS Code function as an Integrated Development Environment (IDE), exactly like its Eclipse counterpart, while also having the advantage of being extendable to other languages thanks to the Language Server Protocol (LSP). \cite{Rask&21}

The Eclipse UML transformation plugin has been reimplemented as a command on VDM VS Code. This was achieved by packaging the main functions \verb|Vdm2UmlMain| and \verb|Uml2VdmMain| into a JAR file that can be run from VS Code. Currently, this JAR file is located on the client side of VDM VS Code, but will be moved to the VDMJ language server in the future to decouple the specification language functions from the VS Code extension.

\subsection{VDMJ}

VDMJ is a command line tool that provides basic tool support for Overture. Some of the features it provides are type checking, proof obligations, and combinatorial test generation among many others. In Overture for Eclipse, VDMJ and its tools are executed through the Eclipse IDE interface. For VDM VS Code however, the VDMJ features are accessed through the LSP, letting VS Code act as an interface for VDMJ. The latest version of the tool, VDMJ 4, lets the user easily reconstruct the VDM Abstract Syntax Trees (AST) for their desired tool related purposes \cite{Battle17}.






\subsection{PlantUML}

PlantUML \cite{plant_info} is a text-based diagram tool that exists as a VS Code extension \cite{plant_vscode}. Any model made in PlantUML can be opened in a separate tab in VS Code and continually updates as the model is changed. The models can be worked on by multiple team members as it integrates with versioning tools like GitHub. This is possible since the PlantUML diagrams are generated from text files and can therefore be located in the software repository alongside the program or model which the diagram represents. 

PlantUML also seeks to allow interoperability between tool vendors as many export formats are supported (including XMI). It also attempts automatic placement of objects in the diagram and allows for potential adjustments from the user, a feature which was not available in Modelio.

One caveat in representing VDM models using PlantUML is that qualified associations, which are commonly used in VDM class diagrams,  but alternatives are available such as using square brackets to indicate the qualifier. A typical class diagram that represents a VDM++ model is presented on Fig. \ref{fig:alarm}.

\begin{figure}[h!]
    \begin{adjustbox}{center}
    \includegraphics[width=0.65\textwidth]{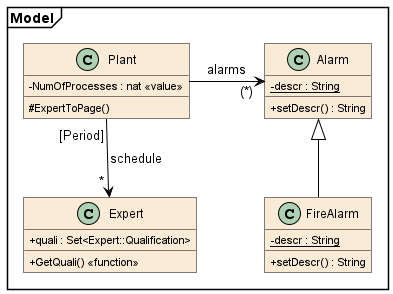}
    \end{adjustbox}
    \caption{Resulting PlantUML}
    \label{fig:alarm}
\end{figure}

\noindent
PlantUML is even cited in books pertaining to critical \cite{CECRIS} and cyber-physical \cite{AMADEOS} systems. These features make the PlantUML extension one of the more relevant options for fulfilling the goals of this project.


\section{Translation Between VDM and UML}
\label{sec:trans}

This section gives an overview on the migration of the UML connection from Overture to VDM VSCode. To implement the connection, the existing Overture code for converting VDM to UML is packaged as a JAR file and called within VS Code. This method does not involve changing the code which is executed to perform the conversions. It involves the creation of a handler that reads the files, checks the validity of these files, and passes the expected arguments to the converting methods. 

This is the first part of the two that this paper will cover. The second part regarding PlantUML is presented in section \ref{sec:plant}.

\subsection{Overture UML Migration to VDM VS Code}

In order to port the existing UML connection to VDM VSCode, the connection has to first be decoupled from the Eclipse IDE. This is a necessary step in allowing Maven to package the command in a JAR file. After the decoupling, the main handlers for the UML transformation can be developed.

\begin{figure}[h!]
    \begin{adjustbox}{center}
    \includegraphics[width=\textwidth]{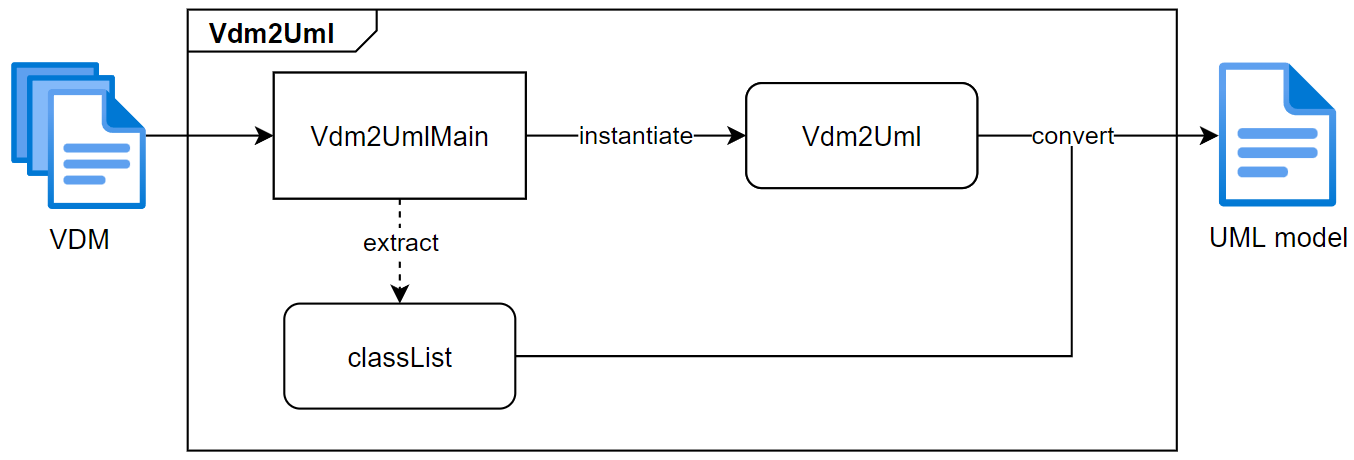}
    \end{adjustbox}
    \caption{VDM to UML Architecture Overview}
    \label{fig:VDM2UML}
\end{figure}
\noindent
\texttt{Vdm2UmlMain} is the transformation handler in the direction of VDM to UML. As illustrated in Fig. \ref{fig:VDM2UML}, it receives the path to the folder containing the VDM files as input. An instance of the \verb|Vdm2Uml| class is constructed, which is the class containing the methods for the transformation.
\verb|Vdm2UmlMain| then extracts a list of classes from the VDM files, called \verb|classList|. This list is given as input to the \verb|Vdm2Uml| method called \verb|convert|. 

\begin{figure}[h!]
    \centering
    \includegraphics[width=\textwidth]{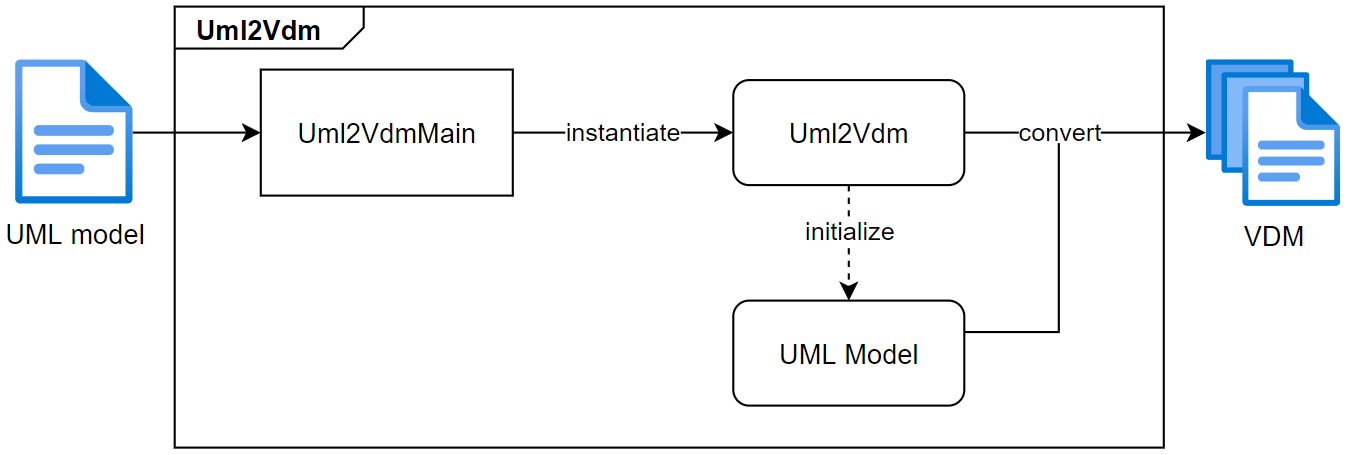}
    \caption{UML to VDM Architecture Overview}
    \label{fig:UML2VDM}
\end{figure}
\noindent
\texttt{Uml2VdmMain} is the main function in the direction of UML to VDM. As illustrated in Fig. \ref{fig:UML2VDM}, it receives the path to the UML file as an input. An instance of the \verb|Uml2Vdm| class is constructed, after which the class is initialised using the \verb|initialize| method with the path to the UML file as an argument. The \verb|convert| method is then called, and the corresponding VDM files are created.

\begin{figure}[h!]
    \centering
    \includegraphics[width=\textwidth]{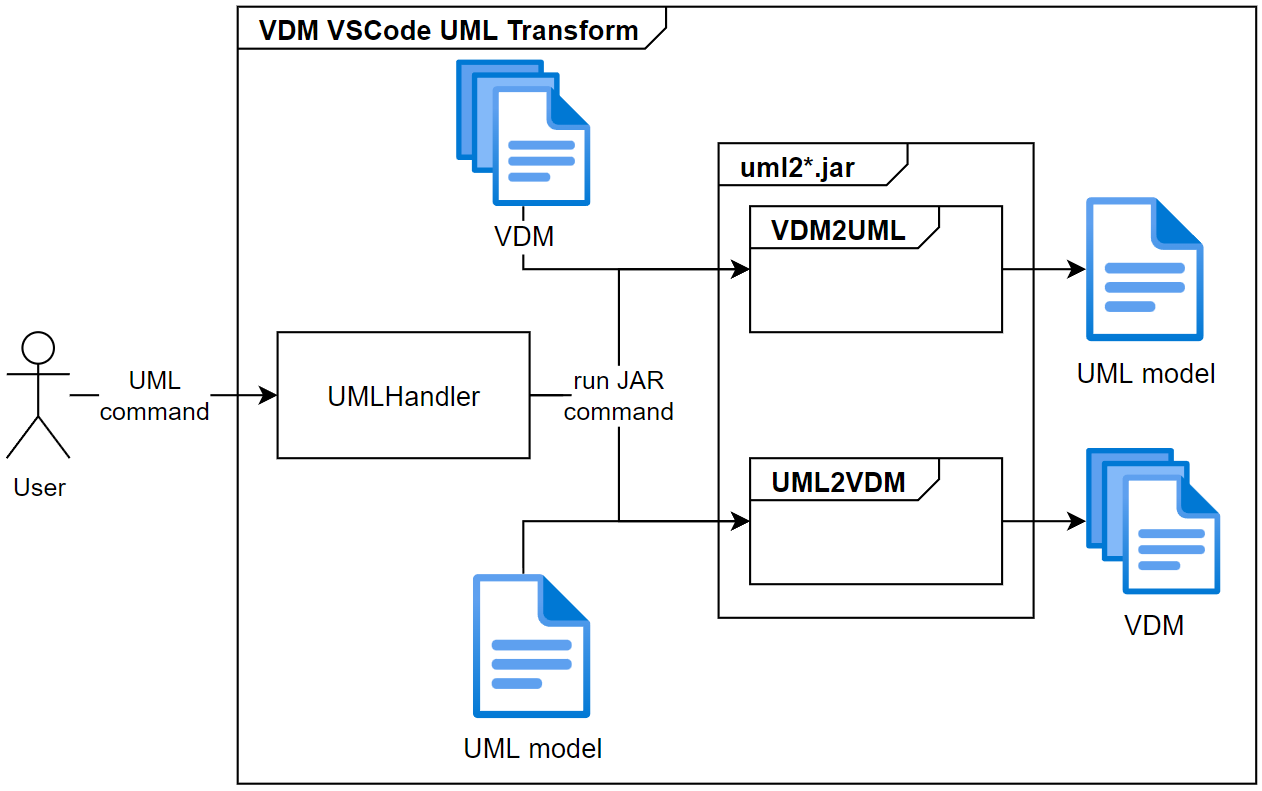}
    \caption{VDM VS Code Architecture Overview}
    \label{fig:VS CodeArch}
\end{figure}
\noindent
The UML transformations are packaged in a JAR file using Maven. The JAR file is integrated into the VDM VS Code environment where it is accessed through a handler called \verb|UMLHandler.ts|. This handler is responsible for executing the JAR file with appropriate arguments. Both main functions are contained in the same JAR. The function executed is determined by which command is issued by the VDM VSCode user. The architecture of the JAR file integration is illustrated in Fig. \ref{fig:VS CodeArch}.

\subsection{Repurposing the Overture Code for UML Transformations}

To understand how the VDM VSCode UML transformation handlers will fill the role of the Overture handlers, while enabling VS Code integration, this section will describe what resources the Overture implementation handlers used in the UML transformations, and whether these resources are fit for reuse.

\subsubsection{Analysis of the Overture UML Handlers}
 To extract the information needed to perform the UML conversions, the handler in the direction of VDM to UML, called \verb|Vdm2UmlCommand|, relies heavily on the \verb|IVdmProject| class.

The \verb|IVdmProject| class is inherited from the \verb|IProject| class, which is an Eclipse-based interface that grants the notion of a project as an implementation resource. A project encapsulates information about a group of files that share a directory. Furthermore, the need for any file handling is resolved as this is also encapsulated by the Eclipse interface.

The \verb|IVdmProject| class is modified to suit VDM and it therefore provides further abilities that are specific to a group of VDM files. In a similar manner, as \verb|IVdmProject| is derived form the \verb|IProject| class, the \verb|IVdmProject| class contains a method for creating an instance of the \verb|IVdmModel| class, which in turn encapsulate the abstract structure of the VDM model. This model is then used to examine whether the model is syntax and parse correct and therefore fit for UML transformation. The code snippet below shows how the \verb|IVdmModel| class is instantiated and used to check the correctness of the model. 

\begin{codebox}
\begin{lstlisting}[language=Java]
final IVdmModel model = vdmProject.getModel();
	if (model.isParseCorrect())
	{
		if (model == null || !model.isTypeCorrect())
		{...}
\end{lstlisting}
\end{codebox} 
\noindent
The Eclipse project classes and the functionality they offer are illustrated in Fig. \ref{fig:project}.

\begin{figure}[h]
    \begin{adjustbox}{center}
    \includegraphics[width=\textwidth]{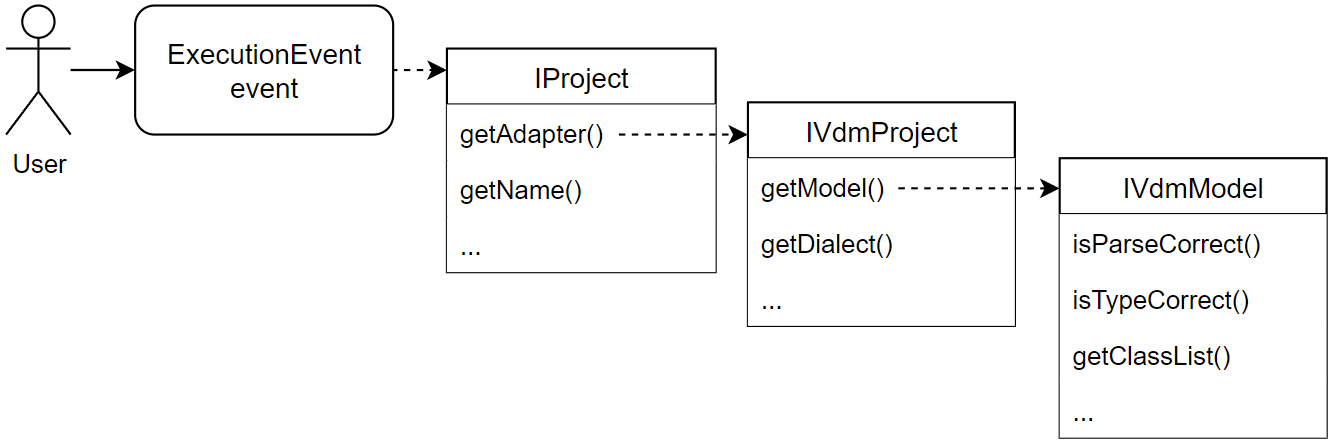}
    \end{adjustbox}
    \caption{Overview of IProject its derived classes, and the methods used in the Overture VDM to UML handler}
    \label{fig:project}
\end{figure}
\noindent
Likewise, the handler in the direction of UML to VDM, called \verb|Uml2VdmCommand| also use Eclipse dependent classes to fulfil its tasks. The class \verb|Uml2Vdm|, which contains the method that performs the conversion needs to be instantiated with the path to the UML file along with information about what dialect the to be generated VDM files should be in. The UML file path is extracted from an instance of the \verb|iFile| class, which encapsulates the Eclipse notion of a file. This class can then be cast to an \verb|IVdmProject| class, which in turn can be used to discover the dialect of the to be generated VDM files.

Considering the goal of migrating the UML connection to VDM VS Code, the handlers' use of the project classes poses a problem, since their usefulness is dependent on the Eclipse IDE. This stems from the fact that the input to the handler has changed from an \verb|IVdmProject| in Overture, to simply being a list of VDM files in VDM VSCode. 

One way to solve this could be to extract information about the current VS Code project, create an intermediate Eclipse project, and then perform the transformation like the Overture UML handlers do. However, this runs into the problem of having to depend on the IDE side. To circumvent this, one might also simply remove the notion of an \verb|IProject| and its derived types. It would then require finding new ways of providing the same functionality as these classes provide. 

\subsubsection{The VDM VSCode UML Handlers}

The VDM VSCode handler achieves the same functionality as the Overture handler, without utilising the notion of an Eclipse project. In a lot of cases, this is done simply by accessing the methods used by the \verb|IProject| and its derived classes. This results in the same functionality, but the methods used are accessed at a lower level. This section shows examples of how this strategy is applied.

In the implementation of the Overture handler, the file handling is done implicitly by Eclipse. However, as this is no longer a possibility, the file handling is done internally by \verb|Vdm2UmlMain|. This is achieved by the \verb|Vdm2UmlMain| method '\verb|main|', taking a set of arguments which denote the directory containing the files to be transformed and the directory where the resulting UML file should be placed. The arguments are distinguished using \verb|-folder| and \verb|-output| flags before the path arguments. 

On Overture, the handler read the VDM dialect using a method from the \verb|IVdmProject| class. The VDM VSCode handler is instead passed the dialect as a flag, along with the command to execute the JAR file.   

While the \verb|IVdmModel| class has an attribute informing whether the model is syntax and type correct, the methods for performing the actual checks are not dependent on the Eclipse project classes, and are therefore fit for reuse. The class list is also provided by the type checker, \verb|typeCheckPp| or \verb|typeCheckRt|, depending on the dialect of the VDM files. The two methods of the class \verb|TypeCheckerUtil| initialise an instance of the \verb|TypeCheckResult| class, which has an attribute containing the class list needed for the conversion. 

In the direction of UML to VDM, instead of the path to the UML file being extracted from the \verb|iFile| instance, the path is instead passed as an argument in the command to execute the transformation on the JAR file. Currently the dialect of the resulting VDM files are assumed to be VDM++. The consequence of this is that a UML file containing a VDM-RT model will be changed to a VDM++ model. This short-coming should be trivial to fix in the future.

With the handlers working and ported to the VDM VSCode extension, the UML transformation commands can be run and tested in VS Code. A screenshot demonstrating how the feature is used is shown in Fig. \ref{fig:VS Code}.

\begin{figure}[h!]
    \centering
    \includegraphics[width=\textwidth]{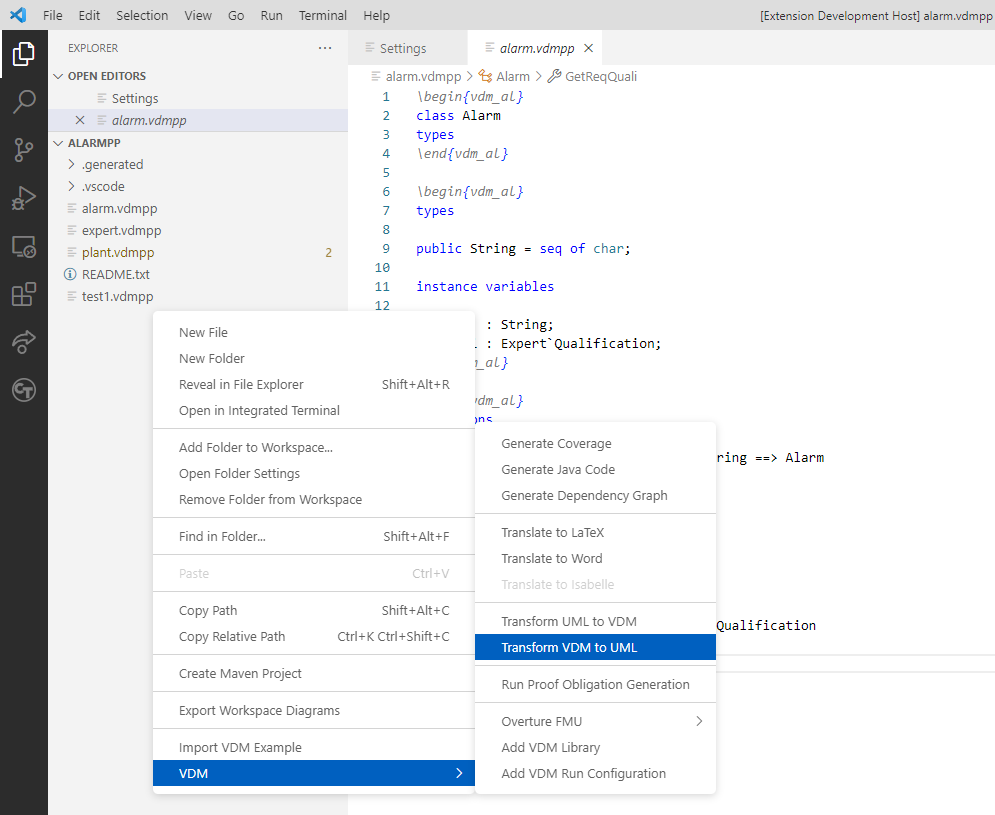}
    \caption{Screenshot of using the command in VS Code.}
    \label{fig:VS Code}
\end{figure}

\clearpage
\section{PlantUML Implementation Planning}
\label{sec:plant}

%
%
%

This section covers the second part of the paper, which involves describing the motivation behind choosing PlantUML as the primary UML visualisation tool, as well as determining the best way to couple PlantUML with VDM VSCode. 

This is almost entirely separate from the porting of the existing UML connection, and will instead dive into the possibility of using the new architecture of the VDM VSCode environment to create a new UML connection.

\subsection{Motivation}

\noindent
Currently, the UML transformations are coupled with Modelio version 2.2.1, as the original Overture UML connection was specifically tailored to this tool. This poses some issues: 

\begin{itemize}
    \item The specific version of XMI used is compatible with only some versions of Modelio, and brings little compatibility with any other tools, restricting interoperability.
    \item There is no spatial information in the standard and Modelio makes no attempt at guessing it, meaning that classes and links must be positioned manually.
    \item Employing a JAR file that uses Eclipse IDE resources is undesirable, since the Overture project is trying to distance itself from Eclipse with the migration to VDM VSCode. Furthermore, having to rely on Eclipse results in making it significantly more difficult for the UML transformations to achieve maintainability and extendibility.
\end{itemize}
\noindent
It would therefore be optimal to use a VS Code extension to visualise UML models that can import and export UML files that are XMI compliant, since this would allow for developing models directly in VS Code. Unfortunately, no such tool seems to exist. 

VDMJ version 4, however, introduces class mapping \cite{Battle17}, allowing developers to easily reconstruct the VDM ASTs for whichever tool-related purposes needed. It may well be possible to utilise class mapping to develop an AST for a textually based UML tool, that is supported in a third party VS Code extension. This would, in turn, allow a direct translator between UML and this tool to be created, solving the problems laid out and thus achieving a direct connection between VDM and UML on the language server, without relying on outdated Eclipse resources.
\newpage

\subsection{Transformation Overview and Considerations}
In order to begin development on the PlantUML transformation using ASTs on the VDMJ language server, some notions must be considered. 

The previous UML transformations also use ASTs, specifically for the object-oriented VDM dialects and a UML AST that would be used for converting to the XMI format \cite{Lausdahl&08}. It is worth considering how much of the implementation of this previous transformation would be reusable for the PlantUML transformation. Since a UML AST does seem to exist, would it be possible to repurpose it for the new VDMJ, and develop translators between UML and PlantUML? Or is it better to take inspiration from the previously used ASTs and instead create a new AST for PlantUML?

The methods for developing ASTs have improved greatly since the Overture UML implementation with the introduction of VDMJ 4 \cite{Battle17}. This has been achieved with class mapping, and it is because of this progress that it seems viable to produce an AST ourselves for the purpose of UML transformations. An overview of how the transformation between VDM++ and PlantUML could work is illustrated in Fig. \ref{fig:puml-vdm++}. The dashed arrows denote the PlantUML to VDM++ direction and the solid lines denote the VDM++ to PlantUML direction.

\begin{figure}[h]
    \begin{adjustbox}{center}
    \includegraphics[width=\textwidth]{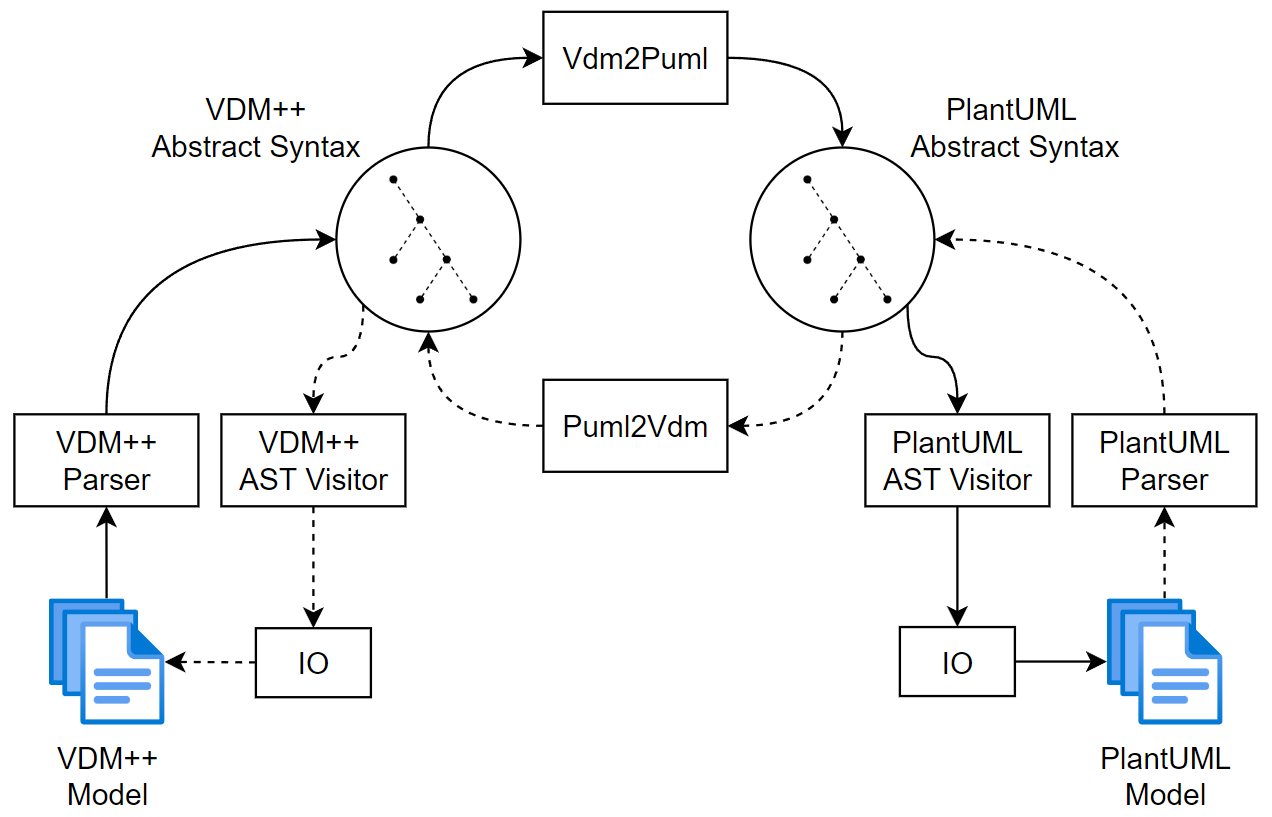}
    \end{adjustbox}
    \caption{Overview of the PlantUML transformations using ASTs.}
    \label{fig:puml-vdm++}
\end{figure}
\noindent
Note that there are several uncertainties as for how the transformations will actually work, since we have yet to properly examine what the process of creating ASTs entails. One uncertainty being how the nodes of the trees are traversed to output the code for both the VDM model and the PlantUML model. It is expected to be done through some visitor method, similar to how it was done for the previous VDM++ AST \cite{Lausdahl&08}. Whether or not the same can be done for the PlantUML AST is also uncertain.

The conversion between the two ASTs is expected to make use of the class mapping mechanism. An example of the class mapping in use can be seen in a simple VDM to C translator \cite{v2c}. The translator makes use of mapping objects from the type checker (TC) AST to a new translator (TR) AST through a .mappings file. This process of reconstructing existing ASTs for the purposes of a new plugin will make the base for the functions that transform between VDM and PlantUML. However, in this example, it seems that one AST is made for the feature that is translating VDM to C. This poses the question of whether it will be necessary to create ASTs corresponding to the languages involved, or corresponding to the translations that happen between the languages.

\subsection{Transformation Rules}

In the VDM++ book called Validated Designs for Object-oriented Systems, 12 mapping rules between UML class diagrams and VDM++ are presented \cite{Fitzgerald&05}. These serve as requirements for the translator between the OO VDM dialects and PlantUML. In Tab. \ref{tab:VDMtoPUML1-3}, \ref{tab:VDMtoPUML4-7}, and \ref{tab:VDMtoPUML8-12}, an example in VDM++ and PlantUML are presented along with the transformation rules.

\clearpage
\begin{table}[h!]

%
%
%
%
\includegraphics[width=\textwidth]{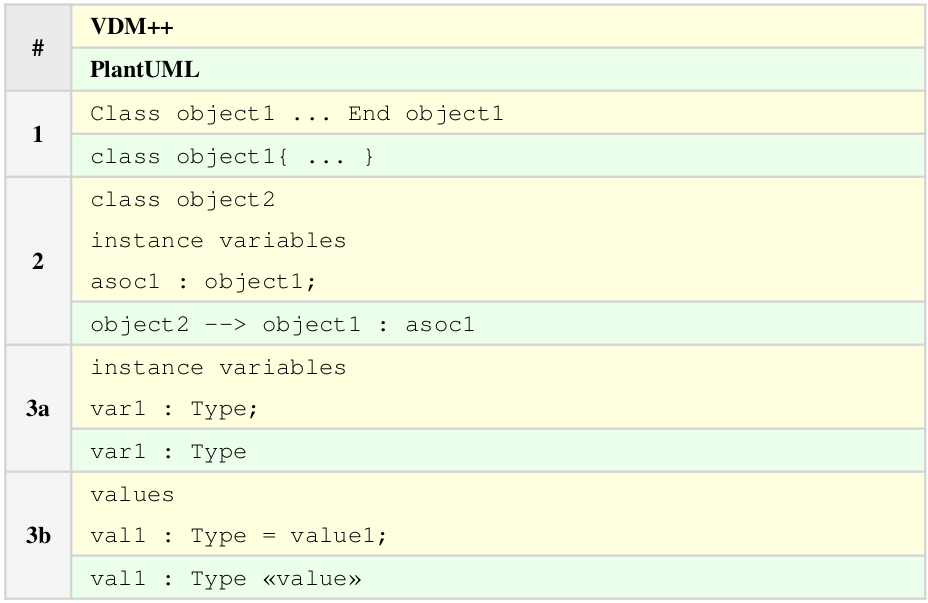}
\caption{VDM++ to PlantUML Transformation Mapping Table \#1-3}
\label{tab:VDMtoPUML1-3}
\end{table}
\vspace{-1.2cm}
\noindent
\subsubsection{1 --- Class Declarations}in both views are very simple, since "there is a one-to-one relationship between classes in UML and classes in VDM++", as stated in the original mapping rules \cite{Fitzgerald&05}. The ellipses in the PlantUML view surrounds the contents of the class and all class members are declared within them.

\subsubsection{2 --- Associations}must each be given a role name that denotes the direction of the association. This is represented in VDM++ as an instance variable with the type defined by the type of the destination class of the association. In PlantUML, the direction of an association is denoted by an arrow, followed by the destination. Associations in PlantUML are declared after class declarations. 

\subsubsection{3a --- Instance Variables}in the VDM++ view are equivalent to attributes in UML classes, and share a similar notation. 

\subsubsection{3b --- Value Definitions} can be differentiated from instance variables using a stereotype to indicate that they should be treated as constants. 

\clearpage
\begin{table}[h!]
    \includegraphics[width=\textwidth]{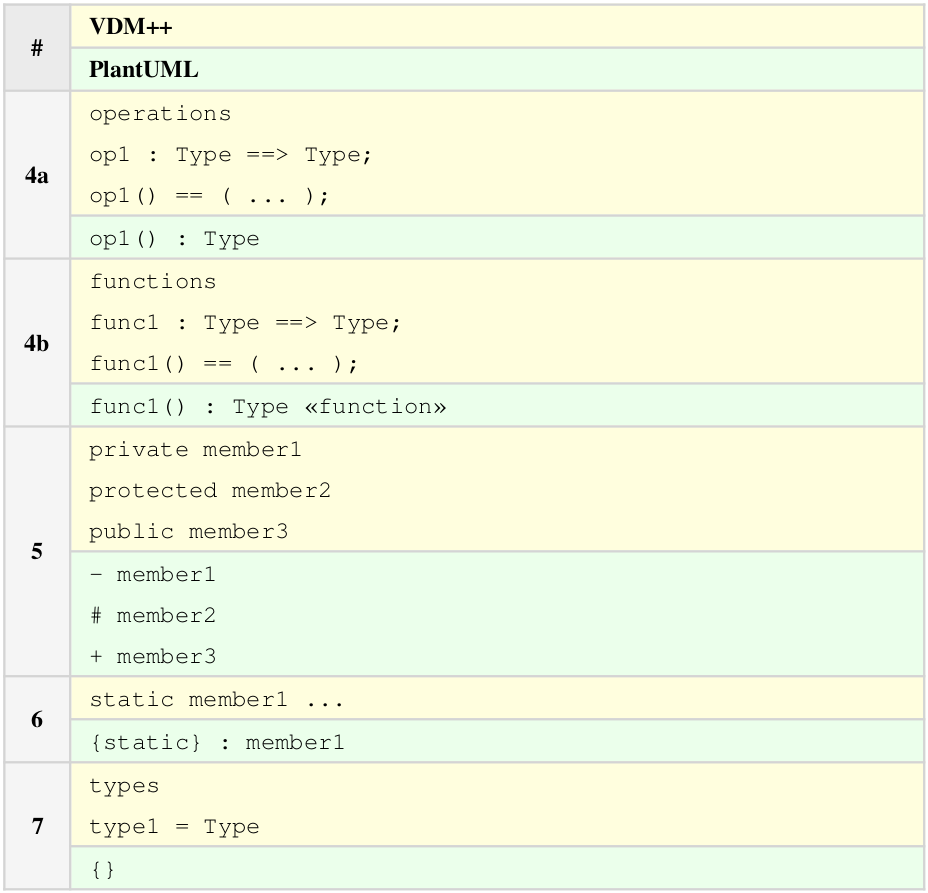}

\caption{VDM++ to PlantUML Transformation Mapping Table \#4-7}
\label{tab:VDMtoPUML4-7}
\end{table}

\vspace{-1.2cm}

\noindent
\subsubsection{4a --- Operations}are defined in both views using a type and a name.

\subsubsection{4b --- Functions} are differentiated from operations using a stereotype in the UML view. The type of function (total, partial, injection, etc.) is also specified in the stereotype.

\subsubsection{5 --- Access Modifiers}are available for all members in both views and are denoted in the member declaration. 

\subsubsection{6 --- The Static keyword}is used in both views, with a slight notational change.

\subsubsection{7 --- Types}have no equivalent in UML, and are therefore omitted. However, it may be possible to use stereotypes to declare a type in PlantUML.

\clearpage
\begin{table}[h!]
    \includegraphics[width=\textwidth]{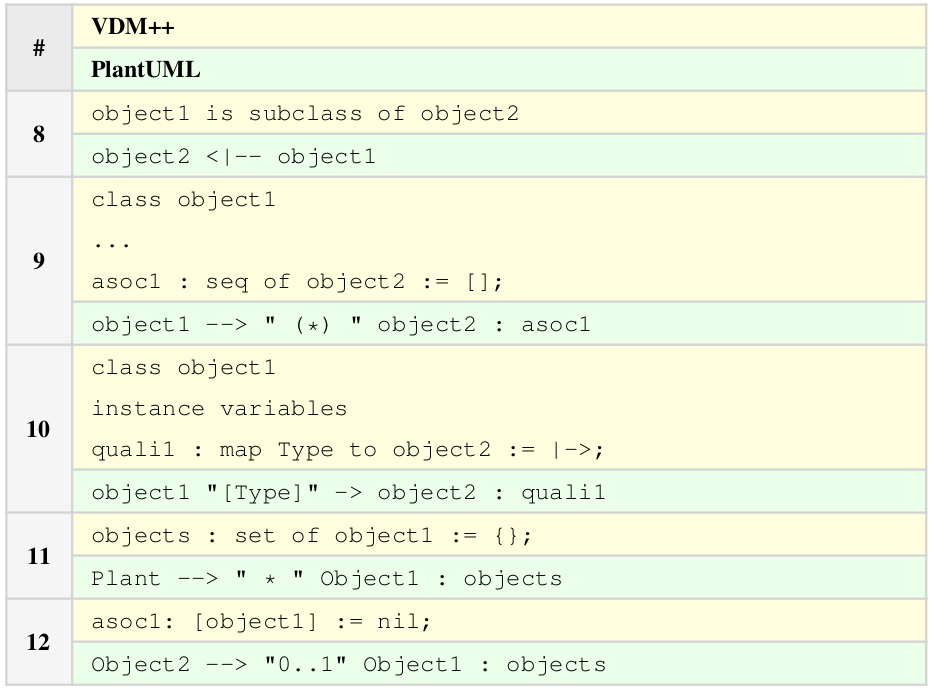}

\caption{VDM++ to PlantUML Transformation Mapping Table \#8-12}
\label{tab:VDMtoPUML8-12}
\end{table}
\vspace{-1.2cm}

\noindent
\subsubsection{8 --- Inheritance}between the two views have a one-to-one mapping. It is defined in PlantUML using an arrow head notation. Inheritance in PlantUML is declared after class declarations. 

\subsubsection{9 --- Association Multiplicity}in UML define the type of the corresponding VDM++ instance variable. An ordered zero to $n$ multiplicity corresponds to a VDM++ sequence with the type of the association destination object. An ordered multiplicity is denoted using parentheses.

\subsubsection{10 --- Qualified Associations}in VDM++ are instance variables that encapsulate the mapping from a qualifier type or qualifier class, to a class. In PlantUML the qualifier appears in square brackets on the side of the associating class, similar to how it is shown in Modelio using a smaller box bordering the class.

\subsubsection{11 --- Sets}in VDM++ come as a result of an unordered zero to $n$ multiplicity using the set type constructor.

\subsubsection{12 --- The Optional Type}constructor is used when a multiplicity of zero or one is used.

\clearpage
\noindent
We hereby see how PlantUML has the capability to fulfil these transformation rules, and is therefore a suitable host for UML visualisations of VDM models. Note that there may very well be edge cases, where some construct exists in VDM but is not defined in UML. It is difficult to consider all such cases, but one way is to do static analysis of the VDM model and warn the user, if they are about to transform some VDM that will not be available in the UML model.

\section{Future Work}
\label{sec:futurework}


Further progress is required before a UML visualisation tool is seamlessly integrated into VDM VS Code. Achieving this involves developing a prototype of the PlantUML translator by following the architecture laid out in this paper. This development will hopefully allow for a smooth bidirectional mapping between the OO dialects of VDM and PlantUML. The current overview only considers the VDM++ dialect, but it will also have to incorporate the VDM-RT dialect in the future.

The creation of the direct translator involves the use of class mapping to reconstruct the ASTs present on the VDMJ tool. The process involved is still alien to us, but help from and collaboration with Nick Battle and the rest of the team is expected in the future.

At the moment, a challenge in representing VDM models using PlantUML is the lack of a qualified association definition, even if workarounds exist \cite{plant_asoc}. A solution to this may involve collaborating with the creators of PlantUML allowing OO VDM models to be fully compatible with PlantUML. The current connection does not support VDM types in its class diagrams, but as PlantUML is malleable to specify custom members, it may be possible to represent VDM types in PlantUML class diagrams. 

Furthermore, there is still work to be done on the current implementation of the connection between VDM VSCode and UML. Transforming VDM files to UML is not supported for models that include libraries. Similarly, when transforming a UML file representing a VDM-RT model to VDM files, the files are converted to the VDM++ dialect. However both these problems are relatively simple to correct. 

In the long term, work can be done to enable continual updates on the bidirectional mapping between VDM and PlantUML. Another area for potential development is to work on the inherent information loss that occurs when transforming. Solving this would involve developing some way of encapsulating the functionality of the classes present in the model and pass this information along, together with the UML diagram information. It is also planned to perform static analysis of the models to make the tool more user friendly by warning whenever some VDM construct does not have a compatible UML counterpart.

\section{Conclusion}
\label{sec:conclusion}

This paper has described how the coupling between VDM and UML was established on VS Code. This was performed by repurposing the already existing connection on the Eclipse version of Overture through decoupling it from the Eclipse IDE and packaging the transformation functionality, along with all its dependencies, in a JAR file and integrating it into VS Code. 
\noindent
Steps have been taken to adapt the textually based open-source diagram tool, PlantUML, for UML visualisations. Together with its VS code extension, PlantUML offers a continually updating view of UML diagrams next to the natural language based code. Progress was then made to create a translator that can transform between VDM++ and PlantUML by deciding on the architecture of the translator and a prototype was made.



\subsubsection*{Acknowledgments}

We would like to thank all the stakeholders that have contributed to the Overture/VDM development but in particular Kenneth Lausdahl for the original UML mapping and Nick Battle for the establishment of VDMJ. 

 \newcommand{\noop}[1]{}

\clearpage
\endgroup

\begingroup
\include{customlangdef}
\lstset{basicstyle=\scriptsize,tabsize=2,frame=trBL,frameround=fttt}
\renewcommand\theHchapter{5-Pierce:\thechapter}
\renewcommand\theHsection{5-Pierce:\thesection}
\locallabels{5-Pierce:}
\setcounter{footnote}{0}
\setcounter{chapter}{0}
\setcounter{lstlisting}{0}

\makeatletter
\def\input@path{{5-Pierce/}}
\makeatother

\graphicspath{{5-Pierce/}}

\title{Speeding Up Design Space Exploration through Compiled Master Algorithms}
\titlerunning{Speeding Up DSE through Compiled Master Algorithms}

\author{Ken Pierce\inst{1} \and
  Kenneth Lausdahl\inst{2} \and
  Mirgita Frasheri\inst{2}
}
\institute{School of Computing, Newcastle University, United Kingdom \\\email{kenneth.pierce@ncl.ac.uk}
\and 
Dept. of Electrical and Computer Engineering, DIGIT, Aarhus University, Aarhus, Denmark
\\\email{kenneth@lausdahl.com,mirgita.frasheri@ece.au.dk}
}
\authorrunning{ }

\maketitle
\begin{abstract}

The design of Cyber-Physical Systems (CPSs) is complex, but can be mitigated through Model-Based Engineering. 
The models representing the components of a CPS are often heterogeneous, combining cyber and physical elements, and can be produced in different formalisms by engineers from diverse disciplines.
To assess system-level properties of a specific design, the joint behaviour of such components can be analysed through co-simulation, whereas different design alternatives can be compared through Design Space Exploration (DSE). 
Due to its combinatorial nature, the DSE can suffer from state-space explosion, where the number of combinations rapidly increases the time required to analyse them.
While careful experiment design and Genetic Algorithms can be used to reduce the number of simulations,
the benefit of speeding up co-simulations is clear. 
In this paper, we present an extension of the FMI-compliant Maestro co-simulation engine that generates a custom co-simulation Master algorithm as C code that can be compiled and run natively, reducing the overheads from the existing Java-based version. 
We apply this to a standard water tank case study and show an initial speed up of five times over the existing Maestro implementation.
\end{abstract}

\section{Introduction}

Cyber-Physical Systems (CPSs) are systems constructed of interacting hardware and software elements, with components networked together and distributed geographically~\cite{Lee08}. The development, deployment and maintenance of CPSs requires multiple disciplines to work together, which is often hampered by the organisational, social and technical barriers between engineering disciplines, including use of different terminologies, techniques and tools. Modelling such systems, and analysing these models with techniques such as simulation, are increasingly used in the development of CPSs. Combining models from different disciplines to create system-level ``multi-models'' has been shown as one way to bring together these disciplines and to permit better analysis of CPSs~\cite{Fitzgerald&14c}.

INTO-CPS~\cite{Larsen&17a} is a tool chain for model-based design of CPSs based around this concept of multi-models, using the Functional Mock-up Interface (FMI) standard~\cite{FMIStandard2.0.1} (FMI). The FMI standard provides a way for individual models to be packaged as Functional Mock-up Units (FMUs), with a standard interface for inputs and outputs. A multi-model therefore represents a CPS through a combination of FMUs representing the various components or parts of the system of interest, and the connections between them (inputs and outputs that influence other FMUs in the multi-model). A multi-model can be analysed in various ways including static checking and model checking. The most common way is through \emph{co-simulation}, where the FMUs are executed simultaneously under the control of a \emph{Master Algorithm (MA)} that is responsible for advancing (simulated) time and passing inputs and outputs between FMUs. The INTO-CPS tool chain is centered around a co-simulation engine called Maestro~\cite{Thule&19} that provides a variety of Master algorithms.

One main benefit of modelling CPSs is that it enables investigation of different designs before committing time and resources to physical prototypes. A design is characterised by its \emph{design parameters}, which are properties of the CPS that affect its behaviour (both physical properties and those of software). The set of all possible design parameters defines a \emph{design space}. \emph{Design Space Exploration} (DSE) is the act of assessing one or more areas of the design space by analysing and ranking a set of designs automatically, with the aim of helping the engineering team to make better informed decisions about which designs are most promising.

DSE is one feature of the INTO-CPS tool chain. INTO-CPS includes a set of Python scripts that allow an engineer to define a design space and to perform DSE through either exhaustive search (analysing all combinations of designs defined by a set of  parameters) or genetic search (iteratively selecting promising designs from an initial set). Even with trivial multi-models that co-simulate quickly, large DSEs with many combinations of parameters rapidly increase the required number of co-simulations making DSE time and resource intensive.

Given the pressures of industry, it is important that decisions on designs be made quickly and effectively. While DSE can give a lot of useful information, examining combinations of even a few parameters can take a long time. This paper describes work on improving the speed of co-simulation by generating an MA to native \CC code, which in turn speeds up co-simulation and allows a given design space to be explored faster, or a larger design space to be explored in a similar time.

The remainder of this paper is structured as follows. Section~\ref{sec:dse} describes the existing DSE framework in INTO-CPS. Section~\ref{sec:codegen} introduces the new code-generation feature. Section~\ref{sec:watertank} provides a refresher on the standard water tank case study. Section~\ref{sec:results} shows the performance of the new feature. Finally, Section~\ref{sec:conc} presents some conclusions and future work.

%
%

\section{Design Space Exploration with INTO-CPS}
\label{sec:dse}


Design Space Exploration (DSE) involves running multiple co-simulations, each of which represents a different design, and analysing the results in some form to provide the engineer with evidence for making design decisions about which designs are most promising~\cite{Gamble&14}. A common way to perform a DSE is to run multiple simulations and sweep across one or more design parameters. Each design is characterised by a set of design parameters, which do not change during a co-simulation, but take different values for each design~\cite{Fitzgerald&14c}. 

Defining a DSE then involves selecting which parameters will be changed, what values will be explored, and in what combination. Design parameters could be numerical, in which case could be defined through minimum, maximum and step size, or they could be defined as an enumerated set of possible values. Combinations can be constrained, such as ensuring that one parameter is always greater than another. The number of possibilities defined by a given combination defines the size of the design space for a given DSE, and hence the number of co-simulations that must be run. 

After a co-simulation has run, the results are  saved as Comma-Separated Values (CSV) files, and can be analysed. This is accomplished by defining one or more objective functions, which represent metrics by which the design is judged. An objective function can be defined programmatically, for example by providing a Python script that computes a score from the CSV and any additional data, such as details of the scenario. Once all co-simulations in a DSE have run, the designs can be ranked based on the objectives. 

A clear issue with DSE is state space, or in this case design space, explosion. As the number of parameters and values increases, the number of combinations increases exponentially. As mentioned below, the main ways to explore design spaces further using the same compute and time resources are:

\begin{enumerate}
    \item To perform careful experiment design in order to focus on the most likely outcomes, for example by selecting combinations that demonstrate the most important parameters~\cite{Gamble&14}; 
    \item To explore the design space iteratively by running only some co-simulations initially, then deciding based on those results which are the most promising designs. This can be achieved automatically through genetic algorithms, for example~\cite{Rose&Fitzgerald21}; or
    \item To speed up the co-simulation itself by optimising the FMUs and/or Master algorithm.
\end{enumerate}

The DSE features of the INTO-CPS tool chain take the form of a set of Python scripts written to use the Maestro co-simulation engine to carry out experiments with multiple co-simulation runs. These scripts were originally developed in the INTO-CPS project~\cite{INTOCPSD4.3d} using Python 2.7. These original Python scripts included an exhaustive search which computes all possible combinations of parameters given (excluding those that don’t meet the constraints). The scripts also contained a basic genetic algorithm to select which designs to run after performing analysis at each step in order to search down and reduce the overall number of simulations 

Since the INTO-CPS project has finished, a number of people have improved upon them. The most recent version now works with Python 3 and includes support for parallelization~\cite{Rose&Fitzgerald21}. Other recent papers have also looked at using libraries to achieve similar results with particle swarm optimization and simulated annealing~\cite{Stanley&Pierce21}. While different designs could also include different FMUs, the current scripts do not support sweeping across different FMUs, since care must be taken to understand what should happen if those have diverse design parameters.

\lstdefinestyle{bashstyle}{
frame=single,
basicstyle=\footnotesize\ttfamily,  caption=dd,label=lst:mabl,
keepspaces=true,
captionpos=b,
language=bash,
morekeywords={},
commentstyle=\color{mygray},
numbers=left,        
numbersep=5pt,}

\section{Code Generation Extension} 
\label{sec:codegen}


Speeding up  co-simulation speed is crucial to the efficiency of DSE as it is directly linked to the size of the design space that can be explored in a feasible manner, as described in Section~\ref{sec:dse}. In general there are a number of ways to speed up a co-simulation:
\begin{enumerate}
    \item Increase the step size (so fewer co-simulation steps are taken);
    \item Reduce the complexity of the simulation (simplify the FMUs and their connections); or
    \item Optimise the Master Algorithm
\end{enumerate}

While the first two options have the largest potential for speeding up co-simulations, they both have the drawback that they require changes to the work already carried out, which may not be possible while retaining the required fidelity. Other circumstances, such as a need to use legacy models, can also limit the ability of the engineer to optimise the individual FMUs. Therefore, this paper focuses on the third option, which is a generic approach that significantly improves the co-simulation speed of the MA itself. 

Maestro\footnote{Maestro was formally known as the INTO-CPS Co-simulation Orchestration Engine or COE.}~\cite{Thule&17} is a tool which implements a co-simulation MA as a Java application with support of multiple orchestration strategies. The earlier versions (\textsf{maestro1}) had good performance at the time, however as shown in Figure~\ref{fig:maestro-perf}, performance degrades over time and incurs a high initial cost for each co-simulation run. Therefore a new internal architecture of Maestro was introduced by Thule et al.~\cite{10.1007/978-3-030-57506-9_5} (\textsf{maestro2}). This new version separates the construction of the MA from the specific co-simulation execution as shown in Figure~\ref{fig:maestro12-flow}. This was implemented as a extensible Java application that could construct an MA and includes an interpreter for running co-simulations. This version matches the behaviour of the initial Maestro implementation, but with better performance without degraded performance over time (seen also in Figure~\ref{fig:maestro-perf}). A main factor for the improved performance is that all relations between signals are resolved during specification generation and not during run-time. Listing~\ref{lst:gen:mabl} illustrates how Maestro can be used to produce the intermediate representation (mabl) of MA and how that can be interpreted.

\begin{figure}[tb]
    \centering
\begin{tikzpicture}[scale=0.66, every node/.style={transform shape},
roundnode/.style={circle, draw=green!60, fill=green!5, very thick, minimum size=7mm},
squarednode/.style={ draw, minimum size=1cm, thick, fill=white, rounded corners},
]
\node[squarednode]      (m1-parse)                              {Process \texttt{mm/coe.json}};
\node[squarednode]      (m1-interpret) [right=of m1-parse]                              {Interpret};
\node (m1) [left=of m1-parse] {Maestro 1};

\node[draw,rectangle,dashed,fit=(m1) (m1-parse) (m1-interpret) ,inner sep=10pt] (a) {};

\draw[->] (m1-parse.east) -- (m1-interpret.west);

\node[squarednode]      (m2-parse)  [below=of m1-parse]          {Process \texttt{mm/coe.json}};
\node[squarednode]      (m2-generate) [right=of m2-parse]        {Generate MABL};
\node[squarednode]      (m2-write)    [below=of m2-generate]     {Write mabl.spec};
\node[squarednode]      (m2-interpret) [right =20mm of m2-generate]  {Interpret specification};
\node[squarednode]      (m2-cmake)  [below =of m2-interpret ]     {Create C++ program};
\node[squarednode]      (m2-compile)  [below =of m2-cmake]     {Compile};
\node[squarednode]      (m2-execute)  [right =of m2-compile]     {Execute};
\node (m2) [left=of m2-parse] {Maestro 2};

\draw[->] (m2-parse) -- (m2-generate);
\draw[->] (m2-generate) -- (m2-write);
\draw[->] (m2-generate) -- (m2-interpret);
\draw[->] (m2-generate) -- (m2-cmake);
\draw[->] (m2-cmake) -- (m2-compile);
\draw[->] (m2-compile) -- (m2-execute);
\draw [->,decoration={
            text along path,
            text={External Parameters},
            text align={center},
            raise=-0.2cm},decorate,draw] (m2-execute.south)arc(-160:135:1.2) ;
\draw [->,dotted] (m2-execute.south)arc(-160:140:1) ;
\node (m2-lower-left)[below right= 25pt and 40pt of m2-execute.south east,anchor=west] {};
\node[draw,rectangle,dashed,fit=(m2) (m2-lower-left)  ,inner sep=10pt] (a) {};
\end{tikzpicture}
    \caption{Maestro 1 and 2 execution flow}
    \label{fig:maestro12-flow}
\end{figure}
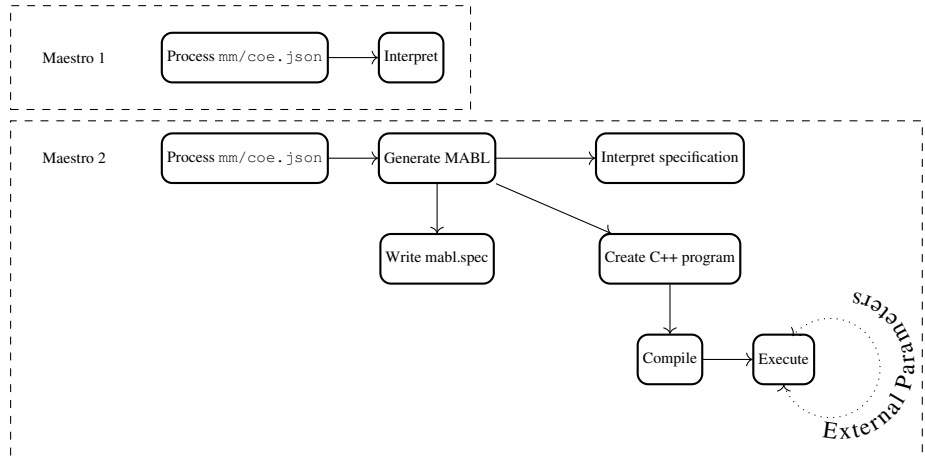

\clearpage

\begin{lstlisting}[style=bashstyle,caption=Creation of MA and interpretation,label=lst:gen:mabl,]
# Generate the mabl spec
java -jar $mabl import sg1 -fmu-search-path FMUs -output . \
    Multi-models/mm/co-sim-51/mm.json \
    Multi-models/mm/co-sim-51/co-sim-51.coe.json

# Interpret the spec using JAVA
java -jar $mabl interpret -runtime spec.runtime.json spec.mabl
\end{lstlisting}

\begin{lstlisting}[style=bashstyle,caption=Generation and execution of native simulation from of MA,label=lst:gen:cpp,]
# Generate the mabl spec
java -jar $mabl import sg1 -fmu-search-path FMUs -output . \
    Multi-models/mm/co-sim-51/co-sim-51.coe.json \
    sim-dse/mm.json 

# Interpret the spec using JAVA
java -jar $mabl interpret -runtime spec.runtime.json spec.mabl

# Generate native cpp simulator
java -jar $mabl export cpp -output cpp \
     -runtime spec.runtime.json \
     spec.mabl

# CMake
cmake -Bcpp/program -Scpp

# Compiling
make -Ccpp/program -j9

# Run native simulation
./cpp/program/sim -runtime spec.runtime.json
\end{lstlisting}

The INTO-CPS project defines two types of files to configure a co-simulation, \texttt{mm} and \texttt{coe}. These are JSON (Javascript Object Notation) files that define the FMUs, their connections, and co-simulation properties. The \texttt{spec.runtime.json} defines properties external  to the co-simulation, such as the path to the output file.

In this paper we extend this work with the ability to make parameters external to the MA. This allows reuse of the MA across DSE runs by storing the parameters in the \texttt{spec.runtime.json} configuration. This enables the MA to be converted to a \CC application, thus removing the overhead of a Java interpreter and only having the compilation overhead once per exploration by effectively generating a custom, native MA for a given co-simulation. This significantly improves the execution time but adds a one-time extra compilation overhead which on average takes longer than starting the Java process used for interpretation. The compilation overhead is directly related to the systems CMake performance capabilities and the download of external libraries, where the latter can be avoided if preinstalled. Once the make files for the native MA are ready for compilation, it is very fast to compile the native MA source file or any further MA source files. Because of this it is clear that the benefit of this native MA increases with the design space size.

The new converter is implemented as an extension to Maestro and thus runs in Java. The converter outputs a CMake project. The project includes a Maestro-specific library for FMI and external dependencies to \texttt{libzip} and \texttt{rapidjson} and a single generated \texttt{co-sim.cxx} file representing the current MA.
The CMake initialisation will fetch all dependencies and the subsequent compilation will compile all into a single executable \texttt{sim}. This process can be seen in Listing~\ref{lst:gen:cpp}. It consumes the same input files as the interpreter on line 7 from Listing~\ref{lst:gen:mabl}. This corresponds the lower right part of Maestro 2 in Figure~\ref{fig:maestro12-flow}. 

\begin{figure}[tb]
\centering
\includegraphics[width=0.8\textwidth]{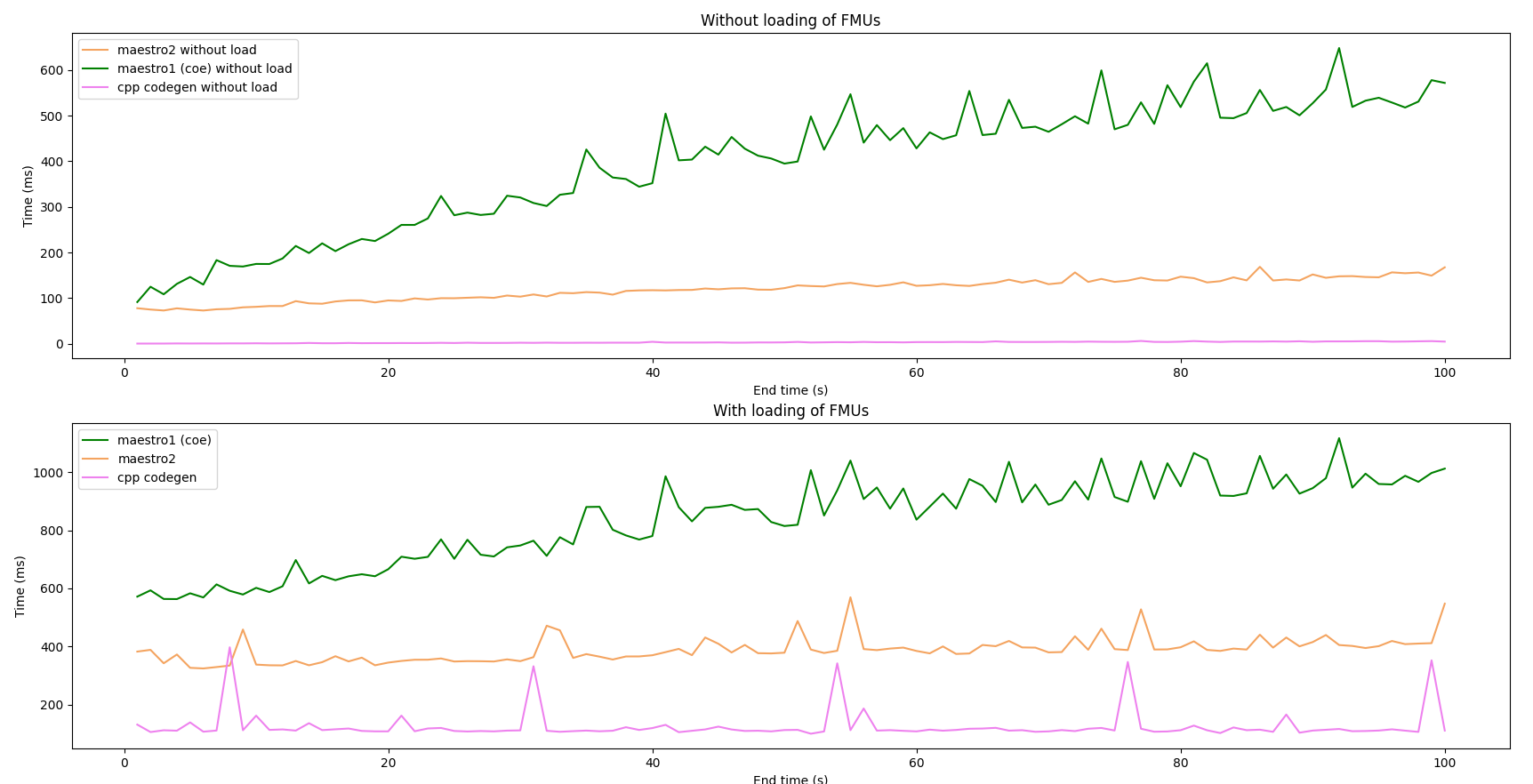}
\caption{Performance comparison between Meastro (COE), the new Meastro and the native \CC program.}
\label{fig:maestro-perf}
\end{figure}

The CMake project supports all major FMI target platforms Linux, MacOS and Windows using MSYS with mingw. For DSE purposes the process can be further improved by reusing a CMake generated project for subsequent explorations to avoid running the slow CMake initialisation and compilation of the external zip library. The only change for a new generation with the same version of the tool will be \texttt{co-sim.cxx}. Re-compilation of \texttt{co-sim.cxx} is significantly faster than re-running CMake.

\section{Case Study: Single-tank Water Tank}
\label{sec:watertank}

The single-tank water tank example~\cite{INTOCPSD3.6} is used as a case study in this paper to demonstrate the speedup of the proposed approach. 
The system, is composed of two FMUs, namely a water tank and a controller FMU.
The water tank FMU models a physical water tank component, composed of a valve with two states (open/close) and a sensor that outputs the level of the water in the tank. 
The controller FMU represents the digital component that controls the behaviour of the water tank through the valve, based on the current level of water. 
\begin{figure}[tb]
    \centering
    \subfloat[\label{fig:watertank}]{\includegraphics[trim={20 0 20 0},clip,scale=1.8]{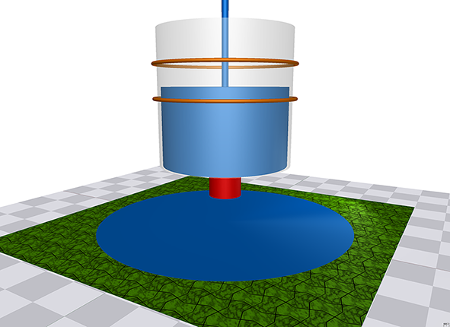}}\qquad
    \subfloat[\label{fig:wt-behaviour}]{\begin{tikzpicture}
\begin{groupplot}[group style={group size=1 by 1, horizontal sep=1.5cm},
                    title={Water-tank Controller Behvaiour}, 
                    set layers,
                    height=5.cm,width=6.cm,
                    xlabel={t}, 
                    title style={ align=center},
                    legend style={at={(axis description cs:0.5,-0.4)},anchor=south,legend columns=-1},
                    minor tick num=1,
                    ]
\nextgroupplot[ylabel={Level},clip=false]
\addplot[no marks, mycolor1] table [x=time, y=crtlInstance.valve, col sep=comma]{outputs_wt_normal.csv};
\addplot[no marks, mycolor4] table [x=time, y=wtInstance.level, col sep=comma] {outputs_wt_normal.csv};
\addlegendentry{$C_{state}$}
\addlegendentry{$WT_{level}$}
\draw [<-, >=stealth',thick] ($(62,1)$)  node[yshift=8pt, xshift=14pt]{$l_{min}$} -- ($ (77,1)$);
\draw [<-, >=stealth', thick] ($(62,2)$)  node[yshift=8pt, xshift=14pt]{$l_{max}$} -- ($ (77,2)$);
\end{groupplot}
\end{tikzpicture}}
    \caption{Visualisation of the water tank (left) and controller behaviour (right) over co-simulation time $t$ (adapted from~\protect\cite{frasheri2021}), where $C_{state}$ refers to the state of the controller, $1$ for open, 5$0$ for closed, and $WT_{level}$ refers to the water level in the tank)}
\end{figure}


The logic of the controller is straightforward: it attempts to keep the water level between a defined minimum and maximum at all times (visible in Figure~\ref{fig:watertank}). 
Should the water level go below a user defined minimum $l_{min}$, the controller closes the valve.
In case the level of water goes above a user defined maximum $l_{max}$, the controller opens the valve.
The behaviour of the system is depicted in Figure~\ref{fig:wt-behaviour} for $l_{min}=1$ and $l_{max}=2$. The minimum and maximum levels are used as design parameters in the DSE, with the constraint that minimum must be strictly below the maximum, $l_{min} < l_{max}$.



\section{Results}
\label{sec:results}

To investigate the speed up with the native MA over the existing Java version of Maestro, a set of DSEs were run with differing design space sizes (i.e.\ number of combinations and hence co-simulations) and simulated durations (i.e.\ number of co-simulation steps needed to complete the co-simulation). The water tank multi-model introduced above served as the design, using native C FMUs for both the water tank (singlewatertank-20sim.fmu) and controller (watertankcontroller-c.fmu). 
Version 0.4.1\footnote{\texttt{\href{https://www.github.com/INTO-CPS-Association/dse\_scripts/releases/tag/0.4.1}{github.com/INTO-CPS-Association/dse\_scripts/releases/tag/0.4.1}} (January 2021)} of the DSE scripts were used as the baseline. For the native MA run, the scripts were modified to call the compiled executable instead of calling the COE. This required a simple change to the \texttt{Common.py} scripts, shown in Listing~\ref{lst:python}. Additional small changes were needed in \texttt{Output\_CSV.py} and \texttt{Output\_HTML.py} files to handle different naming conventions in the results generated by the experimental native MA. 

The changes made however are not robust and did not include calls to generate or compile the code. To fully integrate the option for native compilation, the scripts need to be refactored, ideally with a layer of abstraction to handle the two paradigms transparently to the higher-level DSE algorithm scripts. For example, various options for the co-simulation are passed at run-time in the existing version (such as the co-simulation end time) but are required at compile time in the native \CC version. 

\begin{lstlisting}[basicstyle=\ttfamily, keywordstyle=\bfseries,language=Python,caption=Calling of the native co-simulation in the DSE scripts,label=lst:python]
runTimeJson = {}
runTimeJson["environment_variables"] = 
  parsedMultiModelJson["parameters"]
runTimeJson["DataWriter"] = [  
  {
    "filename": os.path.join(simFolderPath, "results.csv"), 
     "type": "CSV"
  }
]
jsonOutput = json.dumps(runTimeJson, 
  sort_keys=True, indent=4, separators=(",", ":"))
jsonOutputFile = open(filePath, 'w')
jsonOutputFile.write(jsonOutput)
jsonOutputFile.close()
subprocess.run(
  ["../sim-dse/cpp/program/sim.exe", "-runtime", filePath])
\end{lstlisting}

 To try and estimate the scalability of both approaches, a range of orders of magnitude were chosen for both the size of the design space and end time of the co-simulations:  
\begin{itemize}
    \item Design space size: 1, 10, 100, and 500 and 1000 combinations; and
    \item Simulated duration (s): 1, 10, 100, 1000 and 10000 simulated seconds.
\end{itemize}

While 1000 combinations was attempted for the Java version, this taxed the memory and disk capacity of the test computer and was abandoned after several attempts and crashes. The test machine was a laptop with an Intel\textsuperscript{\textregistered} Core\textsuperscript{\texttrademark} i7-5600U Processor (4M Cache, up to 3.20 GHz), 8GB DDR3L-12800 1600 MHz and an M2 solid state drive running Windows 10 Pro (21H2). The native MA was compiled under the MSYS2 environment (a collection of tools for building native Windows app from a Bash-like environment) using gcc 11.2.0-10, make 4.3-1 and cmake 3.23.0-1. 

\begin{table}[tb]
\centering
\bgroup
\def\arraystretch{1.5}%
\setlength{\tabcolsep}{3pt}
\begin{tabular}{cccccccccc} \cline{2-10}
& \multicolumn{8}{c}{Design space size (\# of co-simulations)} \\ \cline{2-10}
& \multicolumn{2}{c}{1} & \multicolumn{2}{c}{10} & \multicolumn{2}{c}{100} & \multicolumn{2}{c}{500} & 1000 \\ \hline
End time (s) & Java & Native & Java & Native & Java & Native & Java & Native & Native \\ \hline
1       & 5.00 & \cellcolor{native}0.46 & 35.60 & \cellcolor{native}4.14 & 264.64 & 43.50 & 1239.41 & 238.85 & 487.42 \\
10      & 5.47 & \cellcolor{native}0.57 & 37.10 & \cellcolor{native}4.13 & 289.33 & 44.42 & 1290.90 & 237.45 & 493.73 \\
100     & 6.50 & \cellcolor{native}0.48 & 39.48 & \cellcolor{native}4.21 & 271.24 & 45.86 & 1361.09 & 251.42 & 507.28 \\
1000    & 10.37 & \cellcolor{native}0.61 & 41.84 & \cellcolor{native}5.88 & 329.07 & 63.00 & 1499.83 & 347.70 & 709.81 \\
10000   & 15.17 & \cellcolor{native}2.16 & 113.59 & \cellcolor{native}22.48 & 885.56 & 239.38 & 3459.32 & 1181.84 & 2406.97 \\ \hline
\end{tabular}
\egroup\\[1ex]
\caption{Total time (s) to explore design spaces with increasing numbers of designs and simulated duration for the existing Java implementation and native version. The shaded columns indicate where a Native run is slower when accounting for the one-time cost of generating and compiling the code.}
\label{tabl:times}
\end{table}

Table~\ref{tabl:times} shows the overall execution times as recorded by the DSE scripts for each combination of design space and co-simulation length. Generation, compilation and re-compilation time are not included in the timings since these are not incorporated into the DSE scripts during these experiments. These incur the following penalties:
\begin{itemize}
    \item Generate code from Maestro: 6 seconds
    \item Configure compilation with cmake: 2 minutes 34 seconds 
    \item Compilation with make: 50 seconds
    \item Re-compilation after changing simulation duration: 6 seconds
\end{itemize}

Therefore, approximately 160 seconds should be added to the times in the Native columns. This means that for a one-off DSE with a design space of less than 100, the existing Java implementation is faster (indicated by the red colour in the columns Table~\ref{tabl:times}). In practice, most engineers will run multiple DSEs on a single multi-model. For example, running a small DSE to check the objectives are being calculated correctly, then running again with different parameter sets and sweeps. In such cases, due to the negligible re-compilation time, the initial cost of generation and compilation is amortised as more DSEs are run.    


While the raw data is useful, it is easier to interpret after some processing. Table~\ref{tabl:speedup} shows the relative speed up of the native MA over the Java by dividing the Java execution time by the native execution time, and rounding to the nearest whole number. This shows that the native MA is significantly faster on single co-simulations, though the benefit is less on the `longer' co-simulations. It is not immediately clear why this is. Similarly, as the design space increases, the speed up settles to around five times faster for the native MA. Given however that it was not possible to complete a 1000-design DSE using the Java version on the test machine, there may well be other factors at play, since memory usage was not examined in this test.       

\begin{table}[tb]
\centering
\bgroup
\def\arraystretch{1.5}%
\setlength{\tabcolsep}{3pt}
\begin{tabular}{cccccc} \cline{2-5}
& \multicolumn{4}{c}{Design space size} \\ \cline{2-5}
End time (s) & 1 & 10 & 100 & 500 \\ \hline
1       & 11x   & 9x & 6x & 5x \\
10      & 10x   & 9x & 7x & 5x \\
100     & 13x   & 9x & 6x & 5x \\
1000    & 17x   & 7x & 5x & 4x \\
10000   & 7x    & 5x & 4x & 3x \\ \hline
\end{tabular}
\egroup\\[1ex]
\caption{Relative speed-up of native MA over the existing Java implementation.}
\label{tabl:speedup}
\end{table}

\begin{figure}[p!]
\centering
\includegraphics[width=\textwidth]{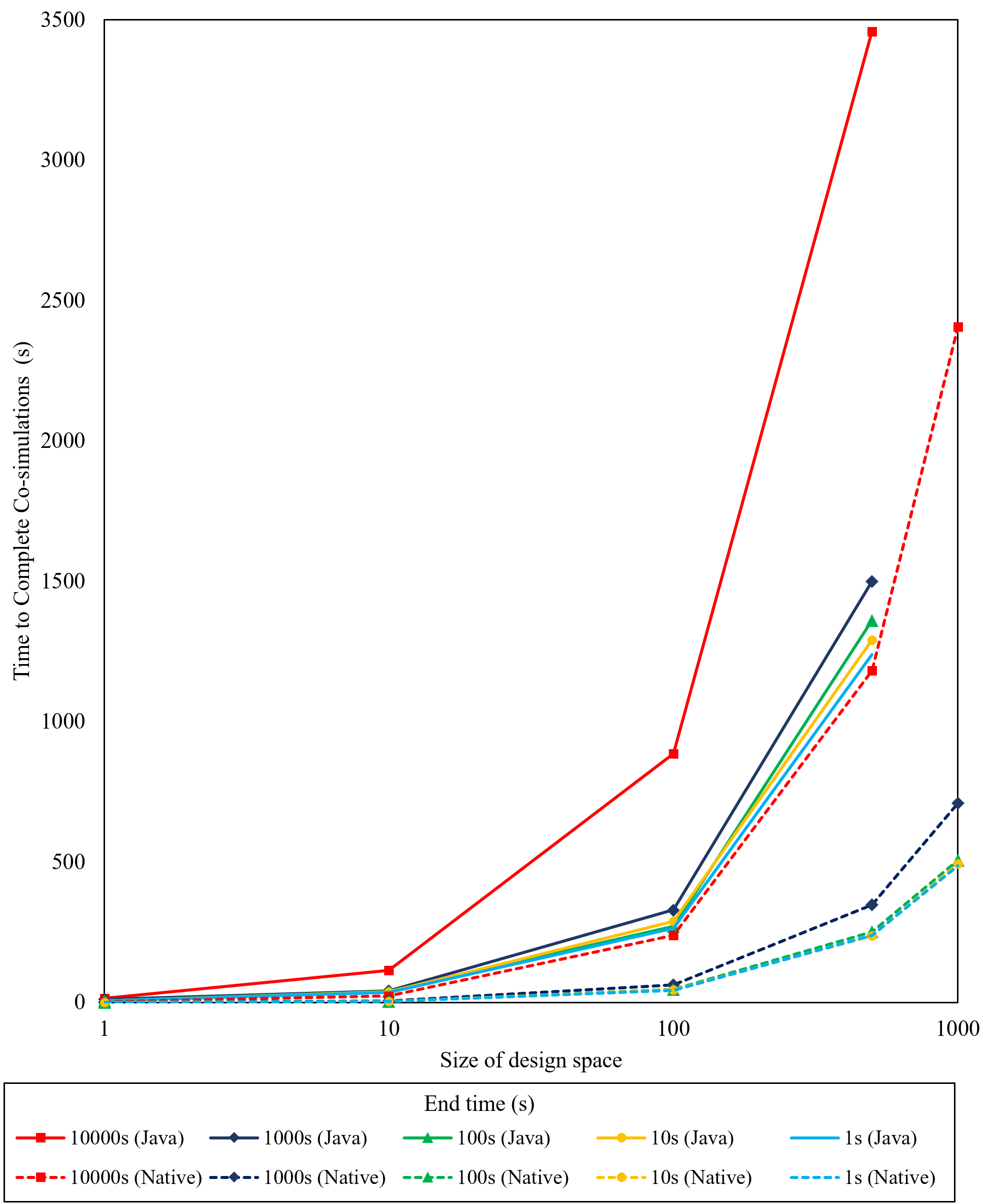}
\caption{Effect of design space size on time to complete DSE (using logarithmic x-axis).}
\label{fig:dse-size-plot}
\end{figure}

\begin{figure}[p!]
\centering
\includegraphics[width=\textwidth]{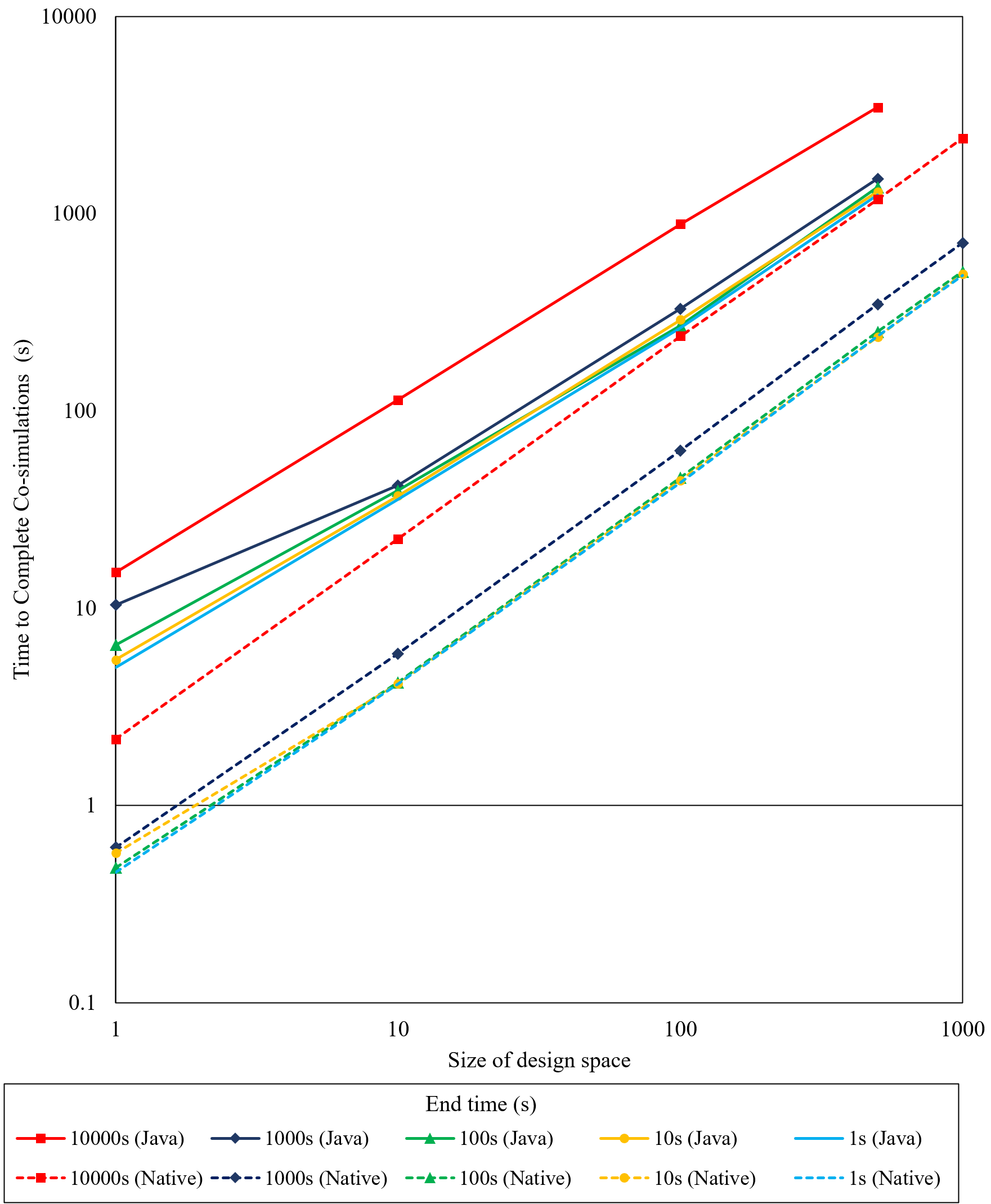}
\caption{Log-log plot showing effect of design space size on time to complete DSE.}
\label{fig:dse-size-plot-log}
\end{figure}

To further explore the data, they can be plotted on graphs. Given that there are two dependent variables, two graphs are plotted. The first is the effect of design space size on the time to complete the DSE, with the various end times as different series, shown in Figure~\ref{fig:dse-size-plot}. The second shows effect of end time on the time to complete a DSE, with the various sizes of design space as different series, shown in Figure~\ref{fig:co-sim-length-plot}. Both figures use a logarithmic x-axis. Similarly, in both figures, the Java version uses a solid line while the native MA uses a dashed line, then each pair has the same tick marks and colours.

Figure~\ref{fig:dse-size-plot} confirms that for any given design space, the native version is faster. It also highlights the growth of the Java version for the largest design space, and suggests that were a 1000-design DSE completed it would reach into the hours range. The log-log plot in Figure~\ref{fig:dse-size-plot-log} suggesting a clear power law relationship, however the data is not robust enough to draw any conclusions. The main takeaway from Figure~\ref{fig:co-sim-length-plot} is the significant uptick in time taken for the longest co-simulation. The log-log plot in Figure~\ref{fig:co-sim-length-plot-log} this is still apparent, so it would be useful to characterise this with more data.

Since this was only a single example with two FMUs, it would be important to try a range of examples and to run each multiple times to remove noise from the results. It is likely that there are some effects of the DSE scripts not being optimised for the native MA as well, since the changes made were experimental, there may well be superfluous files written to disk, for example. 

When considering the various caveats from these initial results however, a cautious conclusion is that the native MA improves DSE by an order of magnitude. That is to say, by switching to the native MA, a DSE can be 10-times larger and the co-simulations can be 10-times longer for the a given time and compute cost. This is certainly of benefit and warrants further work to both characterise the speed up more robustly, and to create DSE scripts that can handle generation, compilation and use of the native MA functionality.

\begin{figure}[p]
\centering
\includegraphics[width=\textwidth]{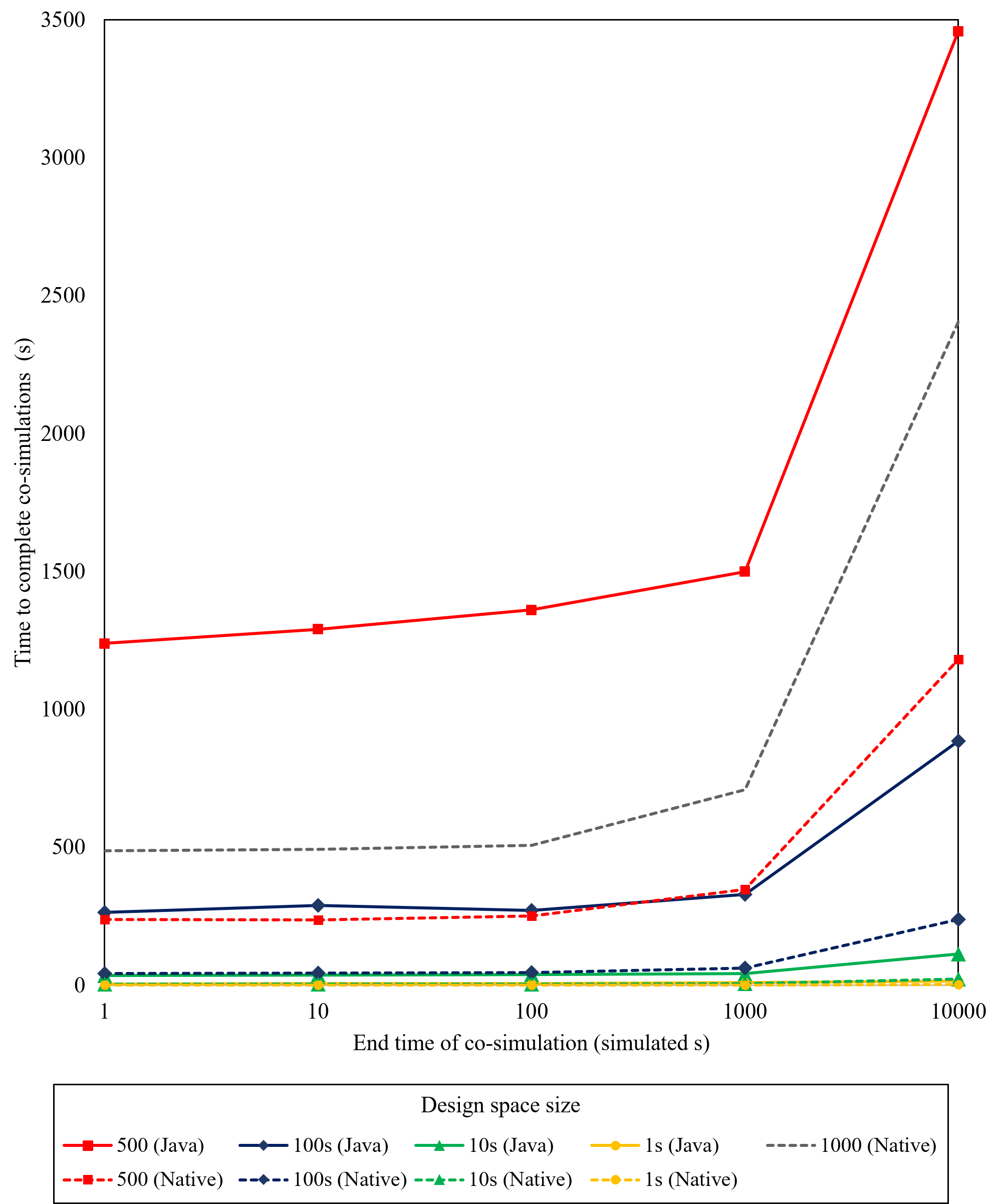}
\caption{Effect of end time on time to complete DSE (using logarithmic x-axis).}
\label{fig:co-sim-length-plot}
\end{figure}

\begin{figure}[p]
\centering
\includegraphics[width=\textwidth]{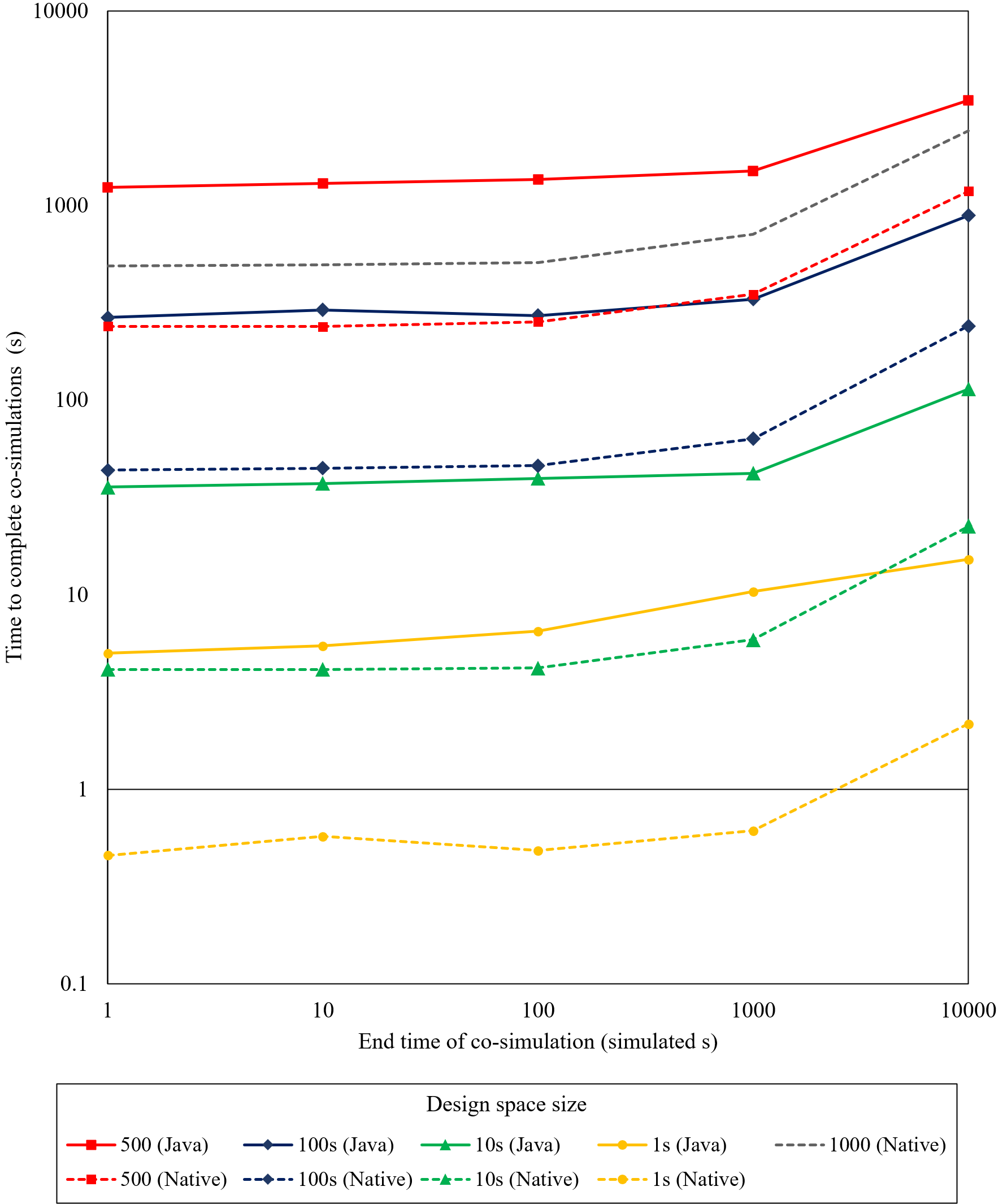}
\caption{Log-log plot showing effect of end time on time to complete DSE (using logarithmic x-axis).}
\label{fig:co-sim-length-plot-log}
\end{figure}

\section{Conclusions and Future Work}
\label{sec:conc}

In this paper we propose an approach that optimises the speed of execution of co-simulations.
This is extremely useful in the context of Design Space Exploration (DSE). In such a context, the number of co-simulations to be explored can be rather high, making the speed of execution critical for time-bounded projects. 
In our work, we consider the Maestro co-simulation engine, an extensible Java application, able to construct an Master Algorithm (MA), and thereafter interpret it to execute the actual co-simulation.
Our approach consists in converting the MA into a custom, native \CC application, thus reducing the overhead of the Java interpreter. 
The presented results show the benefit of such approach, however there are still some limitations related to the non-trivial setup, and the need for refactoring of the DSE scripts to allow for both types of execution, as well as open questions about the nature of the speedup achieved and relative scalability of the approach.
Nevertheless, the achieved speedup creates opportunities that extend beyond DSEs defined by design parameter sweeps.
Consider the case where for each FMU, there are a number of variants, and it is of interest to compare the behaviour of such variants in co-simulation.
This means that there will be several multi-models, on which individuals DSEs should be performed, while also comparing the results across alternatives. 
In such scenarios, the number of co-simulations to be executed would increase rapidly and the approach presented in this paper may make such wider DSEs a realistic prospect.
In future work, we will investigate such scenarios and validate the benefits from the approach proposed in this paper.

\section*{Acknowledgements}

We acknowledge the European Union's support for the INTO-CPS and HUBCAP projects (Grant Agreements 644047 and 872698).
In addition we would like to thank Innovation Foundation Denmark
for funding the AgroRobottiFleet project, and the Poul Due Jensen Foundation that funded our basic research for engineering of digital twins.

 \newcommand{\noop}[1]{}

\clearpage
\endgroup

\end{document}